\documentclass[%
reprint,
superscriptaddress,
nopreprintnumbers,
 amsmath,amssymb,
 showkeys,
 aps,
prx,
floatfix,
]{revtex4-2}

\usepackage[utf8]{inputenc}

\usepackage{graphicx}
\usepackage{dcolumn}
\usepackage{bm}
\usepackage{hyperref}

\usepackage[capitalize]{cleveref}
\usepackage[svgnames]{xcolor}
\usepackage{siunitx}[=v2]
\usepackage{textcomp}
\usepackage{soul}

\DeclareSIUnit\molar{\mole\per\cubic\deci\metre}
\DeclareSIUnit\Molar{\textsc{m}}
\DeclareSIUnit\bit{\textrm{bit}}

\renewcommand{\b}[1]{#1}
\let\oldhl\hl
\renewcommand{\hl}[1]{\PackageWarning{text-highlight}{#1}\oldhl{#1}}

\newcommand{\dfm}[0]{\delta\!f_m}

\begin{document}

\title{Path Weight Sampling: Exact Monte Carlo Computation of the Mutual Information between Stochastic Trajectories}

\author{Manuel Reinhardt}
\affiliation{AMOLF, Science Park 104, 1098 XG, Amsterdam, The Netherlands}
\author{Gašper Tkačik}
\affiliation{Institute of Science and Technology Austria, 3400 Klosterneuburg, Austria}
\author{Pieter Rein ten Wolde}%
\email{tenwolde@amolf.nl}
\affiliation{AMOLF, Science Park 104, 1098 XG, Amsterdam, The Netherlands}

\date{\today}

\begin{abstract}
Most natural and engineered information-processing systems transmit information via signals that vary in time. Computing the information transmission rate or the information encoded in the temporal characteristics of these signals, requires the mutual information between the input and output signals as a function of time, i.e. between the input and output trajectories. Yet, this is notoriously difficult because of the high-dimensional nature of the trajectory space, and all existing techniques require approximations. 
We present an exact Monte Carlo technique called \emph{Path Weight Sampling} (PWS) that, for the first time, makes it possible to compute
the mutual information between input and output trajectories for any stochastic system that is described by a master equation.
The principal idea is to use the master equation to evaluate the exact conditional probability of an individual output trajectory for a given input trajectory, and average this via Monte Carlo sampling in trajectory space to obtain the mutual information. We present three variants of PWS, which all generate the trajectories using the standard stochastic simulation algorithm.
While \emph{Direct PWS} is a brute-force method,
\emph{Rosenbluth-Rosenbluth PWS} exploits the analogy between signal trajectory sampling and polymer sampling,
and \emph{Thermodynamic Integration PWS} is based on a reversible work calculation in trajectory space.
PWS also makes it possible to compute the mutual information between input and output trajectories for systems with hidden internal states as well as systems with feedback from output to input.
Applying PWS to the bacterial chemotaxis system, consisting of 182 coupled chemical reactions, demonstrates not only that the scheme is highly efficient, but also that the number of receptor clusters is much smaller than hitherto believed, while their size is much larger.
\end{abstract}

\keywords{Complex Systems, Statistical Physics, Computational Physics}

\maketitle

\section{Introduction}

Quantifying information transmission is vital for understanding and designing natural and engineered information-processing systems, ranging from biochemical and neural networks, to electronic circuits and optical systems~\cite{2011.Tkacik,2016.Tkacik,2003.MacKay}. Claude Shannon introduced the mutual information and the information rate as the central measures of Information Theory more than 70 years ago~\cite{1948.Shannon}. These measures quantify the fidelity by which a noisy system transmits information from its inputs to its outputs. Yet, computing these quantities exactly remains notoriously difficult, if not impossible.
This is because the inputs and outputs are often not scalar values, but rather temporal \emph{trajectories}.

Most, if not all, information-processing systems transmit signal that vary in time. The canonical measure for quantifying information transmission via time-varying signals is the mutual information rate \cite{1948.Shannon,2006.Cover,2009.Tostevin,2014.Fiedor}. It quantifies the speed at which distinct messages are transmitted through the system, and it depends not only on the accuracy of the input-output mapping but also on the correlations within the input and output signals. 
Computing the mutual information rate thus requires computing the mutual information between the input and output trajectories, not between their signal values at given time points.
The rate at which this trajectory mutual information increases with the trajectory duration in the long-time limit defines the mutual information rate. In the absence of feedback this rate also equals the multi-step transfer entropy \cite{1990.Massey,2000.Schreiber}.

More generally, useful information is often contained in the temporal dynamics of the signal. A prime example is bacterial chemotaxis, where the response does not depend on the current ligand concentration, but rather on whether it has changed in the recent past \cite{1983.Block,1986.Segall}. Moreover, the information from the input may be encoded in the temporal dynamics of the output \cite{1995.Marshall,2013.Purvis,2014.Selimkhanov,2018.Granados}. Quantifying information encoded in these temporal features of the signals requires the mutual information not between two time points, i.e. the instantaneous mutual information, but rather between input and output trajectories \cite{2009.Tostevin}.

Unfortunately, computing the mutual information between trajectories is exceptionally difficult.
The conventional approach requires
non-parametric distribution estimates of the input and output distributions, e.g. via histograms of data obtained
through simulations or experiments \cite{1998.Strong,2003.Paninski,2011.Cheong,2008.Tkacik,2014.Tkacik,2021.Meijers}.
These non-parametric distribution estimates are necessary because the mutual information cannot generally be computed from summary statistics like the mean or variance of the data alone.
However, the high-dimensional nature of trajectories makes it infeasible to obtain enough empirical data to accurately estimate the required probability distributions.
Moreover, this approach requires the discretization of time, which becomes problematic when the information is encoded in the precise timing of signal spikes, as, e.g., in neuronal systems \cite{1999.Rieke}.
Except for the simplest systems with a binary state space \cite{2021.Meijers}, the conventional approach to estimate the mutual information via histograms therefore cannot be transposed to trajectories.

Because there are currently no general schemes available to compute the mutual information between trajectories exactly, approximate methods or simplified models are typically used. 
While empirical distribution estimates can be avoided by employing the K-nearest-neighbors entropy estimator \cite{2002.Kaiser,2004.Kraskov}, this method depends on a choice of metric in trajectory space and can become unreliable for long trajectories \cite{2019.Cepeda-Humerez}. \b{Alternative, decoding-based information estimates can be developed for trajectories \cite{2008.Gao}, but merely provide a lower bound of the mutual information, and it remains unclear how tight these lower bounds are~\cite{1999.Borst,2019.Cepeda-Humerez,2019.Hledik}. Analytical results are avaiable for simple systems \cite{2016.Thomas}, and for linear systems that obey Gaussian statistics, the mutual information between trajectories can be obtained from the covariance matrix \cite{2009.Tostevin}. } However, many information processing systems are complex and non-linear such that the Gaussian approximation does not hold, and analytical solutions do not exist.
A more promising approach to estimate the trajectory mutual information for chemical reaction networks has been developed by \citet{2019.Duso} and generalized in Ref.~\cite{2023.Moor}.
However, the scheme relies on a moment closure approximation and has so far only been applied to very simple networks, seemingly being difficult to extend to complex systems.

\b{Here, we present \emph{Path Weight Sampling} (PWS), an \emph{exact} technique to compute the trajectory mutual information for any system described by a master equation.}
Master equations are widely used to model chemical
reaction networks \cite{1940.Delbrueck,1963.McQuarrie,1964.McQuarrie,2000.Elowitz}, biological population growth \cite{1939.Feller,2010.Park,2012.Cremer}, economic processes \cite{1992.Weidlich,1995.Lux}, and a large variety of other systems \cite{2001.Helbing,2009.Castellano}, making our
scheme of interest to a broad class of problems.

PWS is an exact Monte Carlo scheme, in the sense that it provides an unbiased statistical estimate of the trajectory mutual information.
In PWS, the mutual information is computed as the difference between the marginal output entropy associated with the marginal distribution $\mathcal{P}[\bm{x}]$ of the output trajectories $\bm{x}$, and the conditional output entropy associated with the output distribution $\mathcal{P}[\bm{x}|\bm{s}]$ conditioned on the input trajectory $\bm{s}$. 
Our scheme is inspired by the observation from \citet{2019.Cepeda-Humerez} that the path likelihood, i.e. the probability $\mathcal{P}[\bm{x}|s]$, can be computed exactly from the master equation for a \emph{static} input signal $s$.
This makes it possible to compute the mutual information between a discrete input and a time-varying output via a Monte Carlo averaging procedure of the likelihoods, rather than from an empirical estimate of the intractable high-dimensional probability distribution functions. The scheme of \citet{2019.Cepeda-Humerez} is however limited to discrete input signals that do not vary in time.
Here we show that the path likelihood $\mathcal{P}[\bm{x}|\bm{s}]$ can also be computed for a dynamical input \emph{trajectory} $\bm{s}$, which allows us to compute the conditional output entropy also for time-varying inputs. 
While this solves the problem in part, the marginal output entropy associated with $\mathcal{P}[\bm{x}]$
cannot be computed with the approach of \citeauthor{2019.Cepeda-Humerez}, and thus requires a different scheme.

We show how, for time-varying input signals, the marginal probability $\mathcal{P}[\bm{x}]$ can be obtained as a Monte Carlo average of $\mathcal{P}[\bm{x}|\bm{s}]$ over a large number of input trajectories.
How to do this effectively is the crux of PWS. 
We then use the Monte Carlo estimate for $\mathcal{P}[\bm{x}]$ to compute the marginal output entropy.
We present three variants of PWS, all of which compute the conditional entropy in the same manner, but differ in the way this Monte Carlo averaging procedure for computing the marginal probability $\mathcal{P}[\bm{x}]$ is carried out. 

To compute $\mathcal{P}[\bm{x}]$, \emph{Direct} PWS (DPWS) performs a brute-force average of the path likelihoods $\mathcal{P}[\bm{x}|\bm{s}]$ over the input trajectories $\bm{s}$.
While we show that this scheme works for simple systems, the brute-force Monte Carlo averaging procedure becomes more difficult for larger systems and exponentially harder for longer trajectories. 

Our second and third variant of PWS are based on the realization that the marginal probability $\mathcal{P}[\bm{x}]$ is akin to a partition function. These schemes leverage techniques for computing free energies from statistical physics.
Specifically, the second scheme, \emph{Rosenbluth-Rosenbluth PWS} (RR-PWS), exploits the observation that the computation of $\mathcal{P}[\bm{x}]$ is analogous to the calculation of the (excess) chemical potential of a polymer, for which efficient methods have been developed \cite{1990.Siepmann,1997.Grassberger, 2002.Frenkel}.
The third scheme, \emph{Thermodynamic Integration PWS} (TI-PWS), is based on the classic free energy estimation technique of thermodynamic integration \cite{1984.Frenkel,1998.Gelman,2001.Neal} in conjunction
with a trajectory space MCMC sampler using ideas
from transition path sampling \cite{2002.Bolhuis}.

All three PWS variants make it possible to compute the mutual information between trajectories exactly, without the need of time-discretizing the input and output signals.

The DPWS method is presented in \cref{sec:sec2}, followed by a review of the required concepts from the theory of Markov jump processes and master equations. The other two PWS schemes are then presented as improvements of DPWS in \cref{sec:improvements}.

In \cref{sec:integrating-out} we show that, surprisingly, our PWS
methods additionally make it possible to compute the mutual
information between input and output trajectories of systems with
hidden internal states.  
Hidden states correspond, for example, to
network components that merely relay, process or transform 
the signal from the input to the output. 
Indeed, the downstream system typically responds to the information that is encoded in this output, and not the other internal system components. 
Most information processing systems contain such hidden states, and generally we want to integrate out these latent network components. In addition, we can generalize PWS to systems with feedback from the output to the input as shown in \cref{sec:feedback}.

In \cref{sec:case-studies} we apply PWS to two well-known model systems.
The first is a simple pair of coupled birth-death processes which allows us to test the efficiency of the three PWS variants, as well as to compare the PWS results with analytical results from the Gaussian approximation \cite{2009.Tostevin} and the technique by \citet{2019.Duso}. 
\b{Our second application concerns the bacterial chemotaxis system, which is arguably the best characterized signaling system in biology.
\citet{2021.Mattinglyg3a} recently argued that bacterial chemotaxis in shallow gradients is information limited.
Yet, to compute the information rate from their experimental data they had to employ a Gaussian framework.
PWS makes it possible to asses the accuracy of this approximation.
Our results show that the Gaussian framework is accurate in the regime of shallow concentration gradients as studied in \citet{2021.Mattinglyg3a}.
Yet, comparing the PWS predictions of a model based on previous literature against their experimental results reveals that the composition of the receptor array differs from that hitherto believed.}

\section{Monte Carlo Estimate of the Mutual Information\label{sec:sec2}}

\b{In this section we present the fundamental ideas of PWS. These ideas lie at the heart of DPWS and also form the foundation of the other two more advanced PWS variants which will be explained in subsequent sections.}

\subsection{Statement of the Problem}

\b{All information processing systems repeatedly take an input value $s$ and produce a corresponding output $x$. 
Due to noise, the output produced for the same input can be different every time, 
such that the system samples outputs from the distribution $\mathrm P(x|s)$.}
In the information theoretic sense, the device's capabilities are fully specified by its output distributions for all possible inputs. We consider the inputs as being distributed according to a probability density $\mathrm{P}(s)$ such that the whole setup of signal and device is completely described by the joint probability density $\mathrm{P}(s, x) = \mathrm{P}(s)\,\mathrm P(x|s)$.

\begin{figure}
    \centering
    \includegraphics{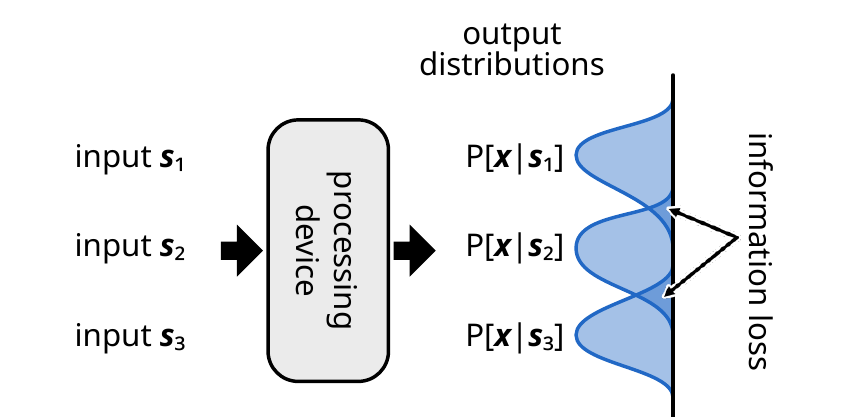}
    \caption{%
    Schematic of information processing under the influence of noise. Overlapping output distributions for different inputs lead to information loss, because the input cannot always be uniquely inferred from the output. The mutual information $\mathrm{I}(\mathcal{S},\mathcal{X})$ quantifies how much information the observation of the output typically retains about the input signal.
    }
    \label{fig:information}
\end{figure}

\b{When the conditional output distributions $\mathrm P(x|s)$ overlap with each other, information is lost because the input can not always be inferred uniquely from the output  (see \cref{fig:information}).} The remaining information that the output carries about the signal on average is quantified by the mutual information between input and output.

Mathematically, the mutual information between a random variable $\mathcal{S}$, representing the input, and a second random variable $\mathcal{X}$, representing the output, is defined as
\begin{equation}
    \mathrm I(\mathcal{S}, \mathcal{X}) = \iint \mathrm{d}s\,\mathrm{d}x\  \mathrm{P}(s, x) \ln \frac{\mathrm{P}(s, x)}{\mathrm{P}(s) \mathrm{P}(x)}\,,
    \label{eq:mutual_information}
\end{equation}
where the marginal output distribution is given by $\mathrm{P}(x) = \int \mathrm{d}s\ \mathrm{P}(s, x)$. The quantity $\mathrm I(\mathcal{S}, \mathcal{X})$ as defined above is a non-negative real number, representing the mutual information between $\mathcal{S}$ and $\mathcal{X}$ in nats. \b{The integrals in \cref{eq:mutual_information} run over all possible realizations of the random variables $\mathcal{S}$ and $\mathcal{X}$. In our case, $\mathcal{S}$ and $\mathcal{X}$ represent stochastic trajectories and so the integrals become path integrals.}

In general, the mutual information can be decomposed into two terms, a conditional and marginal entropy. Due to the symmetry of \cref{eq:mutual_information} with respect to exchange of $\mathcal{S}$ and $\mathcal{X}$, this decomposition can be written as
\begin{equation}
    \mathrm I(\mathcal{S}, \mathcal{X}) = \mathrm{H}(\mathcal{S}) - \mathrm{H}(\mathcal{S}|\mathcal{X}) = \mathrm{H}(\mathcal{X}) - \mathrm{H}(\mathcal{X}|\mathcal{S})\,.
    \label{eq:mutual_information_entropies}
\end{equation}
The (marginal) input entropy $\mathrm{H}(\mathcal{S})$ represents the total uncertainty about the input, and the conditional input entropy $\mathrm{H}(\mathcal{S}|\mathcal{X})$ describes the remaining uncertainty of the input after having observed the output. Thus, the mutual information $\mathrm{I}(\mathcal{S},\mathcal{X})=\mathrm{H}(\mathcal{S}) - \mathrm{H}(\mathcal{S}|\mathcal{X})$ naturally quantifies the reduction in uncertainty about the input through the observation of the output.

When analyzing data from experiments or simulations however, the mutual information is generally estimated via $\mathrm{I}(\mathcal{S},\mathcal{X})=\mathrm{H}(\mathcal{X}) - \mathrm{H}(\mathcal{X}|\mathcal{S})$.
\b{This is because simulation or experimental data generally provide information about the distribution of outputs for a given input, rather than vice versa.}
The accessible entropies are thus the marginal output entropy $\mathrm{H}(\mathcal{X})$ and the conditional output entropy $\mathrm{H}(\mathcal{X}|\mathcal{S})$, which are defined as
\begin{align}
    \mathrm{H}(\mathcal{X}) &= -\int\mathrm{d}x\ \mathrm{P}(x) \ln \mathrm{P}(x)
    \label{eq:marginal-entropy} \\
    \mathrm{H}(\mathcal{X}|\mathcal{S}) &= -\int\mathrm{d}s\ \mathrm{P}(s) \int\mathrm{d}x\  \mathrm{P}(x|s) \ln \mathrm{P}(x|s) \,.
    \label{eq:conditional-entropy}
\end{align}

\b{The conventional way of computing the mutual information involves generating many samples to obtain empirical distribution estimates for $\mathrm{P}(x|s)$ and $\mathrm{P}(x)$ via histograms.
However, the number of samples needs to be substantially larger than the number of histogram bins to reduce the noise in the bin counts.
Obtaining enough samples is effectively impossible for high-dimensional data, like signal trajectories.
Moreover, any nonzero bin size leads to a systematic bias in the entropy estimates, even in one dimension \cite{2003.Paninski}. 
These limitations of the conventional method make it impractical for high-dimensional data, highlighting the need for alternative approaches to accurately compute mutual information for trajectories.}

\subsection{\label{sec:algorithm} Direct PWS}

\begin{figure}
    \centering
    \includegraphics{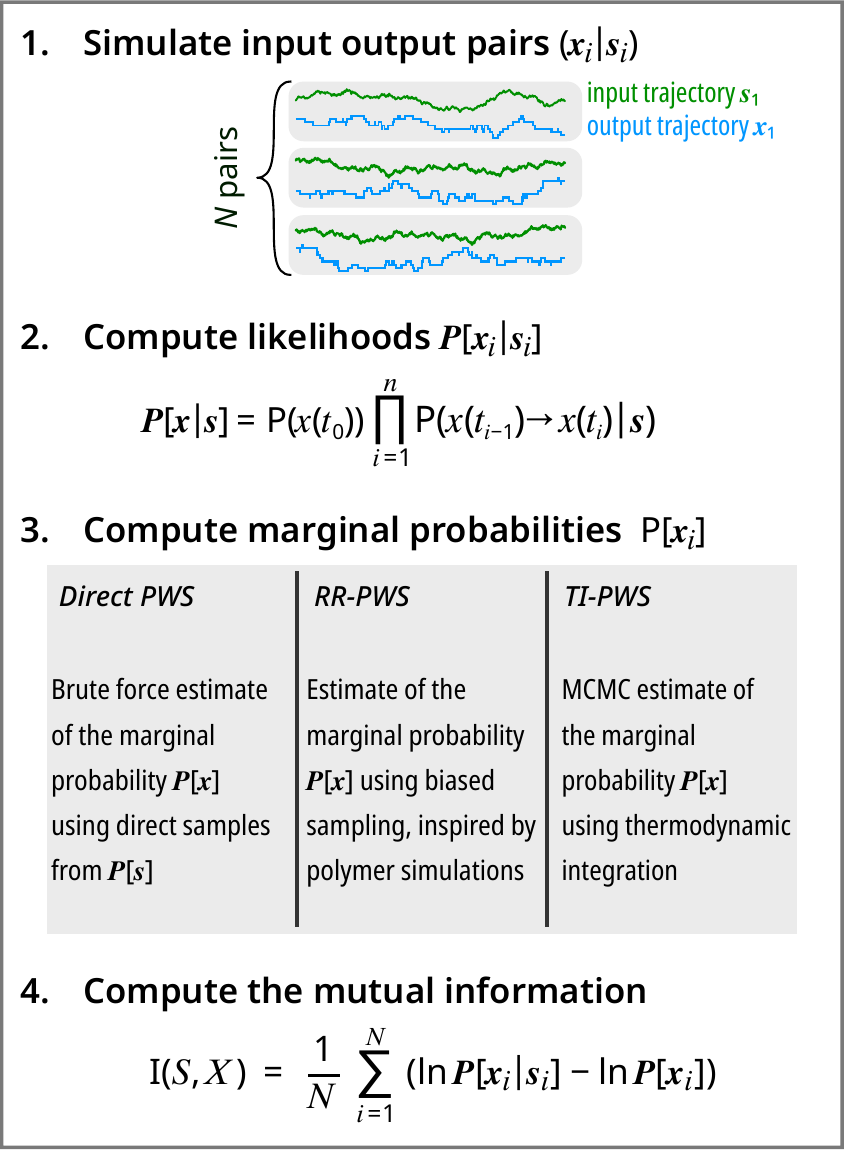}
    \caption{The PWS scheme to compute the mutual information between trajectories in 4 steps. \textbf{1}.~Generate $N$ input-output pairs from $\mathcal{P}[\bm{s},\bm{x}]$. \textbf{2}.~For each input-output pair compute the trajectory likelihood $\mathcal{P}[\bm{x_i}|\bm{s_i}]$ using \cref{eq:traj_prob}. \textbf{3}.~Compute $\mathcal{P}[\bm{x}_i]$ for every output. This step differentiates the different variants of PWS from each other. Direct PWS is presented in \cref{sec:algorithm}, whereas RR-PWS and TI-PWS are presented in \cref{sec:smc,sec:thermodynamic-integration}. \textbf{4}.~Using the likelihoods and the marginal probabilities from the previous steps we can estimate the mutual information using \cref{eq:average-of-differences}. }
    \label{fig:algorithm}
\end{figure}

\b{The central idea of PWS is to compute probability densities for trajectories exactly, sidestepping the problem having to estimate them via histograms. 
We exploit that for systems described by a master equation, the conditional probability of an output trajectory for a given input trajectory can be computed analytically.
With this insight we can derive a procedure to compute the mutual information.}
Specifically, we will show that
\begin{itemize}
    \item for a system described by a master equation, the trajectory likelihood $\mathcal{P}[\bm{x}|\bm{s}]$ is a quantity that can be computed on the fly in a stochastic simulation;
    \item input trajectories can be generated from $\mathcal{P}[\bm{s}]$, output trajectories for a given input $\bm{s}$ can be generated according to $\mathcal{P}[\bm{x}|\bm{s}]$ using standard SSA (Gillespie) simulations;
    \item by combining the two ideas above, we can derive a direct Monte Carlo estimate for the mutual information $\mathrm{I}(\mathcal{S},\mathcal{X})$, as illustrated in \cref{fig:algorithm}.
\end{itemize}
Note that we denote trajectories by bold symbols to distinguish them from scalar quantities.

Our technique is conceptually straightforward.
Using Monte Carlo simulations we can compute averages over the configuration space of trajectories. Suppose we have a function $f[\bm{z}]$ that takes a trajectory $\bm{z}$ and produces a scalar value. The mean of $f[\bm{z}]$ with respect to the trajectory distribution $\mathcal{P}[\bm{z}]$ is then
\begin{equation}
\langle f[\bm{z}] \rangle_{\mathcal{P}[\bm{z}]}\equiv\int\mathcal{D}[\bm{z}]\, \mathcal{P}[\bm{z}]f(\bm{z}) \,.
\end{equation}
We write $\int\mathcal{D}[\bm{z}]$ to denote a path integral over all possible trajectories of a given duration.
We estimate $\langle f[\bm{z}] \rangle_{\mathcal{P}[\bm{z}]}$, by generating a large number of trajectories $\bm{z}_1,\ldots,\bm{z}_N$ from $\mathcal{P}[\bm{z}]$ and evaluating the corresponding Monte Carlo average
\begin{equation}
    \hat{f}_N = \frac{1}{N} \sum^N_{i=1} f(\bm{z}_i)
\end{equation}
which converges to the true mean in the limit $N\to\infty$.

\b{Specifically, we want to estimate the conditional and the marginal entropy to compute the mutual information.} Let us imagine that we generate $N$ input trajectories $\bm{s}_1,\ldots,\bm{s}_N$ from the distribution $\mathcal{P}[\bm{s}]$. Next, for every input $\bm{s}_i$, we generate a set of $K$ outputs $\bm{x}_{i,1},\ldots,\bm{x}_{i,K}$ from $\mathcal{P}[\bm{x}|\bm{s}_i]$. Then, the Monte Carlo estimate for the conditional entropy is
\begin{equation}
\begin{aligned}
    \mathrm{H}(\mathcal{X}|\mathcal{S}) &= -\int \mathcal{D}[\bm{s}]\ \mathcal{P}[\bm{s}]\int \mathcal{D}[\bm{x}]\ \mathcal{P}[\bm{x}|\bm{s}] \ln\mathcal{P}[\bm{x}|\bm{s}] \\ 
    &=-\left\langle \left\langle \ln\mathcal{P}[\bm{x}|\bm{s}] \right\rangle_{\mathcal{P}[\bm{x}|\bm{s}]} \right\rangle_{\mathcal{P}[\bm{s}]} \\
    &\approx -\frac{1}{N}\sum^{N}_{i=1}\frac{1}{K}\sum^{K}_{j=1} \ln\mathcal{P}[\bm{x}_{i,j}|\bm{s}_i] \,.
\end{aligned}
\label{eq:conditional-entropy-estimate}
\end{equation}
\b{Secondly, for a given output $\bm{x}$ we generate $M$ inputs $\bm{s}^\prime_1,\ldots,\bm{s}^\prime_M$ according to $\mathcal{P}[\bm{s}]$,} then we can obtain a Monte Carlo estimate for the marginal probability of the output trajectory $\mathcal{P}[\bm{x}]$:
\begin{equation}
\begin{aligned}
    \mathcal{P}[\bm{x}] &= \int\mathcal{D}[\bm{s}]\  \mathcal{P}[\bm{s}]  \mathcal{P}[\bm{x}|\bm{s}] \\
    &= \left\langle \mathcal{P}[\bm{x}|\bm{s}] \right\rangle_{\mathcal{P}[\bm{s}]} \\
    &\approx \frac{1}{M}\sum^M_{j=1} \mathcal{P}[\bm{x}|\bm{s}^\prime_{j}]\,.
\end{aligned}
    \label{eq:marginal-naive}
\end{equation}
The  estimate for the marginal entropy is then given by 
\begin{equation}
\begin{aligned}
    \mathrm{H}(\mathcal{X}) &= -\int\mathcal{D}[\bm{x}]\ \mathcal{P}[\bm{x}]\ln\mathcal{P}[\bm{x}] \\
    &= -\left\langle \ln\mathcal{P}[\bm{x}] \right\rangle_{\mathcal{P}[\bm{x}]}\\
    &\approx -\frac{1}{N}\sum^{N}_{i=1} \ln\mathcal{P}[\bm{x}_i] \\
    &\approx -\frac{1}{N}\sum^{N}_{i=1} \ln \left[ \frac{1}{M}\sum^M_{j=1} \mathcal{P}[\bm{x}_i|\bm{s}^\prime_{i,j}] \right]\,.
\end{aligned}
    \label{eq:marginal-entropy-estimate}
\end{equation}
In the last step we inserted the result from \cref{eq:marginal-naive}. In this estimate, the trajectories $\bm{x}_1,\ldots,\bm{x}_N$ are sampled from $\mathcal{P}[\bm{x}]$, i.e., by first sampling from $\mathcal{P}[\bm{s}]$ and then from $\mathcal{P}[\bm{x}|\bm{s}]$.
Finally, the mutual information is obtained by taking the entropy difference, i.e., $\mathrm{I}(\mathcal{S},\mathcal{X})=\mathrm{H}(\mathcal{X}) - \mathrm{H}(\mathcal{X}|\mathcal{S})$.

While this is the main idea behind PWS, it is computationally advantageous to change the order of operations in the estimate.
Specifically, computing the difference of two averages, leads to large statistical errors. 
We can obtain an improved estimate by reformulating the mutual information as a single average of differences:
\begin{equation}
\begin{aligned}
    \mathrm{I}(\mathcal{S},\mathcal{X}) &= \int\mathcal{D}[\bm{s}]\int\mathcal{D}[\bm{x}]\ \mathcal{P}[\bm{s},\bm{x}] \ln\frac{\mathcal{P}[\bm{x}|\bm{s}]}{\mathcal{P}[\bm{x}]} \\
    &=  \left\langle
    \ln\mathcal{P}[\bm{x}|\bm{s}] -
    \ln\mathcal{P}[\bm{x}]
    \right\rangle_{\mathcal{P}{[\bm{s},\bm{x}]}}\,.
\end{aligned}
\label{eq:average-of-differences}
\end{equation}

\b{This equation applies to all variants of PWS. They differ, however, in the way $\mathcal{P}[\boldsymbol{x}]$ is computed. In the brute-force version of PWS, called \emph{Direct PWS} (DPWS), we use \cref{eq:marginal-naive} to evaluate the marginal probability $\mathcal{P}[\boldsymbol{x}]$. DPWS indeed involves  two nested Monte Carlo computations, in which $N$ pairs $(\boldsymbol{s}_i, \boldsymbol{x}_i)$ are generated, and for each output $\boldsymbol{x}_i$, $M$ input trajectories $\{\boldsymbol{s}\}$ are generated from scratch to compute  $\mathcal{P}[\boldsymbol{x}]$.
  In \cref{sec:improvements} below, we will present two additional variants of PWS where the brute-force estimate of the marginal probability $\mathcal{P}[\bm{x}]$ is replaced by more elaborate schemes. That said, DPWS is a conceptually simple, straightforward to implement, and exact scheme to compute the mutual information.} 


\b{Having explained the core ideas of our technique above, we will continue this section with a review of the necessary concepts of master equations to implement PWS.} First, in \cref{sec:mjp}, we derive the formula for the conditional probability $\mathcal{P}[\bm{x}|\bm{s}]$ which lies at the heart of our technique. In \cref{sec:mjp,sec:input-statistics}, we discuss how trajectories are generated according to $\mathcal{P}[\bm{x}|\bm{s}]$ and $\mathcal{P}[\bm{s}]$, which are the remaining ingredients required for using DPWS.
Then, in \cref{sec:improvements}, we will present the two other variants of PWS that improve on DPWS.

\subsection{\label{sec:mjp} Driven Markov Jump Process}

Throughout this article we consider systems that can be modeled by a master equation and are being driven by a stochastic signal. 
The master equation specifies the time evolution of the conditional probability distribution $\mathrm{P}(x,t|x_0, t_0)$ which is the probability for the process to reach the discrete state $x\in\Omega$ at time $t$, given that it was at state $x_0\in\Omega$ at the previous time $t_0$. The state space $\Omega$ is multi-dimensional if the system is made up of multiple components and therefore $x$ and $x_0$ can be vectors rather than scalar values. Denoting the transition rate at time $t$ from state~$x$ to another state~$x^\prime \neq x$ by $w_t(x^\prime, x)$, the master equation reads
\begin{equation}
    \frac{\partial\mathrm{P}(x,t)}{\partial t} = \sum_{\substack{x^\prime\in\Omega \\ x^\prime \neq x}} [w_t(x, x^\prime) \mathrm{P}(x^\prime,t) - w_t(x^\prime, x) \mathrm{P}(x,t)] \,,
\end{equation}
where, for brevity, we suppress the dependence on the initial condition, i.e., $\mathrm{P}(x,t)=\mathrm{P}(x,t|x_0,t_0)$.
By defining $Q_t(x^\prime,x) = w_t(x^\prime,x)$ for $x \neq x^\prime$ and $Q_t(x, x) = -\sum_{x^\prime\in\Omega\smallsetminus\{x\}} w_t(x^\prime, x)$ the master equation simplifies to
\begin{equation}
\frac{\partial\mathrm{P}(x,t)}{\partial t} = \sum_{x^\prime\in\Omega} Q_t(x, x^\prime) \mathrm{P}(x^\prime,t)\,. \label{eq:master-equation}
\end{equation}
Note that by definition the diagonal matrix element $Q_t(x,x)$ is the negative exit rate from state $x$, i.e. the total rate at which probability flows away from state $x$. 

Using the master equation we can compute the probability of any trajectory. A trajectory $\bm{x}$ is defined by a list of jump times $t_1,\ldots,t_{n-1}$, together with a sequence of system states $x_0,\ldots,x_{n-1}$. The trajectory starts at time $t_0$ in state $x_0$ and ends at time $t_n$ in state $x_{n-1}$, such that its duration is $T=t_n-t_0$. At each time $t_i$ (for $i=1,\ldots,n-1$) the trajectory describes an instantaneous jump $x_{i-1}\rightarrow x_{i}$. The probability density of $\bm{x}$ is
\begin{equation}
\begin{aligned}
    \mathcal{P}[\bm{x}] &= 
    \mathrm{P}(x_0)\times \left(
    \prod^{n-1}_{i=1} Q_{t_i}\left(x_i, x_{i-1}\right) 
    \right) \\
    &\quad\times\left(
    \prod^{n}_{i=1}
    \exp\int\limits^{t_{i}}_{t_{i-1}} \mathrm{d}t\ Q_t(x_{i-1}, x_{i-1})
    \right),
\end{aligned}
\label{eq:traj-prob-master-eq}
\end{equation}
\b{a product of the probability of the initial state $\mathrm{P}(x_0)$, the rates of the $n-1$ transitions $Q_{t_i}\left(x_i, x_{i-1}\right)$, and the survival probabilities for the waiting times between jumps, given by $\exp\int^{t_{i}}_{t_{i-1}} \mathrm{d}t\ Q_t(x_{i-1}, x_{i-1})$ for $i=1,\ldots,n$. }

\subsubsection{Computing the Likelihood \texorpdfstring{$\mathcal{P}[x|s]$}{P[x|s]} \label{sec:likelihood}}

\b{To compute {\em likelihood} or conditional probability~$\mathcal{P}[\bm{x}|\bm{s}]$ of an output trajectory~$\bm{x}$ for a given input trajectory $\bm{s}$, we note that the input  determines the time-dependent stochastic dynamics of the jump process.
Indeed, the transition rates at time $t$, given by $Q_t(x^\prime,x; {\bm s})$, depend explicitly on the input $s(t)$ at time $t$ and may even depend on the entire history of ${\bm s}$ prior to $t$.}

In the common case that every input trajectory $\bm{s}$ leads to a unique transition rate matrix $Q_t(x^\prime,x;\bm{s})$, i.e. the map $\bm{s}\mapsto Q_t(\cdot,\cdot;\bm{s})$ is injective, the likelihood is directly given by \cref{eq:traj-prob-master-eq}:
\begin{equation}
\begin{aligned}
    \mathcal{P}[\bm{x}|\bm{s}] &= 
    \mathrm{P}(x_0|s_0)\times \left(
    \prod^{n-1}_{i=1} Q_{t_i}\left(x_i, x_{i-1} ;\bm{s}\right) 
    \right) \\
    &\quad\times\left(
    \prod^{n}_{i=1}
    \exp\int\limits^{t_{i}}_{t_{i-1}} \mathrm{d}t\ Q_t(x_{i-1}, x_{i-1};\bm{s})
    \right)
    \label{eq:traj_prob}
\end{aligned}
\end{equation}
where $\mathrm{P}(x_0|s_0)$ is the probability of the initial state $x_0$ of the output given the initial state of the input $s_0=s(t_0)$. 

The evaluation of the trajectory likelihood is at the heart of our Monte Carlo scheme. 
However, numerically computing a large product like \cref{eq:traj_prob} very quickly reaches the limits of floating point arithmetic since the result is often either too large or too close to zero to be representable as a floating point number. 
Thus, to avoid numerical issues, it is vital to perform the computations in log-space, i.e. to compute
\begin{equation}
    \ln\mathcal{P}[\bm{x}|\bm{s}] = \ln \mathrm{P}(x_0|s_0) + \int^T_{t_0}\mathrm{d}t\ \mathcal{L}_t[\bm{s},\bm{x}]
    \label{eq:log_traj_prob}
\end{equation}
where
\begin{equation}
\begin{aligned}
    \mathcal{L}_t[\bm{s},\bm{x}] &= Q_t(x(t),x(t);\bm{s}) \\ &\quad+\sum^{n-1}_{i=1} \delta(t-t_i) \ln Q_t(x_i,x_{i-1};\bm{s})\,.
\end{aligned}
    \label{eq:path_action}
\end{equation}

\b{The computation of the log-likelihood $\ln\mathcal{P}[\bm{x}|\bm{s}]$ for given trajectories $\bm{s}$ and $\bm{x}$ according to \cref{eq:log_traj_prob,eq:path_action}  proceeds as follows:}
\begin{itemize}
    \item At the start of the trajectory we compute the log-probability of the initial condition $\ln\mathrm{P}(x_0|s_0)$,
    \item for every jump $x_{i-1}\rightarrow x_i$ in $\bm{x}$ compute the log jump propensity $\ln Q_{t_i}(x_i,x_{i-1};\bm{s})$, and
    \item for every interval $(t_{i-1},t_i)$ of constant output value $x(t) = x_{i-1}$ between two jumps of $\bm{x}$, we compute $\int^{t_i}_{t_{i-1}}\mathrm{d}t\ Q_t(x_{i-1},x_{i-1}; \bm{s})$.
    This integral can be performed using standard numerical methods such as the trapezoidal rule, which is also exact if $Q_t(x(t),x(t);\bm{s})$ is a piecewise linear function of $t$ as in our examples in \cref{sec:case-studies}.
\end{itemize}
The sum of the three contributions above yields the exact log-likelihood $\ln\mathcal{P}[\bm{x}|\bm{s}]$ as given in \cref{eq:log_traj_prob}.

Thus, notably, the algorithm to compute the log-likelihood $\ln\mathcal{P}[\bm{x}|\bm{s}]$ is both efficient and straightforward to implement, being closely related to the standard Gillespie algorithm. The only quantity in \cref{eq:log_traj_prob} that cannot be directly obtained from the master equation is the log-probability of the initial state, $\ln\mathrm{P}(x_0|s_0)$.
 
\b{Our scheme can be applied to any system with a well-defined (non-equilibrium) initial distribution $\mathrm{P}(s_0, x_0)$ as specified by, e.g. the experimental setup.
Most commonly though, one is interested in studying information transmission for systems in steady state. 
Then, the initial condition $\mathrm{P}(s_0,x_0)$ is the stationary distribution of the Markov process.
Depending on the complexity of the system, this distribution can be found either analytically from the master equation \cite{2007.vanKampen,2017.Weber} (possibly using simplifying approximations \cite{2009.Walczak,2007.Kim}), or computationally from stochastic simulations \cite{1976.Gillespie}.}

\subsubsection{Sampling from \texorpdfstring{$\mathcal{P}[x|s]$}{P[x|s]}\label{sec:gillespie}}

Standard kinetic Monte Carlo simulations naturally produce exact samples of the probability distribution $\mathcal{P}[\bm{x}|\bm{s}]$ as defined in \cref{eq:traj_prob}. That is, for any signal trajectory $\bm{s}$ and initial state $x_0$ drawn from $\mathrm{P}(x_0|s_0)$ we can use the Stochastic Simulation Algorithm~(SSA) or variants thereof to generate a corresponding trajectory $\bm{x}$. The SSA propagates the initial condition $x_0,t_0$ forward in time according to the transition rate matrix $Q_t(\cdot;\bm{s})$. In the standard Direct SSA algorithm~\cite{1976.Gillespie} this is done by alternatingly sampling the waiting time until the next transition, and then selecting the actual transition.

The transition rates $Q_t(x^\prime,x;\bm{s})$ of a driven master equation are necessarily time-dependent since they include the coupling of the jump process to the input trajectory $\bm{s}$, which itself varies in time. While most treatments of the SSA assume that the transition rates are constant in time, this restriction is easily lifted. Consider step $i$ of the Direct SSA which generates the next transition time $t_{i+1} = t_i+\Delta t_i$. For time-varying transition rates the distribution of the stochastic waiting time $\Delta t_i$ is characterized by the survival function
\begin{equation}
    S_i(\tau) = \mathrm{P}(\Delta t_i > \tau) = \exp\int^{t_i+\tau}_{t_i} \mathrm{d}t\ Q_t(x_i, x_i;\bm{s}) \,.
\end{equation}
The waiting time can be sampled using inverse transform sampling, i.e. by generating a uniformly distributed random number $u\in[0,1]$ and computing the waiting time using the inverse survival function $\Delta t_i = S^{-1}_i(u)$. Numerically, computing the inverse of the survival function requires solving the equation
\begin{equation}
    \ln u = \int^{t_i+\Delta t_i}_{t_i} \mathrm{d}t\ Q_t(x_i, x_i;\bm{s})
    \label{eq:inverse-transform-sampling}
\end{equation}
for the waiting time $\Delta t_i$. Depending on the complexity of $Q_t(x_i, x_i|\bm{s})$, this equation can either be solved analytically or numerically, e.g. using Newton's method.
Hence, this method to generate stochastic trajectories is only truly exact if we can solve \cref{eq:inverse-transform-sampling} analytically, as in the example in \cref{sec:birth-death}.
Additionally, in some cases more efficient variants of the SSA with time dependent rates could be used \cite{1997.Prados,2015.Thanh}.

\subsection{Input Statistics\label{sec:input-statistics}}

For our mutual information estimate, we need to be able to draw samples from the input distribution $\mathcal P[\bm s]$. Our algorithm poses no restrictions on $\mathcal{P}[\bm{s}]$ other than the possibility to generate sample trajectories.

For example, the input signal may be described by a continuous-time jump process, as in \cref{sec:birth-death}.
One benefit is that it is possible to generate exact realizations of such a process (using the SSA) and to exactly compute the likelihood $\mathcal{P}[\bm{x}|\bm{s}]$ using \cref{eq:log_traj_prob}.
Specifically, the likelihood can be exactly evaluated because the transition rates $Q_t(\cdot,\cdot;\bm{s})$ for any input trajectory $\bm{s}$, while time-dependent, are \emph{piece-wise constant}. 
This implies that the integral in \cref{eq:log_traj_prob} can be evaluated analytically without approximations. Similarly, for piece-wise constant transition rates, the inverse function of \cref{eq:inverse-transform-sampling} can be evaluated directly such that we can sample exact trajectories from the driven jump process.
As a result, when both input and output are described by a master equation, PWS is a completely exact Monte Carlo scheme to compute the mutual information.

However, the techniques described here do \emph{not} require the input signal $\bm{s}$ to be described by a continuous-time jump process, or even to be Markovian. 
The input signal can be any stochastic process for which trajectories can be generated numerically.
This includes continuous stochastic processes that are found as solutions to stochastic differential equations \cite{1992.Kloeden}. \b{The application in \cref{sec:chemotaxis} provides an example.}

\section{\label{sec:improvements} Variants of PWS}

The DPWS scheme presented in the previous section makes it possible to compute the mutual information between trajectories of a stochastic input process and the output of a Markov jump process. 
\b{However, the number of possible trajectories increases exponentially with trajectory length, leading to a corresponding increase in the variance of the DPWS estimate.
Hence, for long trajectories the DPWS estimate may prove to be computationally too expensive.
To address this issue, we describe two improved variants of PWS in this section, both based on free-energy estimators from statistical physics.
}

\subsection{Marginalization Integrals in Trajectory Space\label{sec:marginalization}}

The computationally most expensive part of our scheme in \cref{sec:algorithm} is the evaluation of the marginalization integral $\mathcal{P}[\bm{x}_i]=\int\mathcal{D}[\bm{s}] \mathcal{P}[\bm{s},\bm{x}_i]$ which needs to be performed for every sample $\bm{x}_1,\ldots,\bm{x}_N$. Consequently the computational efficiency of this marginalization is essential for the overall performance.

Marginalization is a general term to denote an operation where one or more variables are integrated out of a joint probability distribution. For instance, we obtain the marginal probability distribution $\mathcal{P}[\bm{x}]$ from $\mathcal{P}[\bm{s},\bm{x}]$ by computing the integral
\begin{equation}
    \mathcal{P}[\bm{x}] = \int\mathcal{D}[\bm{s}]\ \mathcal{P}[\bm{s},\bm{x}] = \int\mathcal{D}[\bm{s}]\ \mathcal{P}[\bm{s}]\mathcal{P}[\bm{x}|\bm{s}]\,.
    \label{eq:generic-marginalization}
\end{equation}

In DPWS, we use \cref{eq:marginal-naive} to compute $\mathcal{P}[\bm{x}]$ which involves generating independent input trajectories from $\mathcal{P}[\bm{s}]$. 
However, this this is not the optimal Monte Carlo technique to perform the marginalization.
The generated input trajectories are independent from the output trajectory $\bm{x}$. 
Thus, we ignore the causal connection between $\bm{s}$ and $\bm{x}$, 
and we typically end up sampling trajectories $\bm{s}^\star$ whose likelihoods $\mathcal{P}[\bm{x}|\bm{s}^\star]$ are very small.
Then, most sampled trajectories have small integral weights, 
and only very few samples provide a significant contribution to the average.
The variance of the result is then very large because the effective sample size is much smaller than the total sample size. 
The use of $\mathcal{P}[\bm{s}]$ as the sampling distribution is thus only practical in cases where the dependence of the output on the input is not too strong. It follows, perhaps paradoxically, that this sampling scheme works best when the mutual information is not too large
\footnote{%
Indeed, the mutual information $\mathrm{I}(\mathcal{S}, \mathcal{X})$ precisely quantifies how strong the statistical dependence is between the trajectory-valued random variables $\mathcal{S}$ and $\mathcal{X}$. From its definition $\mathrm{I}(\mathcal{S}, \mathcal{X})=\mathrm{H}(\mathcal{S}) - \mathrm{H}(\mathcal{S}|\mathcal{X})$ we can understand more clearly how this affects the efficiency of the Monte Carlo estimate. Roughly speaking, $\mathrm{H}(\mathcal{S})$ is related to the number of distinct trajectories $\bm{s}$ that can arise from the dynamics given by $\mathcal{P}[\bm{s}]$, while $\mathrm{H}(\mathcal{S}|\mathcal{X})$ is related to the number of distinct trajectories $\bm{s}$ that could have lead to a specific output $\bm{x}$, on average. Therefore, if the mutual information is very large, the difference between these two numbers is very large, and consequently the number of overall distinct trajectories is much larger than the number of distinct trajectories compatible with output $\bm{x}$. 
Now, if we generate trajectories according to the dynamics given by $\mathcal{P}[\bm{s}]$, with overwhelming probability we generate a trajectory $\bm{s}$ which is not compatible with the output trajectory $\bm{x}$, and therefore $\mathcal{P}[\bm{x}|\bm{s}]\approx 0$. 
Hence, the effective number of samples $M_\text{eff}$ is much smaller than the actual number of generated trajectories $M$, i.e. $M_\text{eff} \ll M$.
We therefore only expect the estimate in \cref{eq:marginal-naive} to be reliable when computing the mutual information for systems where it is not too high. Thus, strikingly, the difficulty of computing the mutual information is proportional to the magnitude of the mutual information itself.%
}.

This is a well known Monte Carlo sampling problem and a large number of techniques have been developed to solve it. The two variants of our scheme,  RR-PWS and TI-PWS, both make use of ideas from statistical physics for the efficient computation of free energies.

\begin{table}[b]
\centering
\renewcommand{\arraystretch}{2.0}
\renewcommand{\tabcolsep}{0.5cm}
 \begin{tabular}{c | c}
 \hline
 $\mathcal{P}[\bm{s},\bm{x}]$ & $e^{-\mathcal{U}[\bm{s}, \bm{x}]}$  \\
 \hline
 $\mathcal{P}[\bm{s}]$ & $\frac{1}{\mathcal{Z}_0[\bm{x}]} e^{-\mathcal{U}_0[\bm{s}]}$ \\
  \hline
 $\mathcal{P}[\bm{s}|\bm{x}]$ & $\frac{1}{\mathcal{Z}[\bm{x}]} e^{-\mathcal{U}[\bm{s}, \bm{x}]}$ \\
  \hline
 $1$ & $\mathcal{Z}_0[\bm{x}]$ \\
  \hline
 $\mathcal{P}[\bm{x}]$ & $\mathcal{Z}[\bm{x}]$ \\
  \hline
 $\mathcal{P}[\bm{x}|\bm{s}]$ & $e^{-\Delta\mathcal{U}[\bm{s}, \bm{x}]}$ \\
 \hline
 \end{tabular}
 \caption{\b{Translation to the notation of statistical physics. The definitions of $\mathcal{U}$ and $\mathcal{U}_0$ that are used here are given in \cref{eq:h0,eq:h1}.}}
 \label{tab:translation}
\end{table}

To understand how we can make use of these ideas to compute the marginal probability $\mathcal{P}[\bm{x}]$, it is convenient to rephrase the marginalization integral in \cref{eq:generic-marginalization} in the language of statistical physics. In this language, $\mathcal{P}[\bm{x}]$ corresponds to the normalization constant, or partition function, of a Boltzmann distribution for the potential
\begin{equation}
    \label{eq:h1} \mathcal{U}[\bm{s},\bm{x}] = -\ln\mathcal{P}[\bm{s},\bm{x}] \,.
\end{equation}
In \cref{eq:h1}, we interpret $\bm{s}$ as a variable in the configuration space whereas $\bm{x}$ is an auxiliary variable, i.e. a parameter. Note that both $\bm{s}$ and $\bm{x}$ still represent trajectories. For this potential, the partition function is given by
\begin{equation}
    \mathcal{Z}[\bm{x}] = \int\mathcal{D}[\bm{s}]\; e^{-\mathcal{U}[\bm{s},\bm{x}]} \,.
    \label{eq:partition-function}
\end{equation}
The integral only runs over the configuration space, i.e. we integrate only with respect to $\bm{s}$ but not $\bm{x}$, which remains a parameter of the partition function.
The partition function is precisely equal to the marginal probability of the output, i.e. $\mathcal{Z}[\bm{x}] = \mathcal{P}[\bm{x}]$, as can be verified by inserting the expression for the $\mathcal{U}[\bm{s},\bm{x}]$.
Further, the free energy is given by
\begin{equation}
    \mathcal{F}[\bm{x}] = -\ln \mathcal{Z}[\bm{x}] = -\ln \mathcal{P}[\bm{x}]
    \label{eq:free-energy}
\end{equation}
which shows that the computation of the free energy of the trajectory ensemble corresponding to $\mathcal{U}[\bm{s}, \bm{x}]$ is equivalent to the computation of (the logarithm of) the marginal probability $\mathcal{P}[\bm{x}]$.

Note that above we omitted any factors of $k_{\mathrm{B}}T$ since temperature is irrelevant here. 
Also note that while the distribution $\exp(-\mathcal{U}[\bm{s},\bm{x}])$ looks like the equilibrium distribution of a canonical ensemble from statistical mechanics, this does not imply that we can only study systems in thermal equilibrium. Indeed, PWS is used to study information transmission in systems driven out of equilibrium by the input signal. Thus, the notation introduced in this section is nothing else but a mathematical reformulation of the marginalization integral to make the analogy to statistical physics apparent and we assign no additional meaning of the potentials and free energies introduced here.

In statistical physics it is well known that the free energy cannot be directly measured from a simulation. Instead, one estimates the free-energy difference
\begin{equation}
    \Delta\mathcal{F}[\bm{x}] = \mathcal{F}[\bm{x}] - \mathcal{F}_0[\bm{x}] = -\ln \frac{\mathcal{Z}[\bm{x}]}{\mathcal{Z}_0[\bm{x}]}
    \label{eq:free-energy-difference}
\end{equation}
between the system and a reference system with known free energy $\mathcal{F}_0[\bm{x}]$. The reference system is described by the potential $\mathcal{U}_0[\bm{s}, \bm{x}]$ with the corresponding partition function $\mathcal{Z}_0[\bm{x}]$. 
In our case, a natural choice of reference potential is
\begin{equation}
    \label{eq:h0} \mathcal{U}_0[\bm{s},\bm{x}]=-\ln\mathcal{P}[\bm{s}]
\end{equation}
with  the corresponding partition function
\begin{equation}
        \mathcal{Z}_0[\bm{x}]=\int\mathcal{D}[\bm{s}] \mathcal{P}[\bm{s}]=1\,.
\end{equation}
This means that since $\mathcal{P}[\bm{s}]$ is a normalized probability density function, the reference free energy is zero ($\mathcal{F}_0[\bm{x}]=-\ln\mathcal{Z}_0[\bm{x}]=0$). Hence, for the above choice of reference system, the free-energy difference is
\begin{equation}
    \Delta\mathcal{F}[\bm{x}]= \mathcal{F}[\bm{x}] = -\ln\mathcal{P}[\bm{x}]\,.
    \label{eq:free-energy-difference-equals-lnp}
\end{equation}

Note that in our case the reference potential $\mathcal{U}_0[\bm{s},\bm{x}]=-\ln\mathcal{P}[\bm{s}]$ does not depend on the output trajectory $\bm{x}$, i.e. $\mathcal{U}_0[\bm{s},\bm{x}]\equiv\mathcal{U}_0[\bm{s}]$. It describes a \emph{non-interacting} version of our input-output system where the input trajectories evolve independently of the fixed output trajectory $\bm{x}$. 

What is the interaction between the output $\bm{x}$ and the input trajectory ensemble?
We define the interaction potential $\Delta\mathcal{U}[\bm{s}, \bm{x}]$ through
\begin{equation}
    \mathcal{U}[\bm{s}, \bm{x}] = \mathcal{U}_0[\bm{s}] + \Delta\mathcal{U}[\bm{s}, \bm{x}] \,.
    \label{eq:interaction-potential}
\end{equation}
The interaction potential makes it apparent that the distribution of $\bm{s}$ corresponding to the potential $\mathcal{U}[\bm{s}, \bm{x}]$ is biased by $\bm{x}$ with respect to the distribution corresponding to the reference potential $\mathcal{U}_0[\bm{s}]$.
By inserting the expressions for $\mathcal{U}_0[\bm{s}]$ and $\mathcal{U}[\bm{s}, \bm{x}]$ into \cref{eq:interaction-potential} we see that
\begin{equation}
\begin{aligned}
    \Delta\mathcal{U}[\bm{s}, \bm{x}] &= -\ln\mathcal{P}[\bm{x}|\bm{s}] \\
    &= -\ln\mathrm{P}(x_0|s_0)-\int^T_0 \mathrm{d}t\ \mathcal{L}_t[\bm{s}, \bm{x}]
\end{aligned}
\label{eq:boltzmann-weight}
\end{equation}
where $\mathcal{L}_t[\bm{s}, \bm{x}]$ is given by \cref{eq:log_traj_prob}. This expression illustrates that the interaction of the output trajectory $\bm{x}$ with the ensemble of input trajectories is characterized by the trajectory likelihood $\mathcal{P}[\bm{x}|\bm{s}]$.
Because we can compute the trajectory likelihood from the master equation, we can compute the interaction potential.

In this section we have introduced notation (summarized in \cref{tab:translation}) to show that computing a marginalization integral is equivalent to the computation of a free-energy difference. This picture allows us to distinguish two input trajectory ensembles, the \emph{non-interacting} ensemble distributed according to $\exp(-\mathcal{U}_0[\bm{s}])=\mathcal{P}[\bm{s}]$, and the \emph{interacting} ensemble with input distribution proportional to $\exp(-\mathcal{U}[\bm{s},\bm{x}])\propto\mathcal{P}[\bm{s}|\bm{x}]$. For example, the brute force estimate of $\mathcal{P}[\bm{x}]$ used in DPWS can be written as
\begin{align}
     \mathcal{P}[\bm{x}] = \frac{\mathcal{Z}[\bm{x}]}{\mathcal{Z}_0[\bm{x}]} &= \langle e^{-\Delta\mathcal{U}[\bm{s}, \bm{x}]} \rangle_0 
     \label{eq:boltzmann-average}
\end{align}
where the notation $\langle\cdots\rangle_0$ refers to an average with respect to the non-interacting ensemble.
By inserting the expressions for $\mathcal{U}_0$ and $\Delta\mathcal{U}$, it is easy to verify that this estimate is equivalent to \cref{eq:marginal-naive}. 
As explained in \cref{sec:algorithm}, to compute \cref{eq:boltzmann-average} using Monte Carlo, it is only necessary to sample from the non-interacting system $\mathcal{U}_0[\bm{s}]$ and to compute the Boltzmann weight $\Delta\mathcal{U}[\bm{s}, \bm{x}]$ (i.e. to sample from $\mathcal{P}[\bm{s}]$ and to compute the log-likelihood $\ln\mathcal{P}[\bm{x}|\bm{s}]$).
This is indeed the DPWS scheme. However, by noting the correspondence between signal trajectories and polymers and that \cref{eq:free-energy-difference} has the same form as the expression for the (excess) chemical potential of a polymer, which is the free-energy difference between the polymer of interest and the ideal chain \cite{1990.Siepmann,1994.Mueller}, more efficient schemes can be developed, as we show next.

\subsection{\label{sec:smc}RR-PWS}

\begin{figure*}
    \centering
    \includegraphics{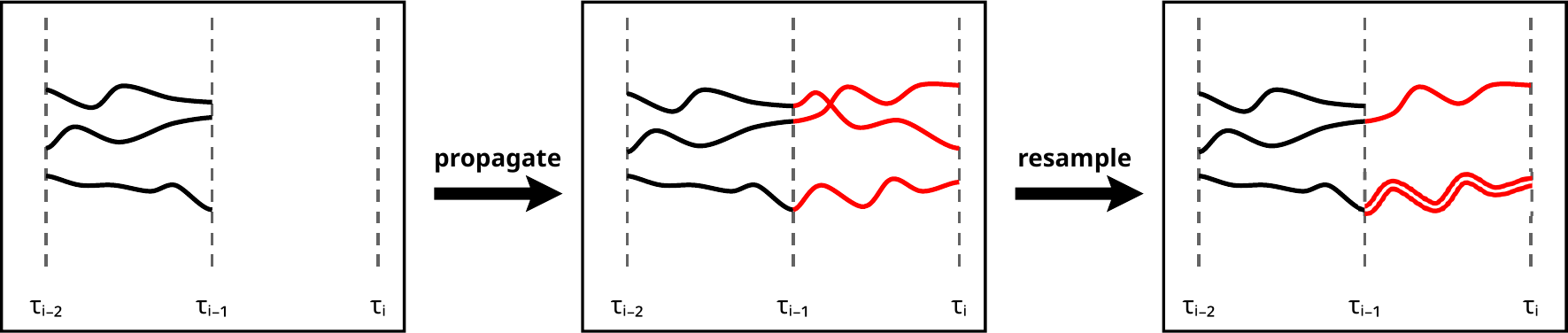}
    \caption{Illustration of one step of the bootstrap particle filter in RR-PWS. We start with a set of trajectories $\bm{s}^k_{[0,i-1]}$ with time span $[\tau_0,\tau_{i-1}]$ (left panel). In the next step we propagate these trajectories forward in time to $\tau_i$, according to $\mathcal{P}[\bm{s}]$ (central panel). Then we resample the trajectories according to the Boltzmann weights of their most recent segments, effectively eliminating or duplicating individual segments. An example outcome of the resampling step is shown in the right panel where the bottom trajectory was duplicated and one of the top trajectories was eliminated. These steps are repeated for each segment, until a set of input trajectories of the desired length is generated. The intermediate resampling steps bias the trajectory distribution from $\mathcal{P}[\bm{s}]$ towards $\mathcal{P}[\bm{s}|\bm{x}]$. }
    \label{fig:smc}
\end{figure*}

In Rosenbluth-Rosenbluth PWS we compute the free-energy difference $\Delta\mathcal{F}$ between the ideal system $\mathcal{U}_0$ and $\mathcal{U}$ in a \emph{single} simulation just like in the brute force method.
However, instead of generating $\bm{s}$ trajectories in an uncorrelated fashion according to $\exp(-\mathcal{U}_0[\bm{s}])=\mathcal{P}[\bm{s}]$, we bias our sampling distribution towards $\exp(-\mathcal{U}[\bm{s}, \bm{x}])\propto\mathcal{P}[\bm{s}|\bm{x}]$ to reduce the sampling problems found in DPWS.

The classical scheme for biasing the sampling distribution in polymer physics is due to \citet{1955.Rosenbluth} in their study of self-avoiding chains. A substantial improvement of the Rosenbluth algorithm was achieved by Grassberger, by generating polymers using pruning and enrichment steps, thereby eliminating configurations that do not significantly contribute to the average. This scheme is known as the pruned-enriched Rosenbluth method, or PERM \cite{1997.Grassberger}. While PERM is much more powerful than the standard Rosenbluth algorithm, its main drawback is that it requires careful tuning of the pruning and enrichment schedule to achieve optimal convergence. Therefore we have opted to use a technique that is similar in spirit to PERM but requires less tuning, the bootstrap particle filter \cite{1993.Gordon}.
We will describe how to use PWS with a particle filter below. That said, we want to stress that the particle filter can easily be replaced by PERM or other related methods \cite{2004.Prellberg}. Also schemes inspired by variants of Forward Flux Sampling \cite{2006.Allen,2012.Becker} could be developed.

In the methods discussed above, a polymer is grown monomer by monomer. In a continuous-time Markov process this translates to trajectories being grown segment by segment. To define the segments, we introduce a time discretization $0<\tau_1<\tau_2<\cdots<\tau_{n-1}<T$. Thus, each trajectory $\bm{s}$ of duration $T$ consists of $n$ segments where we denote the segment between $\tau_i$ and $\tau_j$ by $\bm{s}_{[i,j]}$ (we define $\tau_0=0$ and $\tau_n=T$). The particle filter uses the following procedure to grow an ensemble of trajectories segment by segment:

\begin{enumerate}
    \item Generate $M$ starting points $s^1_0, \ldots, s^M_0$ according to the initial condition of the input signal $\mathrm{P}(s_0)$.
    \item Iterate for $i=1,\ldots,n$:
    \begin{enumerate}
    \item Starting with an ensemble of $M$ partial trajectories of duration $\tau_{i-1}$ (if $i=1$ an ensemble of starting points) which we label $\bm{s}^k_{[0,i-1]}$ for $k=1,\ldots,M$:
    \begin{equation}
        \left(\bm{s}^1_{[0,i-1]}, \ldots, \bm{s}^M_{[0,i-1]}\right)\,,
    \end{equation}
    propagate each trajectory (or each starting point) forward in time from $\tau_{i-1}$ to $\tau_{i}$. Propagation is performed according to the natural dynamics of $\bm{s}$, i.e. generating a new segment $\bm{s}^k_{[i-1,i]}$ with probability
    \begin{equation}
        p^\text{gen}_i(k) = \mathcal{P}\left[\bm{s}^k_{[i-1,i]}|\bm{s}^k_{[0,i-1]}\right] = e^{-\mathcal{U}_0\left[\bm{s}^k_{[i-1,i]}\right]}
    \end{equation}
    for $k=1,\ldots,M$.
    \item Compute the Boltzmann weight 
    \begin{equation}
    U^k_i = \Delta\mathcal{U}[\bm{s}^k_{[i-1,i]},\bm{x}_{[i-1,i]}]
    \end{equation}
    of each new segment. This Boltzmann weight of a segment from $\tau_{i-1}$ to $\tau_i$ can be expressed as
    \begin{equation}
        U^k_i =
        -\delta_{1i} \ln\mathrm{P}(x_0|s_0) -\int^{\tau_i}_{\tau_{i-1}} \mathrm{d}t\ \mathcal{L}_t[\bm{s}^k_{[i-1,i]}, \bm{x}_{[i-1,i]}]\,,
        \label{eq:weight_segment}
    \end{equation}
    see \cref{eq:boltzmann-weight}, and is therefore straightforward to compute from the master equation.
    \item Sample $M$ times from the distribution
    \begin{equation}
        p^\text{select}_i(k) = \frac{e^{-U^k_i}}{w_i}
        \label{eq:index-prob}
    \end{equation}
    where the Rosenbluth weight $w_i$ is defined as
    \begin{equation}
        w_i = \sum^M_{k=1} e^{-U^k_i}\,.
    \end{equation}
    This sampling procedure yields $M$ randomly drawn indices $\ell^1_i, \ldots, \ell^M_i$. 
    \b{Each $\ell^k_i$ is an index that lies in the range from $1,\ldots,M$ and that points to one of the trajectories that have been generated up to $\tau_i$.
    To continue the sampling procedure, we relabel the indices such that the resampled set of trajectories is defined by
    $\tilde{\bm{s}}^k_{[0,i]} \gets \bm{s}^{\ell^k_i}_{[0,i]}$ for $k=1,\ldots,M$.}
    The list $\left( \tilde{\bm{s}}^1_{[0,i]}, \ldots, \tilde{\bm{s}}^M_{[0,i]} \right)$ is subsequently used as the input for the next iteration of the algorithm.
\end{enumerate}
\end{enumerate}

The normalized Rosenbluth factor of the final ensemble is then given by
\begin{equation}
    \mathcal{W} = \prod^n_{i=1} \frac{w_i}{M} \,.
    \label{eq:normalized-rosenbluth-factor}
\end{equation}
As shown in \cref{sec:smc-correctness}, we can derive an \emph{unbiased} estimate for the desired ratio $\mathcal{Z}[\bm{x}]/\mathcal{Z}_0[\bm{x}] = \mathcal{P}[\bm{x}]$ based on the Rosenbluth factor:
\begin{equation}
    \hat{\mathcal{P}}[\bm{x}] = \mathrm{P}(x_0)\ \mathcal{W}
    \label{eq:smc-marginal}
\end{equation}
with $\mathrm{P}(x_0)$ being the probability of the initial output $x_0$.
The particle filter can therefore be integrated into the DPWS algorithm to compute the marginal density $\mathcal{P}[\bm{x}]$, substituting the brute-force estimate given in \cref{eq:marginal-naive}. We call the resulting algorithm to compute the mutual information \emph{RR-PWS}.

We now provide an intuitive explanation for the scheme presented above. First note that steps 1 and 2(a) of the procedure above involve just propagating $M$ trajectories in parallel, according to $\mathcal{P}[\bm{s}]=\exp(-\mathcal{U}_0[\bm{s}])$. The interesting steps are 2(b-c) where we eliminate or duplicate some of the trajectories according to the Boltzmann weights of the most recent segment. Note, that in general the list of indices $(\ell^1_i,\ldots,\ell^M_i)$ that are sampled in step 2(c) will contain duplicates ($\ell^k_i=\ell^{k^\prime}_i$ for $k\neq k^\prime$), thus cloning the corresponding trajectory. 
Concomitantly, the indices $\ell^1_i, \ldots, \ell^M_i$ may not include every original index $1,\ldots,M$, therefore eliminating some trajectories. Since indices of trajectories with high Boltzmann weight are more likely to be sampled from \cref{eq:index-prob}, this ensures that we are only spending computational effort on propagating those trajectories whose Boltzmann weight is not too small. Hence, at its heart the particle filter is an algorithm for producing samples that tend to be distributed according to $\exp(-\mathcal{U}_0[\bm{s}])\exp(-\Delta\mathcal{U}[\bm{s},\bm{x}])=\exp(-\mathcal{U}[\bm{s}, \bm{x}])$, i.e. according to the Boltzmann distribution of the \emph{interacting} ensemble, see also \cref{sec:smc-correctness}. For illustration of the algorithm, one iteration of the particle filter is presented schematically in \cref{fig:smc}.

\b{For the efficiency of the particle filter it is important to carefully choose the number of segments $n$. When the segments are very short ($n$ large), the accumulated weights (\cref{eq:weight_segment}) tend to differ very little between the newly generated segments $\bm{s}^k_{[i-1,i]}$. Hence, the pruning and enrichment of the segments is dominated by noise.
In contrast, when the segments are very long, the distribution of Boltzmann weights $U^k_i$ becomes very wide. Then only few segments contribute substantially to the corresponding Rosenbluth weight $w_i$.
Hence, to carefully choose $n$, we need a measure that quantifies the variance of the trajectory weights. To this ens, we follow \citet{2017.Martino} and introduce an effective sample size (ESS)
\begin{equation}
    M^\text{(eff)}_i = \frac{w_i^2}{\sum^M_{k=1} \left(e^{-U^k_i}\right)^2},
\end{equation}
which lies in the range $1 \leq M^\text{(eff)}_i \leq M$; $M^\text{(eff)}_i = 1$ if one trajectory has a much higher weight than all the others and $M^\text{(eff)}_i = M$ if all trajectories have the same weight.
As a rule of thumb, we resample only when the $M^\text{(eff)}_i$ drops below $M/2$. Additionally, as recommended in Ref.~\cite{2011.Doucet}, we use the \emph{systematic sampling} algorithm to randomly draw the indices in step 2(c) which helps to reduce the variance; we find, however, the improvement over simple sampling is very minor.}
Using these techniques, the only parameter that needs to be chosen by hand for the particle filter is the ensemble size $M$.

\subsection{\label{sec:thermodynamic-integration}TI-PWS}

Our third scheme, thermodynamic integration PWS (TI-PWS), is based on the analogy of marginalization integrals with free-energy computations.
As before, we view the problem of computing the marginal probability $\mathcal{P}[\bm{x}]$ as equivalent to that of computing the free-energy difference between ensembles defined by the potentials $\mathcal{U}_0[\bm{s}, \bm{x}]$ and $\mathcal{U}[\bm{s}, \bm{x}]$, respectively.
For TI-PWS, we define a potential $\mathcal{U}_\theta[\bm{s},\bm{x}]$ with a
continuous parameter $\theta\in[0,1]$ that allows us to transform the
ensemble from $\mathcal{U}_0$ to $\mathcal{U}=\mathcal{U}_1$.
The corresponding partition function is
\begin{equation}
\mathcal{Z}_\theta[\bm{x}]=\int\mathcal{D}[\bm{s}]\ e^{-\mathcal{U}_\theta[\bm{s},\bm{x}]} \,.
\end{equation}
For instance, for $0\leq\theta\leq 1$, we can define our potential as 
\begin{equation}
    \mathcal{U}_\theta[\bm{s},\bm{x}]=\mathcal{U}_0[\bm{s}, \bm{x}]+\theta\,\Delta\mathcal{U}[\bm{s},\bm{x}]\,,
    \label{eq:ti-hamiltonian}
\end{equation}
such that $e^{-\mathcal{U}_\theta[\bm{s},\bm{x}]}=\mathcal{P}[\bm{s}]\mathcal{P}[\bm{x}|\bm{s}]^\theta$. Note that this is the simplest choice for a continuous transformation between $\mathcal{U}_0$ and $\mathcal{U}_1$, but by no means the only one. For reasons of computational efficiency, it can be beneficial to choose a different path between $\mathcal{U}_0$ and $\mathcal{U}_1$, depending on the specific system~\cite{1998.Gelman}. Here we will not consider other paths however, and derive the thermodynamic integration estimate for the potential given in \cref{eq:ti-hamiltonian}.

To derive the thermodynamic integration estimate for the free-energy difference, we first compute the derivative of $\ln\mathcal{Z}_\theta[\bm{x}]$ with respect to~$\theta$:
\begin{equation}
\begin{aligned}
    \frac{\partial}{\partial \theta} \ln\mathcal{Z}_\theta[\bm{x}] &= \frac{1}{\mathcal{Z}_\theta[\bm{x}]} \frac{\partial}{\partial \theta} \int\mathcal{D}[\bm{s}]\  e^{-\mathcal{U}_\theta[\bm{s},\bm{x}]} \\
    &= -\left\langle \frac{\partial \mathcal{U}_\theta[\bm{s},\bm{x}]}{\partial\theta} \right\rangle_\theta\\
    &= -\left\langle
    \Delta\mathcal{U}[\bm{s},\bm{x}]
    \right\rangle_\theta\,.
\end{aligned}
\label{eq:z-derivative}
\end{equation}
Thus, the derivative of $\ln\mathcal{Z}_\theta[\bm{x}]$ is an average of the Boltzmann weight with respect to $\mathcal{P}_\theta[\bm{s}|\bm{x}]$ which is the ensemble distribution of $\bm{s}$ given by
\begin{equation}
    \mathcal{P}_\theta[\bm{s}|\bm{x}] = \frac{1}{\mathcal{Z}_\theta[\bm{x}]} e^{-\mathcal{U}_\theta[\bm{s}, \bm{x}]}\,.
\end{equation}
Integrating \cref{eq:z-derivative} with respect to $\theta$ leads to the formula for the free-energy difference
\begin{equation}
    \Delta\mathcal{F}[\bm{x}] = -\int^1_0 \mathrm{d}\theta\ \left\langle 
    \Delta\mathcal{U}[\bm{s}, \bm{x}]
    \right\rangle_\theta
    \label{eq:ti-estimate}
\end{equation}
which is the fundamental identity underlying thermodynamic integration.

To compute the free-energy difference using \cref{eq:ti-estimate}, we evaluate the $\theta$-integral numerically using Gaussian quadrature, while the inner average
$\left\langle \Delta\mathcal{U}[\bm{s}, \bm{x}] \right\rangle_\theta$
is computed using MCMC simulations. 
To perform MCMC simulations in trajectory space we use ideas from transition path sampling (TPS) \cite{2002.Bolhuis}. 
As discussed in \cref{sec:mcmc}, the efficiency of MCMC samplers strongly depends on the proposal moves that are employed.
While better proposal moves could be conceived, we only use the \emph{forward shooting} and \emph{backward shooting} moves of TPS \cite{2002.Bolhuis} to obtain the results in \cref{sec:case-studies}.
These moves regrow either the end, or the beginning of a trajectory, respectively. A proposal is accepted according to the Metropolis criterion~\cite{1953.Metropolis}.

\section{Integrating Out Internal Components \label{sec:integrating-out}}

So far the output trajectory $\bm{x}$ has been considered to correspond to the trajectory of the system in the \emph{full} state space $\Omega$. Concomitantly, the method presented is a scheme for computing the mutual information between the input signal $\bm{s}$ and the trajectory $\bm{x}$, comprising the time evolution of all the $n$ components in the system, $X^1, X^2,\ldots, X^n$. 
Each component $X^i$ itself has a corresponding trajectory $\bm{x}^i$, such that the full trajectory can be represented as a vector $\bm{x}=(\bm{x}^1,\ldots,\bm{x}^n)$. It is indeed also the conditional probability $\mathcal{P}[\bm{x}|\bm{s}]=\mathcal{P}[\bm{x}^1,\ldots,\bm{x}^n|\bm{s}]$ and the marginal probability $\mathcal{P}[\bm{x}]=\mathcal{P}[\bm{x}^1,\ldots,\bm{x}^n]$ of this vector in the full state space that can be directly computed from the master equation. In fact, it is this vector, which captures the states of all the components in the system, that carries the most information on the input signal $\bm{s}$, and thus has the largest mutual information. 
However, typically the downstream system cannot read out the states of all the components $X^1, \ldots, X^n$. 
Often, the downstream system reads out only a few components or often even just one component, the ``output component'' $X^r$.  
The other components then mainly serve to transmit the information from the input $\bm{s}$ to this readout $X^r$. 
From the perspective of the downstream system, the other components are hidden. 
The natural quantity to measure the precision of information processing is then the mutual information $\mathrm{I}(\mathcal{S};\mathcal{X}^r)$ between the input $\bm{s}$ and the output component's trajectory $\bm{x}^r$, not $\mathrm{I}(\mathcal{S};\mathcal{X})$. The question then becomes how to compute $\mathcal{P}[\bm{x}^r]$ and $\mathcal{P}[\bm{x}^r|\bm{s}]$, from which $\mathrm{I}(\mathcal{S};\mathcal{X}^r)$ can be obtained. Here, we present a scheme to achieve this.

As an example, consider a chemical reaction network with species $X^1,\ldots,X^{n}$. Without loss of generality, we will assume that the $n$-th component is the output component, $X^r=X^n$. 
The other species $X^1,\ldots,X^{n-1}$ are thus not part of the output, but only relay information from the input signal $\bm{s}$ to the output signal $\bm{x}^n$.
To determine the mutual information $\mathrm{I}(\mathcal{S},\mathcal{X})$ we need $\mathcal{P}[\bm{x}^n|\bm{s}]$, where $\bm{x}^n$ is the trajectory of only the readout component~$X^n$. 
However, from the master equation we can only obtain an expression for the full conditional probability $\mathcal{P}[\bm{x}^1,\ldots,\bm{x}^n|\bm{s}]$ of all components. To compute the value of $\mathcal{P}[\bm{x}^n|\bm{s}]$, we must perform the marginalization integral
\begin{equation}
    \mathcal{P}[\bm{x}^n|\bm{s}] = \int\mathcal{D}[\bm{x}^1] \cdots \int\mathcal{D}[\bm{x}^{n-1}]\; \mathcal{P}[\bm{x}^1,\ldots,\bm{x}^n|\bm{s}]\,.
    \label{eq:marginalization_integral}
\end{equation}
We can compute this integral using a Monte Carlo scheme as described below and use the resulting estimate for $\mathcal{P}[\bm{x}^n|\bm{s}]$ to compute the mutual information using our technique presented in \cref{sec:algorithm}.

The marginalization of \cref{eq:marginalization_integral} entails integrating out degrees of freedom from a known joint probability distribution.
In \cref{eq:marginal-naive} we solved the analogous problem of obtaining the marginal probability $\mathcal{P}[\bm{x}]$ by integrating out the input trajectories through the integral $\mathcal{P}[\bm{x}]=\int\mathrm{d}\bm{s}\ \mathcal{P}[\bm{s},\bm{x}]=\int\mathrm{d}\bm{s}\ \mathcal{P}[\bm{s}]\mathcal{P}[\bm{x}|\bm{s}]$. 
As described in \cref{sec:algorithm}, the integral from \cref{eq:marginal-naive} can be computed via a Monte Carlo estimate by sampling many input trajectories from $\mathcal{P}[\bm{s}]$ and taking the average of the corresponding conditional probabilities $\mathcal{P}[\bm{x}|\bm{s}_i]$. We will show that in the case where there is no feedback from the readout component back to the other components, a completely analogous Monte Carlo estimate can be derived for \cref{eq:marginalization_integral}. 
We describe this below.
Additionally, in the absence of feedback, the techniques presented in \cref{sec:improvements} above can be employed to develop computationally more efficient schemes.

More specifically, we can evaluate \cref{eq:marginalization_integral} via a direct Monte Carlo estimate under the condition that the stochastic dynamics of the other components $X^1,\ldots,X^{n-1}$ are not influenced by $X^n$ (i.e., no feedback from the readout). Using the identity
\begin{equation}
    \mathcal{P}[\bm{x}^1,\ldots,\bm{x}^n|\bm{s}] = \mathcal{P}[\bm{x}^1,\ldots,\bm{x}^{n-1}|\bm{s}]\ \mathcal{P}[\bm{x}^n|\bm{x}^1_i,\ldots,\bm{x}^{n-1}_i,\bm{s}]
\end{equation}
to rewrite the integrand in \cref{eq:marginalization_integral}, we are able to represent the conditional probability $\mathcal{P}[\bm{x}^n|\bm{s}]$ as an average over the readout component's trajectory probability 
\begin{equation}
    \mathcal{P}[\bm{x}^n|\bm{s}] = 
    \left\langle \mathcal{P}[\bm{x}^n|\bm{x}^1_i,\ldots,\bm{x}^{n-1}_i,\bm{s}] \right\rangle_{\mathcal{P}[\bm{x}^1,\ldots,\bm{x}^{n-1}|\bm{s}]} \,.
    \label{eq:marginalization_average}
\end{equation}
Thus, assuming that we can evaluate the conditional probability of the readout given all the other components, $\mathcal{P}[\bm{x}^n|\bm{x}^1_i,\ldots,\bm{x}^{n-1}_i,\bm{s}]$, we arrive at the estimate
\begin{equation}
    \mathcal{P}[\bm{x}^n|\bm{s}] 
    \approx \frac{1}{M}\sum^M_{i=1} \mathcal{P}[\bm{x}^n|\bm{x}^1_i,\ldots,\bm{x}^{n-1}_i,\bm{s}]
\label{eq:marginalization_mc}
\end{equation}
where the samples $\bm{x}^1_i,\ldots,\bm{x}^{n-1}_i$ for $i=1,\ldots,M$ are drawn from $\mathcal{P}[\bm{x}^1,\ldots,\bm{x}^{n-1}|\bm{s}]$. Notice that the derivation of this Monte Carlo estimate is fully analogous to the estimate in \cref{eq:marginal-naive}, but instead of integrating out the input trajectory $\bm{s}$ we integrate out the component trajectories $\bm{x}^1,\ldots,\bm{x}^{n-1}$.

To obtain $\mathcal{P}[\bm{x}^n|\bm{x}^1_i,\ldots,\bm{x}^{n-1}_i,\bm{s}]$ in \cref{eq:marginalization_average,eq:marginalization_mc}, we note that, in absence of feedback, we can describe the stochastic dynamics
of the readout component $X^n$ as a jump process with time-dependent transition rates whose time-dependence arises from the trajectories of the other components $\bm{x}^1,\ldots,\bm{x}^{n-1}$ and the input input $\bm{s}$. In effect, this is a driven jump process for $X^n$, driven by all upstream components $X^1,\ldots,X^{n-1}$ and the input signal. Specifically, denoting $\bm{u}=(\bm{x}^1,\ldots,\bm{x}^{n-1},\bm{s})$ as the joint trajectory representing the history of all upstream components as well as the input signal, we can, as explained in \cref{sec:mjp}, write the time dependent transition rate matrix $Q_t(\cdot|\bm{u})$ for the stochastic dynamics of $X^n$ and use \cref{eq:traj_prob} to compute $\mathcal{P}[\bm{x}^n|\bm{u}]=\mathcal{P}[\bm{x}^n|\bm{x}^1_i,\ldots,\bm{x}^{n-1}_i,\bm{s}]$. Using \cref{eq:marginalization_mc}, this then allows us to compute $\mathcal{P}[\bm{x}^n|\bm{s}]$.

Finally, to compute the mutual information $\mathrm{I}(\mathcal{S};\mathcal{X}^n)$, e.g. using the estimate in \cref{eq:average-of-differences}, we additionally need to evaluate the marginal output probability $\mathcal{P}[\bm{x}^n]$. This requires us to perform one additional integration over the space of input trajectories $\bm{s}$:
\begin{equation}
\begin{aligned}
    \mathcal{P}[\bm{x}^n] &= \int\mathcal{D}[\bm{s}]\ \mathcal{P}[\bm{s}] \mathcal{P}[\bm{x}^n|\bm{s}] \\
    &= \left\langle \mathcal{P}[\bm{x}^n|\bm{s}] \right\rangle_{\mathcal{P}[\bm{s}]} \,.
\end{aligned}
\end{equation}
The corresponding Monte Carlo estimate is
\begin{equation}
    \begin{aligned}
    \mathcal{P}[\bm{x}^n] &\approx \frac{1}{N}\sum^N_{i=1} \mathcal{P}[\bm{x}^n|\bm{s}_i] \\
    &\approx \frac{1}{N}\sum^N_{i=1}\frac{1}{M}\sum^M_{j=1} \mathcal{P}[\bm{x}^n|\bm{x}^1_{ij}, \ldots, \bm{x}^{n-1}_{ij}, \bm{s}_i]
    \end{aligned}
\end{equation}
where the input trajectories $\bm{s}_i$ follow $\mathcal{P}[\bm{s}]$ and the intermediate components $(\bm{x}^1_{ij},\ldots,\bm{x}^{n-1}_{ij})$, for $i=1,\ldots,N$ and $j=1,\ldots,M$, follow $\mathcal{P}[\bm{x}^1,\ldots,\bm{x}^{n-1}|\bm{s}_i]$. 

In summary, the scheme to obtain $\mathcal{P}[\bm{x}^n|\bm{u}]$ in the presence of hidden intermediate components is analogous to that used for computing $\mathcal{P}[\bm{x}]$ from $\mathcal{P}[\bm{x}|\bm{s}]$. In both cases, one needs to marginalize a distribution function by integrating out components. 
Indeed, the schemes presented here and in \cref{sec:algorithm} are bona fide schemes to compute the mutual information between the input $\bm{s}$ and either the trajectory of the output component $\bm{x}^n$ or the full output $\bm{x}$. 
However, when the trajectories are sufficiently long or the stochastic dynamics are sufficiently complex, then the free-energy schemes of \cref{sec:improvements} may be necessary to enhance the efficiency of computing the marginalized distribution, $\mathcal{P}[\bm{x}]$ or $\mathcal{P}[\bm{x}^n|\bm{s}]$.

\section{\label{sec:case-studies}Results}

To demonstrate the power of our framework and illustrate how the techniques of the previous sections can be used in practice, we apply PWS to two instructive chemical reaction networks.
 We first consider a linearly coupled birth-death process. 
This system has already been studied previously using a Gaussian model 
\cite{2009.Tostevin}, and by \citet{2019.Duso} using an approximate technique, and we compare our results to those of these studies. 
This simple birth-death system serves to illustrate the main ideas of our approach and also highlights that linear systems can be distinctly non-Gaussian. 
The second example has been chosen to demonstrate the practical applicability of our technique. We use RR-PWS to compute the mutual information rate in the bacterial chemotaxis system, which is a prime example of a complex information processing system consisting of many reactions. \b{Then we compare the computed rate against recent experiments.}

The code used to produce the PWS estimates was written in the Julia programming language~\cite{2017.Bezanson} and has been made freely available~\cite{manuel_reinhardt_2021_6334035,*pws_github}.
For performing stochastic simulations we use the DifferentialEquations.jl package~\cite{2017.Rackauckas} and biochemical reaction models are set up with help from the ModelingToolkit.jl package~\cite{2021.Ma}. 

\subsection{Coupled Birth-Death Processes\label{sec:birth-death}}

As a first example system we consider a simple birth-death process $\emptyset\rightleftharpoons\mathrm{X}$ of species~$\mathrm{X}$ which is created at rate $\rho(t)$ and decays with constant rate $\mu$ \b{per copy of $\mathrm{X}$}. This system receives information from an input signal that modulates the birth rate $\rho(t)$.
For simplicity, we assume it is given by
\begin{equation}
    \rho(t)=\rho_0 s(t)
\end{equation}
where $\rho_0$ is a constant and $s(t)$ is the input copy number at time $t$. This is a simple model for gene expression, where the rate of production of a protein $\mathrm{X}$ is controlled by a transcription factor $\mathrm{S}$, and $\mathrm{X}$ itself has a characteristic decay rate.
The input trajectories $s(t)$ themselves are generated via a separate birth-death process $\emptyset\rightleftharpoons\mathrm{S}$ with production rate $\kappa$ and decay rate $\lambda$. 

We compute the trajectory mutual information for this system as a function of the trajectory duration $T$ of the input and output trajectories. 
\b{For $T\rightarrow\infty$, the trajectory mutual information is expected to increase linearly with $T$, since, on average, every additional output segment contains the same additional amount of information on the input trajectory.
Because we are interested in the mutual information in steady state, the initial states $(s_0,x_0)$ were drawn from the stationary distribution $\mathrm{P}(s_0,x_0)$.}
This distribution was obtained using a Gaussian approximation.
This does not influence the asymptotic rate of increase of the mutual information, but leads to a nonzero mutual information already for $T=0$.

\begin{figure}
    \includegraphics{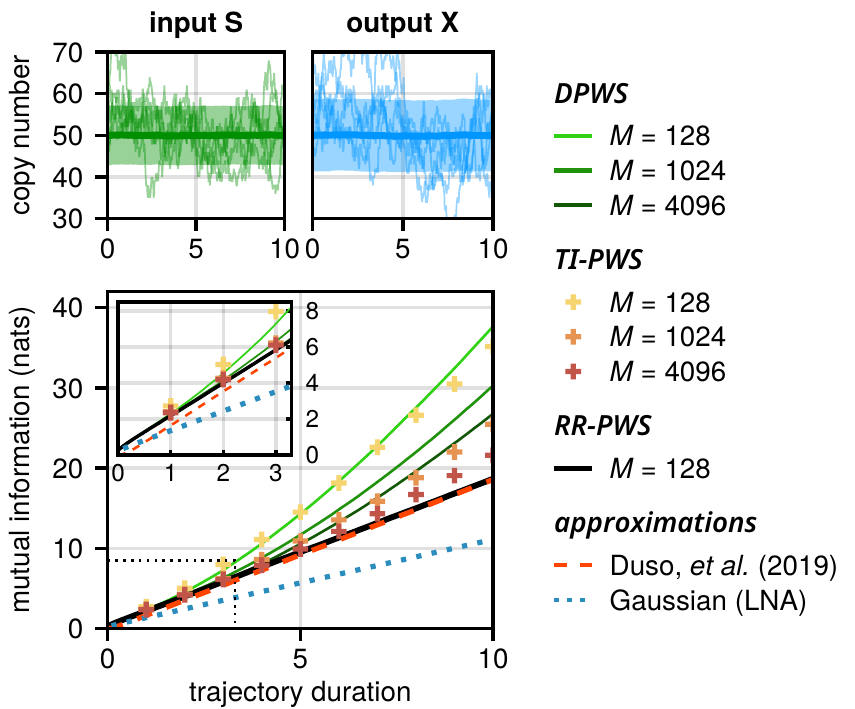}
    \caption{Comparing different schemes to compute the mutual information as a function of trajectory duration for a simple coupled birth-death process with rates $\kappa = 50, \lambda=1, \rho_0=10, \mu = 10$ and steady-state initial condition. 
    \b{The top panels show example trajectories of input and output as well as the mean (solid line) and standard deviation (shaded region).}
    Below, the mutual information is shown as a function of trajectory duration.
    The inset shows an enlarged version of the dotted rectangle near the origin. 
    For short trajectories all PWS estimates agree. Yet, for longer trajectories, DPWS and TI-PWS require a much larger number of input trajectories $M$ for computing $\mathcal{P}[\bm{x}]$ than RR-PWS to converge.
    Results for the three PWS variants are compared with the \citet{2019.Duso} estimate, and with the linear noise approximation from Ref. \cite{2009.Tostevin}. 
    We find excellent agreement between the Duso scheme and RR-PWS.
    The Gaussian linear noise approximation systematically underestimates the mutual information. All PWS estimates, as well as the Duso approximation were computed using $N=10^4$ samples from $\mathcal{P}[\bm{s},\bm{x}]$. 
    }
    \label{fig:gene-expr}
\end{figure}

\b{\Cref{fig:gene-expr} shows the mutual information as a function of the trajectory duration $T$. 
We compare the three PWS variants and two approximate schemes.
One is that of \citet{2019.Duso}.
To apply it, we used the code publicly provided by the authors \footnote{\url{https://github.com/zechnerlab/PathMI/tree/302f03e51ad195adc6be39fa9618886c76590cc4}},
and to avoid making modifications to this code, we chose a fixed initial condition $(s_0=x_0=50)$ which causes the mutual information to be zero for $T=0$.
The figure also shows the analytical result of a Gaussian model \cite{2009.Tostevin}, obtained using the linear-noise approximation (see \cref{sec:lna}).}

\b{We find that the efficiency of the respective PWS variants depends on the duration of the input-output trajectories.
For short trajectories all PWS variants  yield very similar estimates for the mutual information.
However, for longer trajectories the estimates of DPWS and, to a smaller degree, TI-PWS diverge, because of poor sampling of the trajectory space in the estimate of $\mathcal{P}[\bm{x}]$.
For longer trajectories, the estimate becomes increasingly dominated by rare trajectories, which make an exceptionally large contribution to the average of $\mathcal{P}[\bm{x}]$.
Missing these rare trajectories with a high weight tends to increases the marginal entropy  $  \mathrm{H}(\mathcal{X})$ (see \cref{eq:marginal-entropy-estimate}),  and thereby the mutual information; indeed, the estimates of DPWS and TI-PWS are higher than that of RR-PWS.}
For brute-force DPWS, the error decreases as we increase the number $M$ of input trajectories per output trajectory used to estimate  $\mathcal{P}[\bm{x}]$.
Similarly, for TI-PWS the error decreases as we use more MCMC samples for the marginalization scheme. 
For the RR-PWS, however, already for $M=128$ the estimate has converged; we verified that a further increase of $M$ does not change the results.

We also find excellent agreement between the RR-PWS estimate and the approximate result of \citet{2019.Duso}. Only very small deviations are visible in \cref{fig:gene-expr}. These deviations are mostly caused by the different choice for the initial conditions. 
In RR-PWS, the initial conditions are drawn from the stationary distribution, while in the Duso scheme they are fixed, such that the mutual information computed with RR-PWS is finite while that computed with the Duso scheme is zero.
Yet, as the trajectory duration $T$ increases, the Duso estimate slowly ``catches up'' with the RR-PWS result.

\b{\cref{fig:gene-expr} also shows that although the Gaussian model matches the PWS result for $T=0$, it systematically underestimates the mutual information for trajectories of finite duration $T>0$.
Interestingly, this is not a consequence of small copy-number fluctuations: increasing the average copy number does not significantly improve the Gaussian estimate.
We leave a detailed analysis of this observation for future work.}

\begin{figure}
    \centering
    \includegraphics{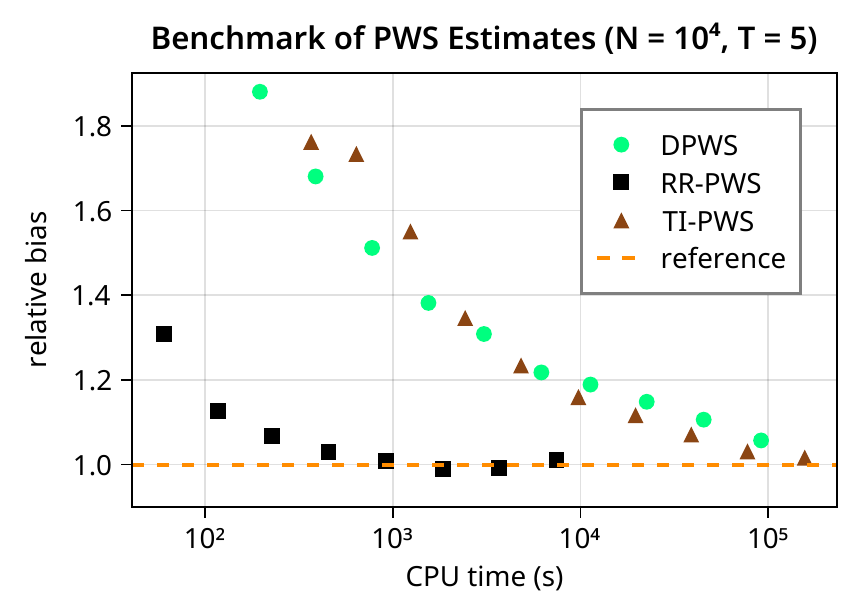}
    \caption{Comparing estimation bias for the different PWS variants in relation to their CPU time requirements. Each dot represents a single mutual information estimate with $N=10^4$ samples for trajectories of duration $T=5$.
    Almost all the CPU time of a PWS estimate is spent on the computation of the marginal probability $\mathcal{P}[\bm{x}]$.
    The bias of the marginal probability estimate can be reduced by using a larger number $M$ of sampled input trajectories to compute the marginalization integral, which also increases the required CPU time.
    The RR-PWS estimate converges much faster than the estimate of DPWS and TI-PWS.
    For DPWS and TI-PWS, the dots represents estimates ranging from $M=2^5$ to $M=2^{14}$, for RR-PWS ranging from $M=2^3$ to $M=2^{10}$.
    As the baseline of zero bias we use the converged result from the RR-PWS estimates.}
    \label{fig:timing}
\end{figure}

The different approaches for computing the marginal probability $\mathcal{P}[\bm{x}]$ lead to different computational efficiencies of the respective PWS schemes.
In \cref{fig:timing}, as a benchmark, we show the magnitude of the error of the different PWS estimates in relation to the required CPU time. Indeed, as expected, the computation of 
the marginal probability poses problems for long trajectories when using the brute force DPWS scheme.
More interestingly,
while TI-PWS improves the estimate of the mutual information, the improvement is not dramatic. 
Unlike the brute-force scheme, thermodynamic integration does make it possible to generate input trajectories $\bm{s}$ that are correlated with the output trajectories $\bm{x}$, 
but it still overestimates the mutual information for long trajectories unless a very large number of MCMC samples are used. 

The RR-PWS implementation evidently outperforms the other estimates for this system. 
The regular resampling steps ensure that we mostly sample input trajectories $\bm{s}$ with non-vanishing likelihood $\mathcal{P}[\bm{x}|\bm{s}]$, thereby avoiding the sampling problem from DPWS. 
Moreover, sequential Monte Carlo techniques such as RR-PWS and FFS \cite{2006.Allen} have a considerable advantage over MCMC techniques in trajectory sampling. 
With MCMC path sampling, we frequently make small changes to an existing trajectory 
such that the system moves slowly in path space,
leading to poor statistics. In contrast, in RR-PWS we generate new trajectories from scratch, segment by segment, and
these explore the trajectory space much faster.

\b{The coupled birth-death process represents a simple yet non-trivial system capable of information transmission. In the next section we apply PWS to a more complex and realistic biochemical signaling network.}

\subsection{Bacterial Chemotaxis\label{sec:chemotaxis}}

\begin{figure}
    \centering
    \includegraphics{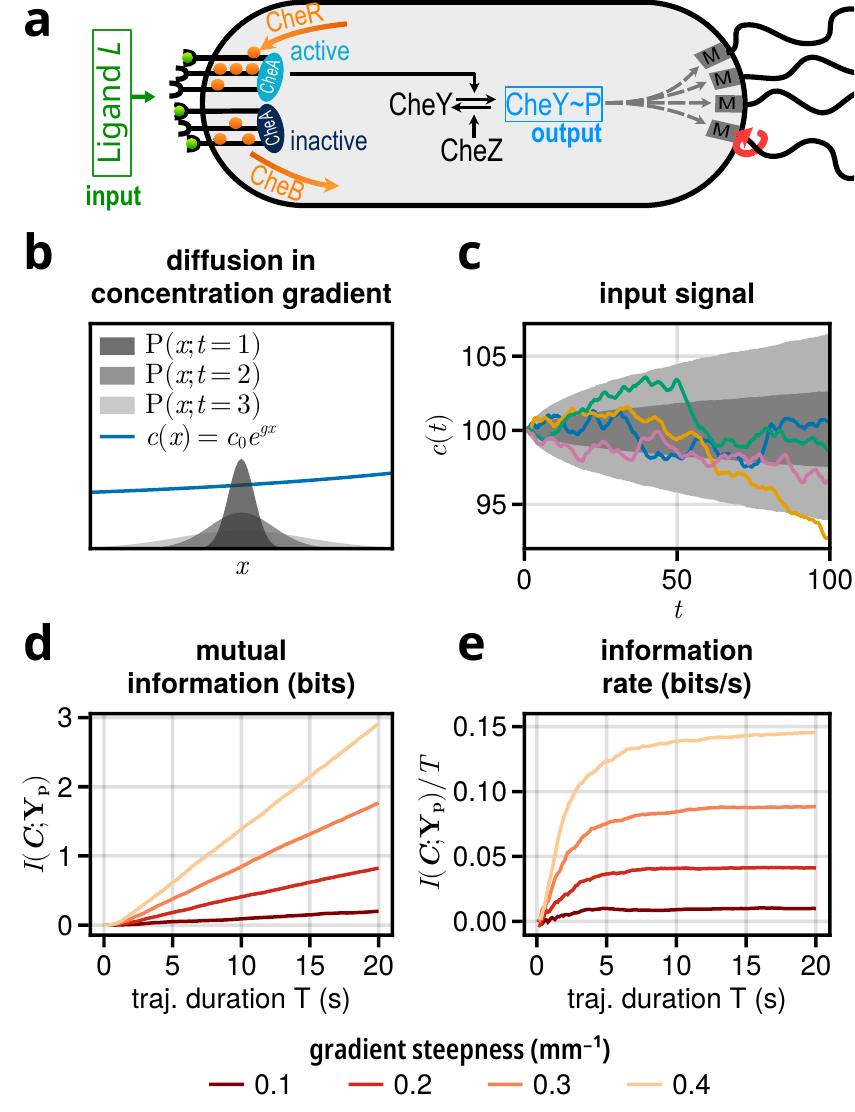}
    \caption{
    The information transmission rate in the bacterial chemotaxis system.
    \textbf{a}~Cartoon of the chemotaxis network of \emph{E. coli}. Receptors form clusters with an associated CheA kinase. A cluster can either be active or inactive, depending on the number of bound ligands (green dots) and methylated sites (orange dots). 
    Active CheA can phosphorylate CheY; phosphorylated CheY controls the rotation direction of the flagellar motors and thereby the movement of the bacterium.
    \b{
    \textbf{b}~In a shallow gradient $c(x)$, the bacterium diffuses nearly freely in the $x$-direction. The variance of the position increases with time, the hallmark of a random walk. The input signal is the concentration $c(t)=c(x(t))$ as experienced by the bacterium at time $t$.
    \textbf{c}~This movement gives rise to the input statistics of the signal. The shaded regions indicate the 75th and 95th percentiles, example trajectories are displayed in color. 
    This signal is non-stationary as its variance always keeps growing.
    \textbf{d}~Mutual information $I(\mathbf{C}_T, \mathbf{Y}_T)$ between input trajectories $c(t)$ and output trajectories $y_p(t)$ as a function of trajectory duration $T$. In each RR-PWS simulation, $N = 7200$ Monte Carlo samples were used ($M = 256$ for the particle filter).
    \textbf{e}~The information transmission rate is defined as $I(\mathbf{C}_T, \mathbf{Y}_T)/T$ in the limit $T\to\infty$.}
    }
    \label{fig:chemotaxis}
\end{figure}

The chemotaxis system of the bacterium \textit{Escherichia coli} is a complex information processing system. It is responsible for detecting nutrient gradients in the cell's environment and using that information to guide the bacterium's movement.
Briefly, \emph{E. coli} navigates through its environment by performing a biased random walk, successively alternating between so-called {\em runs}, during which it swims with a nearly constant speed, and {\em tumbles}, during which it randomly chooses a new direction \cite{1977.Berg}.
The rates of switching between these two states are controlled by the chemotaxis sensing system (\cref{fig:chemotaxis}a).
This system consists of receptors on the cell surface that detect the ligand, and a downstream signaling network that processes this information by taking the time-derivative of the signal, the ligand concentration. This derivative is taken via two antagonistic reactions, which occur on two distinct timescales. Attractant binding rapidly deactivates the receptor, while slow methylation counteracts this reaction by reactivating the receptor, leading to near perfect adaptation \cite{1997.Barkai,2006.Endres,2012.Lan,2015.Sartori}. Lastly, active receptors phosphorylate the downstream messenger protein CheY, which controls the tumbling propensity by binding the flagellar motors that propel the bacterium.

In our model, the receptors are grouped in clusters (\cref{sec:chemotaxis-model}).
Each receptor can switch between an active and an inactive conformational state, but, in the spirit of the Monod-Wyman-Changeux model \cite{1965.Monod}, the energetic cost of having two different conformations in the same cluster is prohibitively large.
We can then speak of each cluster as either being active or inactive. 
Each receptor in a cluster can bind ligand and be (de)methylated, which, together, control the probability that the cluster is active.
\b{
In the simulations, receptor (de)methylation is modeled explicitly, because the (de)methylation reactions are slow.
In contrast, the timescale of receptor-ligand (un)binding is much faster than the other timescales in the system, i.e., those of the input dynamics, CheY (de)phosphorylation, and receptor (de)methylation.
The receptor-ligand binding dynamics can therefore be integrated out without affecting information transmission, in order to avoid wasting CPU time (\cref{sec:chemotaxis-model}). 
In addition, the receptor clusters can phosphorylate CheY, while phosphorylated CheY is dephosphorylated at a constant rate.
The dynamics of the kinase CheA and the phosphatase CheZ which drive (de)phosphorylation are not modeled explicitly.
Table \ref{tab:chemotaxis-parameters} in \cref{sec:chemotaxis-model} gives the parameter values of our chemotaxis model, which are all based on values reported in the literature.
For what follows below, the key parameters are the number of receptors per cluster, which is taken to be $N=6$ based on Refs.~\cite{2010.Shimizu,2020.Kamino}, while the number of clusters is $N_{\rm c} = N_{\rm r} / N = 400$, where $10^3 < N_{\rm r} < 10^4$ is an estimate for the total number of receptors based on Ref.~\cite{2004.Li}.
The translation of our model into a master equation is explained in \cref{sec:chemotaxis-model}.}

\b{We first asked whether this model based on the current literature can reproduce the information transmission rate as recently measured by \citet{2021.Mattinglyg3a}. In what follows, we call this model the ``literature-based'' model.}

\b{The information transmission rate depends not only on the biochemical chemotaxis network, but also on the dynamics of the input signal.
It is therefore important that the dynamics of this signal in our model agree with those in the experiments of \citet{2021.Mattinglyg3a}.
In our model, the input signal is the time-dependent ligand concentration $c(t)$ that is experienced by the swimming bacterium. 
In the experiments, the cells swim in a very shallow chemical gradient, such that their swimming behavior is, to a good approximation, identical to that in the absence of a gradient.
The cell's movement can thus be modeled as a (persistent) random walk. 
Assuming that the gradient is oriented in $x$-direction, the autocorrelation of the $x$-component of the velocity was experimentally found to be well described by
$V(t)\equiv \langle \delta v_x(t) \delta v_x(0) \rangle \simeq
\sigma^2_v e^{-\lambda |t|}$, where $\sigma^2_v$ is the variance of
the fluctuations in the velocity, and $\lambda^{-1}$ is the correlation time of these fluctuations \cite{2021.Mattinglyg3a}.
We therefore model the cell's velocity as $\dot{v}_x = -\lambda v_x + \xi$, which, with $\langle \xi(t) \xi(t^\prime)\rangle = 2 \sigma^2_v \lambda \delta (t-t^\prime)$, gives rise to the measured correlation function $V(t)$.
These cells swim in a shallow, exponential gradient $c(x)\propto e^{gx}$ with steepness~$g$. 
The dynamics of the ligand concentration as experienced by the bacteria are then given by $\dot{c} = g c v_x$.
The cell's own swimming dynamics described by $v_x$ thus give rise to the input signal $c(t)$ of the biochemical network. It constitutes an exponential random walk, as illustrated in \cref{fig:chemotaxis}b,~c (see also \cref{sec:chemotaxis_input}).}

\b{In our model, the output is the concentration of phosphorylated CheY, while in the experiments of \citet{2021.Mattinglyg3a} it is the average activity of the receptor clusters as obtained via FRET measurements.
We argue that this difference does not significantly affect the obtained information rates, and thus, that it is valid to compare our results to the experiments.
In particular, since the copy number of CheY is much larger than the number of receptor clusters, the fluctuations in CheY are dominated by the extrinsic fluctuations coming from the receptor activity noise rather than from the intrinsic fluctuations associated with CheY (de)phosphorylation.
To a good approximation, the copy number of phosphorylated CheY, ${\rm Yp}(t)$, is thus a deterministic function of the average receptor activity $a(t)$.
Mathematically, the mutual information $I(X;Y)$ between two stochastic variables $X$ and $Y$ is the same as the mutual information $I(f(X);g(Y))$ for deterministic and monotonic functions $f$ and $g$. 
It follows that the mutual information between $c(t)$ and ${\rm Yp}(t)$, is nearly the same as that between $c(t)$ and the receptor activity $a(t)$.
It is therefore meaningful to compare the information transmission rates as predicted by our PWS simulations to those measured by \citet{2021.Mattinglyg3a}.}

\b{We use RR-PWS to exactly compute the mutual information for the literature-based model.
Specifically, we measure the mutual information $I(\mathbf{C}, \mathbf{Y_p}; T)$ between the input trajectory of the ligand concentration $c(t)$ and the output trajectory of phosphorylated CheY, $y_p(t)$, and where each trajectory is of duration $T$.
With RR-PWS it is possible to compute $I(\mathbf{C}, \mathbf{Y_p}; \tau)$ for all $\tau \leq T$ within a single PWS simulation of duration $T$ by saving intermediate results after each sampled segment, see \cref{sec:smc}.
The receptor states are hidden internal states, and we use
the technique of \cref{sec:integrating-out} to integrate them out.}

\begin{figure}
    \centering
    \includegraphics{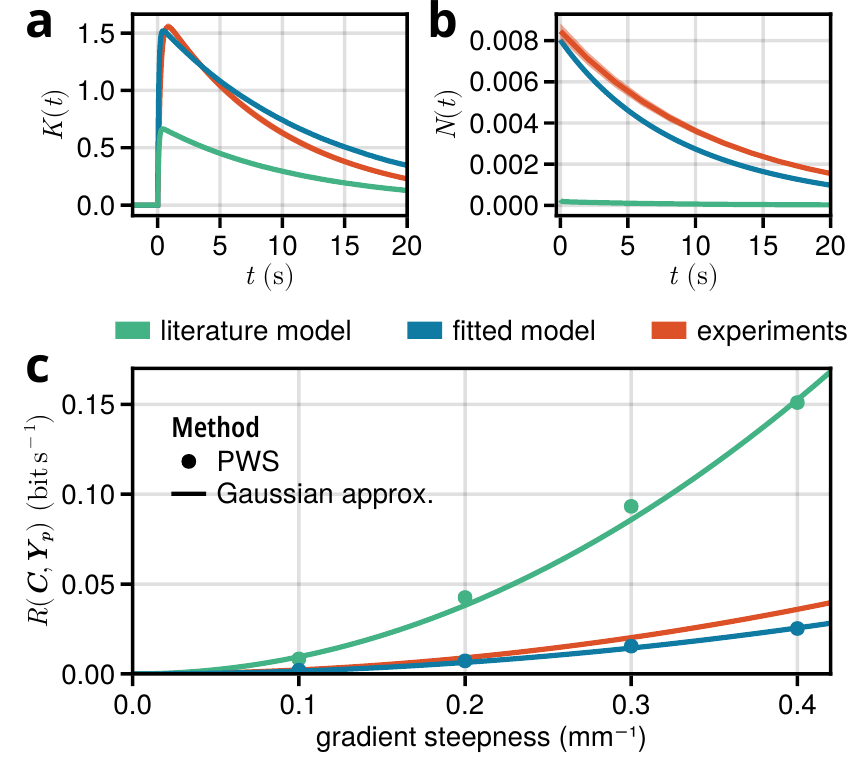}
    \caption{\b{Comparison of theoretical models with experimental data for bacterial chemotaxis system. Panels \textbf{a} and \textbf{b} show the response and noise kernels, respectively, for the model based on literature parameters (green), parameters fitted to experiments (blue), and experiments from \citet{2021.Mattinglyg3a} (orange). In panel \textbf{c}, the information transmission rate is shown for each model as a function of gradient steepness, with results from the Gaussian approximation shown alongside exact PWS calculations. The fitted model closely matches the experiments, while the literature-based model over-estimates information transmission rate by a factor of $\approx 4$ despite having a lower response amplitude (panel \textbf{a}). 
    This is because the literature-based model has a large number of independents receptor clusters $N_c$, resulting in much lower noise in the output (panel \textbf{b}). In all cases, the Gaussian approximation matches the exact PWS results, providing support for the accuracy of the measurements by \citet{2021.Mattinglyg3a}.}}
    \label{fig:chemotaxis_comparison}
\end{figure}

\b{\Cref{fig:chemotaxis}e shows the PWS estimate of the information transmission rate for cells swimming in gradients of different steepnesses $g$. The information transmission rate is obtained from the PWS estimate of the trajectory mutual information $I(\mathbf{C}, \mathbf{Y_p}; T)$, different trajectory durations $T$. As seen in \cref{fig:chemotaxis}d, for short trajectories the mutual information increases non-linearly with trajectory duration $T$, but in the long-duration limit the slope becomes constant. This asymptotic rate of increase of the mutual information with $T$ is the desired information transmission rate $R(\mathbf{C}, \mathbf{Y_p})$. The precise definition is given by
\begin{equation}
    R(\mathbf{C}, \mathbf{Y_p}) = \lim_{T\to\infty} \frac{I(\mathbf{C}, \mathbf{Y_p}; T)}{T} \,.
\end{equation}
}

\b{We then compared our results for the information transmission rate of the literature-based model to those of \citet{2021.Mattinglyg3a}. 
\Cref{fig:chemotaxis_comparison}c shows that the model predictions differ from the experiments by a factor of $\approx 4$. 
Despite this discrepancy, we believe that the agreement between experiment and theory is, in fact, remarkable, because these predictions were made {\em ab initio}: the model was developed based on the existing literature and we did not fit our model to the data of \citeauthor{2021.Mattinglyg3a}}

\b{Yet, the question about the origin of the discrepancy remains.
The difference between their measurements and our predictions could be attributed either to the inaccuracy of our model or to the approximation that \citeauthor{2021.Mattinglyg3a} had to employ to compute the information transmission rate from experimental data. Concerning the latter hypothesis, due to the curse of dimensionality and experimental constraints, \citeauthor{2021.Mattinglyg3a} could not directly obtain the information transmission rate from measured time traces of the input and output of the system.
Instead, they measured three different kernels that describe the system in the linear regime. 
Specifically, they obtained the response $K(t)$ of the kinase activity to a step-change in input signal, the autocorrelation function of the input signal $V(t)$, and the autocorrelation $N(t)$ of the kinase activity in a constant background concentration.
Then they used a Gaussian model to compute the information transmission rate from these measured functions $K(t)$, $V(t)$, and $N(t)$ \cite{2009.Tostevin, 2021.Mattinglyg3a} (see also \cref{sec:lna}). 
This Gaussian model is based on a linear noise assumption and cannot perfectly capture the true non-linear dynamics of the biochemical network. 
This could be the cause for the observed discrepancies in the information rate.
We have indeed already seen in \cref{sec:birth-death} that there can be substantial differences between exact computations and the Gaussian approximation for the trajectory mutual information.}

\b{To uncover the reason for the discrepancy we first tested whether our literature-based model reproduces the experimentally measured kernels.
If the kernels do not match, then, clearly, the discrepancy in the information rate may be caused by the difference between our model and the experimental system, as opposed to the inaccuracy of the Gaussian framework.
Our input correlation function, $V(t)$, is, by construction, the same as that of \citet{2021.Mattinglyg3a}. However, we find that the response kernel $K(t)$ and the autocorrelation function of the noise $N(t)$ of our system are different.
\Cref{fig:chemotaxis_comparison}a,~b shows that our model reproduces the timescales of $N(t)$ and $K(t)$ as measured experimentally.
This is perhaps not surprising, because the decay of both $N(t)$ and $K(t)$ is set by the (de)methylation rate, which has been well-characterized experimentally.
Yet, the figure also shows that our model significantly underestimates the amplitudes of both $N(t)$ and $K(t)$.}

\b{This raises the question of whether other parameter values would allow our model to better reproduce the measured kernels $K(t)$ and $N(t)$, and, secondly, whether this would resolve the discrepancy in information rate between our simulations and the experiments.}

\b{The amplitude $\sigma^2_N$ of the output noise correlation function $N(t)$ is bounded by the number of receptor clusters $N_{\rm c}$. In particular, the variance of the receptor activity is $\sigma^2_N = \sigma^2_a / N_{\rm c} \leq 1 / 4 N_{\rm c}$, where $\sigma^2_a\leq 1/4$ is the variance of the activity of a single receptor cluster.
Comparing this bound to the measured receptor noise strength $\sigma^2_N$ reveals that $N_{\rm c}$ needs to be much smaller than our original model assumes: the number of clusters needs to be as small as $N_{\rm c} \lesssim 10$. 
Indeed, \cref{fig:chemotaxis_comparison}b shows that with $N_{\rm c}=9$, our model quantitatively fits the correlation function $N(t)$ of the receptor activity in a constant background concentration, as measured experimentally \cite{2021.Mattinglyg3a}.}

\b{The amplitude of $K(t)$, i.e. the gain, depends on the ratio $K_{\rm D}^{\rm A}/ K_{\rm D}^{\rm I}$ of the dissociation constants of the receptor for ligand binding in its active or inactive state, respectively, as well as on the number of receptors per cluster, $N$.
Both dissociation constants have been well characterized experimentally \cite{2011.Lazova,2010.Shimizu}, but the number of receptors per cluster has only been inferred indirectly from experiments \cite{2020.Kamino,2010.Shimizu}.
The higher gain as measured experimentally by \citet{2021.Mattinglyg3a} indicates that $N$ is larger than assumed in our model: with $N=15$ our model can quantitatively fit $K(t)$ (\cref{fig:chemotaxis_comparison}a).}

\b{
We thus find that by reducing the number of clusters from $N_{\rm c} = 400$ to $N_{\rm c}=9$ while simultaneously increasing their size from $N=6$ to $N=15$, our model is able to quantitatively fit both $N(t)$ and $K(t)$ \cite{2021.Mattinglyg3a} (\cref{fig:chemotaxis_comparison}). 
This suggests that the number of independent receptor clusters is smaller than hitherto believed, while their size is larger.}

\b{
Finally, how accurately can our revised model reproduce the measured information rate, and how accurate is the Gaussian framework for the experimental system in the regime studied by \citet{2021.Mattinglyg3a}?
In the revised model, called the ``fitted model'', with $N_{\rm c}=9$ and $N=15$, all key quantities for computing the information transmission rate within the Gaussian framework, $V(t)$, $N(t)$ and $K(t)$, are nearly identical to the experiments of \citet{2021.Mattinglyg3a}, see \cref{fig:chemotaxis_comparison}.
Within the Gaussian framework (see \cref{sec:lna}), the information transmission rate in our model is thus expected to be very similar to the experimentally measured one, and \cref{fig:chemotaxis_comparison}c shows that this is indeed the case.
To quantify the accuracy of the Gaussian framework, we then recomputed the information transmission rate for the revised model, using exact PWS (see \cref{sec:pws_fitted}).
We found that the result matches the Gaussian prediction very well.
For these shallow and static chemical gradients, the Gaussian model is thus highly accurate.
Our analysis validates \emph{a posteriori} the Gaussian framework adopted by \citet{2021.Mattinglyg3a}.}

\section{Discussion}

\b{In this manuscript, we have developed a general, practical, and flexible method that makes it possible to compute the mutual information between trajectories exactly. 
PWS is a Monte Carlo scheme based on the exact computation of trajectory probabilities. 
We showed how to compute exact trajectory probabilities from the master equation and thus how to use PWS for any system described by a master equation.}
Since the master equation is employed in many fields and in particular provides an exact stochastic model for well-mixed chemical reaction dynamics, PWS is very broadly applicable.

\b{The application of PWS to the bacterial chemotaxis system shows how crucial it is to have a simulation technique that is exact. 
Without the latter it would be impossible to determine whether the difference between our predictions and the Mattingly data \cite{2021.Mattinglyg3a} is due to the inaccuracy of the model, the inaccuracy of the numerical technique to simulate the model, or the approximations used by Mattingly and coworkers in analyzing the data. 
In contrast, because PWS is exact, we knew the difference between theory and experiment is either due to the inaccuracy of the model or the approximations used to analyze the data.
By then employing the same Gaussian framework to analyze the behavior of the model and the experimental system, we were able to establish that the difference is due to the inaccuracy of our original model.}

\b{Our analysis indicates that the size of the receptor clusters in the {\it E. coli} chemotaxis system, $N\approx 15$, is larger than that based on previous estimates, $N\sim 6$ \cite{2007.Mello,2010.Shimizu,2020.Kamino}.
The early estimates of the cluster size were based on bulk dose-response measurements with a relatively slow ligand exchange, yielding $N \approx 6$ \cite{2007.Mello,2010.Shimizu}. More recent dose-response measurements, at the single cell level and with faster ligand exchange, yield an average that is higher,  $\langle N\rangle\approx 8$, and with a broad distribution around it, arising from cell-to-cell variability \cite{2020.Kamino}. Our estimate, $N\approx 15$, based on fitting the response kernel $K(t)$ to that measured by \citet{2021.Mattinglyg3a}, therefore appears reasonable.
  At the same time, the number of clusters, obtained by fitting the noise correlation function $N(t)$ to the data of \citet{2021.Mattinglyg3a} is surprisingly low, $N_c \sim 10$, given the total number of receptors, $N_r \sim 10^3\text{--}10^4$ \cite{2004.Li}. 
Interestingly, recent experiments indicate that the receptor array is poised near the critical point \cite{2023.Keegstra}, where receptor switching becomes correlated over large distances. This effectively partitions receptors into a few large domains, which may explain our fitted values for $N$ and $N_c$.}

\b{It has been suggested that information processing systems are positioned close to a critical point to maximize information transmission \cite{2015.Tkacik,2021.Meijers}, although it has been argued that the sensing error of the {\it E. coli} chemotaxis system is minimized for independent receptors \cite{2011.Skoge}.
\citeauthor{2021.Mattinglyg3a} have demonstrated that the chemotactic drift speed in shallow exponential gradients is limited by the information transmission rate \cite{2021.Mattinglyg3a}, but whether the system has been optimized for information transmission, and how the latter affects chemotactic performance in other spatio-temporal concentration profiles, remain interesting questions for future work.}

While we have focused on the computation of the mutual information between trajectories of systems governed by a master equation, this concept can be extended to other types of stochastic processes.
The crux of PWS is the exact evaluation of the path likelihood $\mathcal{P}[\bm{x}|\bm{s}]$ from the master equation.
Therefore, in order to use PWS with a different stochastic process, we require the ability to compute the path likelihood $\mathcal{P}[\bm{x}|\bm{s}]$ for its sample paths.
Remarkably, for \emph{stochastic diffusion processes}, which are described by a Langevin equation in a continuous state space, the trajectory probability can be computed by replacing the kernel $\mathcal{L}_t(\bm{s}, \bm{x})$ in \cref{eq:log_traj_prob} with the Onsager-Machlup function \cite{1953.Onsager}.
In particular, while the path probability is not well-defined for continuous sample paths, \citet{2008.Adib} shows that the Onsager-Machlup function yields a correct expression for the path probability of time-discretized diffusion trajectories.
Therefore, PWS can be extended to handle systems described by a Langevin equation.
In such a PWS scheme, the Gillespie simulations are replaced by standard numerical integration techniques for Langevin equations (see e.g. Ref.~\cite{1992.Kloeden}), and the trajectory likelihood is evaluated on-the-fly using the Onsager-Machlup function.

\b{PWS cannot be used to directly obtain the mutual information between trajectories from experimental data, in contrast to model-free (yet approximate) methods such as K-nearest-neighbors estimators \cite{2002.Kaiser,2004.Kraskov},  decoding-based information estimates \cite{2008.Gao}, or schemes that compute the mutual information from the data within the Gaussian framework \cite{2021.Mattinglyg3a}. PWS requires a (generative) model based on a master equation or Langevin equation. However, an increasingly popular approach is to estimate the mutual information from experimental data  using hidden Markov models (HMM) \cite{1966.Baum,2010.Paninski,2021.Tang}. While the construction of these HMM models is beyond the scope of this manuscript, PWS makes it possible to compute the mutual information between time-varying inputs and outputs exactly for these models.
}

We have applied PWS to compute the mutual information (rate) in steady state, but PWS can be used equally well to study systems out of steady state.
For such systems a \hbox{(non-)equilibrium} initial condition $\mathrm{P}(s_0, x_0)$ must be specified in addition to a well-defined non-stationary probability distribution of input trajectories $\mathcal{P}[\bm{s}]$.
These distributions are defined by the (experimental) setup and lead to a well-defined output distribution $\mathcal{P}[\bm{x}]$ when the system is coupled to the input.
Thus, a steady state is no prerequisite for the application of PWS to study the trajectory mutual information.

Throughout the manuscript, we have considered systems in which the output does not feed back onto the input. In systems with feedback, the current output influences future input, which means that we cannot straightforwardly generate input trajectories according to $\mathcal{P}[\bm{s}]$. Moreover, $\mathcal{P}[\bm{x}|\bm{s}]$, the central quantity of all three PWS methods, cannot be obtained straightforwardly. 
Nonetheless, PWS can be extended to systems with feedback as shown in \mbox{\cref{sec:feedback}}. As in free-energy calculations, the trick is to define a reference system for which the marginal probability distribution---and thus the ``free energy''---is known and then compute the ``free-energy difference'' between that reference system and the system of interest.

Aside from DPWS, we developed two additional variants of PWS, capitalizing on the connection between information theory and statistical physics. Specifically, the computation of the mutual information requires the evaluation of the marginal probability of individual output trajectories $\mathcal{P}[\bm{x}]$. This corresponds to the computation of a partition function in statistical physics. RR-PWS and TI-PWS are based on techniques from polymer and rare-event sampling to make the computation of the marginal trajectory probability more efficient.

The different PWS variants share some characteristics yet also differ in others. 
DPWS and RR-PWS are static Monte Carlo schemes in which the trajectories are generated independently from the previous ones. 
These methods are similar to static polymer sampling schemes like PERM \cite{1997.Grassberger} and rare-event methods like DFFS or BG-FFS \cite{2006.Allen}. In contrast, TI-PWS is a dynamic Monte Carlo scheme, where a new trajectory is generated from the previous trajectory. In this regard, this method is similar to the CBMC scheme for polymer simulations \cite{1992.Siepmann} and the TPS \cite{2002.Bolhuis}, TIS \cite{2003.Erp}, and RB-FFS \cite{2006.Allen} schemes to harvest transition paths. The benefit of static schemes is that the newly generated trajectories are uncorrelated from the previous ones, which means that they are less likely to get stuck in certain regions of path space. Concomitantly, they tend to diffuse faster through the configuration space. 
Indeed, TI-PWS suffers from a problem that is also often encountered in TPS or TIS, which is that the middle sections of the trajectories move only slowly in their perpendicular direction. Tricks that have been applied to TPS and TIS to solve this problem, such as parallel tempering, could also be of use here \cite{2001.Vlugt}.

Another distinction is that RR-PWS generates all the trajectories in the ensemble simultaneously yet segment by segment, like DFFS, while DPWS and TI-PWS generate only one full trajectory at the time, similar to RB-FFS, BG-FFS, and also TPS and TIS.
Consequently, RR-PWS, like DFFS, faces the risk of \emph{genetic drift},  which means that, after sufficiently many resampling steps, most paths of the ensemble will originate from the same initial seed.
Thus, when continuing to sample new segments, the old segments that are far in the past become essentially fixed, which makes it possible to miss important paths in the RR-PWS sampling procedure.
As in DFFS, the risk of genetic drift in RR-PWS can be mitigated by increasing the initial number of path segments.
Although we did not employ this trick here, we found that RR-PWS was by far the most powerful scheme of the three variants studied.  

Nonetheless, we expect that DPWS and TI-PWS become more efficient in systems that respond to the input signal with a
  significant delay $\tau$.  In these cases, the weight of a particular
  output trajectory depends on the degree to which the dynamics of the
  output trajectory correlates with the dynamics of the intput
  trajectory a time $\tau$ earlier.  Because in RR-PWS a new segment
  of an output trajectory is generated based on the corresponding
  segment of the input trajectory that spans the same time-interval, it may therefore miss these
  correlations between the dynamics of the output and that of the
  input a time $\tau$ earlier. In contrast, DPWS and TI-PWS generate full
  trajectories one at the time, and are therefore more likely to
  capture these correlations.

Overall, PWS is a general framework for computing the mutual information between trajectories. We presented three variants of PWS for systems described by a master equation. Apart from these, we expect that other variants could be developed to improve efficiency for particular applications.
Because of its flexibility and simplicity, we envision that PWS will become an important and reliable tool for studying information transmission in dynamic stochastic systems.

\begin{acknowledgments}
We thank Bela Mulder, Tom Shimizu, Fotios Avgidis, Peter Bolhuis, and Daan Frenkel for useful discussions and a careful reading of the manuscript, and we thank Age Tjalma for support with obtaining the Gaussian approximation of the chemotaxis system.
This work is part of the Dutch Research Council (NWO) and was performed at the research institute AMOLF.
This project has received funding from the European Research Council (ERC) under the European Union’s Horizon 2020 research and innovation program (grant agreement No.~885065), 
and was financially supported by NWO through the “Building a Synthetic Cell (BaSyC)” Gravitation grant (024.003.019).
\end{acknowledgments}

\appendix

\section{\label{sec:smc-correctness}Justification for the Particle Filter \texorpdfstring{from \cref{sec:smc}}{}}

Here we justify the marginal probability estimate shown in \cref{eq:smc-marginal}, i.e. we show that the bootstrap particle filter used in \cref{sec:smc} provides a consistent estimate for the ratio of partition functions $\mathcal{P}[\bm{x}] = \mathcal{Z}[\bm{x}]/\mathcal{Z}_0[\bm{x}]$. The result that this estimate is also \emph{unbiased} is more difficult to establish; a proof is given by \citet{1997.Moral}.

To make the derivations below easy to follow, we structure our justification of the particle filter into three steps.
We first give a brief description of how a resampling procedure can generally be used to generate samples according to a target distribution when only samples from a different distribution are available. Secondly, we use these insights to explain how the resampling procedure used in the particle filter generates trajectories that are distributed approximately according to $\mathcal{P}[\bm{s}|\bm{x}]$, even though we only generate trajectories according to $\mathcal{P}[\bm{s}]$. Finally, we use this result to show that the particle filter provides a consistent estimate for $\mathcal{P}[\bm{x}]$.

\subsection{Sampling and Resampling}

Sampling and then resampling is a strategy to use samples $\bm{s}^1,\ldots,\bm{s}^M$ from a given prior distribution $f[\bm{s}]$ to generate \emph{approximate} samples from a different distribution of interest, with density proportional to the product $h[\bm{s}]=f[\bm{s}]g[\bm{s}]$. In general, $h[\bm{s}]$ is not normalized, and we denote the corresponding normalized probability density by $\hat{h}[\bm{s}]=h[\bm{s}]/\int\mathcal{D}[\bm{s}]h[\bm{s}]$. To generate samples from $\hat{h}[\bm{s}]$, we assign each of the existing samples from $f[\bm{s}]$ a normalized weight
\begin{equation}
    W^k = \frac{g[\bm{s}^k]}{\sum^M_{j=1}g[\bm{s}^j]}\,.
    \label{eq:weights-appendix}
\end{equation}
Then, by sampling from the discrete set $\{\bm{s}^1,\ldots,\bm{s}^M\}$ according to the weights $W^1,\ldots,W^M$, we pick samples that are approximately distributed according to $\hat{h}[\bm{s}]$. Indeed, for $M\rightarrow\infty$ the distribution of the resulting samples approaches the density $\hat{h}[\bm{s}]$ \cite{1992.Smith}. We use resampling at each iteration of the algorithm of \cref{sec:smc} to regularly prune those trajectories with low overall contribution to the marginalization integral.

\subsection{Distribution of Trajectories in the Particle Filter}

In the bootstrap particle filter, at each iteration, we start with a set of trajectories $\bm{s}^1_{[0,i-1]},\ldots,\bm{s}^M_{[0,i-1]}$ which we assume are approximately distributed according to $\mathcal{P}[\bm{s}_{[0,i-1]}|\bm{x}_{[0,i-1]}]$. In each iteration of the particle filter, the goal is to produce a set of trajectories approximately distributed according to $\mathcal{P}[\bm{s}_{[0,i]}|\bm{x}_{[0,i]}]$. Clearly, by iterating such a procedure, we can generate a set of trajectories distributed approximately according to $\mathcal{P}[\bm{s}_{[0,n]}|\bm{x}_{[0,n]}]$ for any $n>1$. Note that we always carefully use the phrase \emph{approximately distributed} because, as explained above, for finite $M$, a resampling procedure cannot generate \emph{exact} samples from a probability distribution (yet the estimate for $\mathcal{P}[\bm{x}]$ remains unbiased regardless of how good these approximations are). We now take a closer look at one iteration of the particle filter.

We start with the set of trajectories with a time span $[\tau_0,\tau_{i-1}]$, denoted by $\left\{\bm{s}^1_{[0,i-1]},\ldots,\bm{s}^M_{[0,i-1]}\right\}$.
These trajectories are then propagated forward to time $\tau_i$, by adding a new segment $\bm{s}^k_{[i-1,i]}$ to the trajectory $\bm{s}^k_{[0,i-1]}$ for $k=1,\ldots,M$. Each new segment is generated from the distribution $\mathcal{P}[\bm{s}^k_{[i-1,i]}|\bm{s}^k_{[0,i-1]}]$ such that the propagation step results in a set of trajectories $\{\bm{s}^1_{[0,i]},\ldots,\bm{s}^M_{[0,i]}\}$, distributed according to $f[\bm{s}_{[0,i]}]=\mathcal{P}[\bm{s}_{[0,i]}|\bm{x}_{[0,i-1]}]$. 

Next, we resample from the set of trajectories, with the goal of producing a set of trajectories distributed according to the target density $\hat{h}[\bm{s}] = \mathcal{P}[\bm{s}_{[0,i]}|\bm{x}_{[0,i]}]$. Thus, we have to find the appropriate weighting function $g[\bm{s}_{[0,i]}]$ in order to approximately produce samples according to the target distribution.
By choosing $g[\bm{s}_{[0,i]}] = \exp\left\{ -\Delta\mathcal{U}[\bm{s}_{[i-1,i]}, \bm{x}_{[i-1,i]}] \right\} = \mathcal{P}[\bm{x}_{[i-1,i]}|\bm{x}_{[0,i-1]},\bm{s}_{[0,i]}]$, we generate normalized weights
\begin{equation}
    W^k_i = \frac{\mathcal{P}[\bm{x}_{[i-1,i]}|\bm{x}_{[0,i-1]},\bm{s}^k_{[0,i]}]}{\sum^M_{j=1} \mathcal{P}[\bm{x}_{[i-1,i]}|\bm{x}_{[0,i-1]},\bm{s}^j_{[0,i]}]}\,,
    \label{eq:normalized-weights}
\end{equation}
cf. \cref{eq:weights-appendix}. Note that this is the same choice of weighting function as in \cref{sec:smc}, \cref{eq:index-prob}. By comparison with the notation used there, we see that the Boltzmann factors $U^k_i$ and Rosenbluth weights $w_i$ were defined such that we can express the normalized weight equivalently as
\begin{equation}
    W^k_i = \frac{e^{-U^k_i}}{w_i} \,.
\end{equation}

Why is this choice of weighting function the correct one?
First, observe that resampling with the normalized weights of \cref{eq:normalized-weights} produces samples approximately distributed according to
\begin{equation}
\begin{aligned}
    h[\bm{s}_{[0,i]}] &= f[\bm{s}_{[0,i]}] g[\bm{s}_{[0,i]}] \\
    &= \mathcal{P}[\bm{s}_{[0,i]}|\bm{x}_{[0,i-1]}]\  \mathcal{P}[\bm{x}_{[i-1,i]}|\bm{x}_{[0,i-1]},\bm{s}_{[0,i]}]\,.
\end{aligned}
\label{eq:h-unnormalized}
\end{equation}
What remains to be shown is that this density $h[\bm{s}_{[0,i]}]$, when normalized, becomes the desired target distribution $\mathcal{P}[\bm{s}_{[0,i]}| \bm{x}_{[0,i]}]$.

To do so, we need to rewrite the expression for $g[\bm{s}_{[0,i]}]=\mathcal{P}[\bm{x}_{[i-1,i]}|\bm{x}_{[0,i-1]},\bm{s}_{[0,i]}]$ using Bayes' theorem
\begin{equation}
    g[\bm{s}_{[0,i]}]=\frac{ \mathcal{P}[\bm{s}_{[0,i]}|\bm{x}_{[0,i-1]},\bm{x}_{[i-1,i]}]\ \mathcal{P}[\bm{x}_{[i-1,i]}|\bm{x}_{[0,i-1]}]}{\mathcal{P}[\bm{s}_{[0,i]}|\bm{x}_{[0,i-1]}]}\,.
\end{equation}
Notice that the first term of the numerator can be written as $\mathcal{P}[\bm{s}_{[0,i]}|\bm{x}_{[0,i]}]$. After inserting this result into \cref{eq:h-unnormalized}, we obtain
\begin{equation}
    h[\bm{s}_{[0,i]}] = \mathcal{P}[\bm{s}_{[0,i]}|\bm{x}_{[0,i]}]\ \mathcal{P}[\bm{x}_{[i-1,i]}|\bm{x}_{[0,i-1]}]\,.
\end{equation}
The second term in this product is a constant, since $\bm{x}$ is fixed. The first term is a normalized probability density for $\bm{s}_{[0,i]}$. Therefore we find that the normalized density corresponding to $h[\bm{s}_{[0,i]}]$ is
\begin{equation}
    \hat{h}[\bm{s}_{[0,i]}] = \mathcal{P}[\bm{s}_{[0,i]}|\bm{x}_{[0,i]}]\,.
\end{equation}
Consequently, this is the distribution that is approximated by the set of trajectories at the end of the $i$-th iteration of the particle filter, which is what we wanted to show. At its heart, the particle filter is therefore an algorithm to produce samples that are approximately distributed according to $\mathcal{P}[\bm{s}|\bm{x}]$.

\subsection{Marginal Probability Estimate}

We now use these insights to derive an estimate of the marginal density $\mathcal{P}[\bm{x}]$. We start by noting that the marginal density of the $i$-th output segment, $\mathcal{P}[\bm{x}_{[i-1,i]}|\bm{x}_{[0,i-1]}]$, is given by
\begin{equation}
\begin{aligned}
    &\mathcal{P}[\bm{x}_{[i-1,i]}|\bm{x}_{[0,i-1]}] \\
    &=\int\mathcal{D}[\bm{s}_{[0,i]}]\ 
    \mathcal{P}[\bm{x}_{[i-1,i]},\bm{s}_{[0,i]}|\bm{x}_{[0,i-1]}]\\
    &=\int\mathcal{D}[\bm{s}_{[0,i]}]\ 
    \mathcal{P}[\bm{s}_{[0,i]}|\bm{x}_{[0,i-1]}]\ 
    g[\bm{s}_{[0,i]}]\,.
    \end{aligned}
\end{equation}
The third line follows from the definition of $g[\bm{s}_{[0,i]}]=\mathcal{P}[\bm{x}_{[i-1,i]}|\bm{x}_{[0,i-1]},\bm{s}_{[0,i]}]$. Hence, we find that the probability $\mathcal{P}[\bm{x}_{[i-1,i]}|\bm{x}_{[0,i-1]}]$ can be expressed as the average
\begin{equation}
   \mathcal{P}[\bm{x}_{[i-1,i]}|\bm{x}_{[0,i-1]}] = \left\langle
    g[\bm{s}_{[0,i]}]
    \right\rangle_{\mathcal{P}[\bm{s}_{[0,i]}|\bm{x}_{[0,i-1]}]}\,.
    \label{eq:marginal-segment-average}
\end{equation}
In principle, this average can be computed using a Monte Carlo scheme, using trajectories generated from $\mathcal{P}[\bm{s}_{[0,i]}|\bm{x}_{[0,i-1]}]$.
Notice that at each iteration of the particle filter, we \emph{do} dispose of a set of trajectories $\bm{s}^1_{[0,i]},\ldots,\bm{s}^M_{[0,i]}$ which are approximately distributed according to $\mathcal{P}[\bm{s}_{[0,i]}|\bm{x}_{[0,i-1]}]$ above. Therefore, we can compute the average \cref{eq:marginal-segment-average} directly from the trajectories that are present for each iteration of the particle filter. With the notation from \cref{sec:smc}, using $g[\bm{s}^k_{[0,i]}]=\exp(-U^k_i)$, we thus obtain the estimate
\begin{equation}
    \mathcal{P}[\bm{x}_{[i-1,i]}|\bm{x}_{[0,i-1]}] \approx
    \frac{1}{M}\sum^M_{k=1} e^{-U^k_i} = \frac{w_i}{M}\,.
\end{equation}
The probability of the entire output trajectory $\mathcal{P}[\bm{x}]$ is given by the product
\begin{equation}
    \mathcal{P}[\bm{x}] = \mathrm{P}(x_0) \mathcal{P}[\bm{x}_{[0,1]}|x_0]\cdots \mathcal{P}[\bm{x}_{[n-1,n]}|\bm{x}_{[0,n-1]}]
\end{equation}
where $\mathrm{P}(x_0)$ is the probability of the initial output state $x_0$ which is assumed to be known. 
In conclusion, we arrive at the following estimate for the marginal output probability
\begin{equation}
    \hat{\mathcal{P}}[\bm{x}] = \mathrm{P}(x_0) \prod^n_{i=1} \frac{w_i}{M}
\end{equation}
which is precisely \cref{eq:smc-marginal}.

\section{\label{sec:mcmc}MCMC Sampling in Trajectory Space}

Thermodynamic Integration PWS in \cref{sec:thermodynamic-integration} relies on the computation of averages with respect to $\mathcal{P}_\theta[\bm{s}|\bm{x}]\propto\exp(-\mathcal{U}_\theta[\bm{s}, \bm{x}])$. Sampling from these distributions using the SSA (Gillespie) algorithm is not possible. Instead, in this section, we show different ways of how to implement a Markov Chain Monte Carlo (MCMC) sampler in trajectory space to generate correctly distributed trajectories.

We can build an MCMC sampler in trajectory space using the Metropolis-Hastings algorithm. To create a Markov Chain in trajectory space, we need to find a suitable proposal kernel, that generates a new trajectory $\bm{s}^\prime$ from a given trajectory $\bm{s}$ with probability $T(\bm{s}\rightarrow\bm{s}^\prime)$. We accept the proposal using the Metropolis criterion with probability
\begin{equation}
    A(\bm{s}^\prime,\bm{s})=\min\left( 1, e^{\mathcal{U}_\theta[\bm{s}, \bm{x}] - \mathcal{U}_\theta[\bm{s}^\prime, \bm{x}]}\frac{T(\bm{s}^\prime\rightarrow\bm{s})}{T(\bm{s}\rightarrow\bm{s}^\prime)} \right)
    \label{eq:metropolis-acceptance}
\end{equation}
to create a chain of trajectories with stationary distribution given by $\mathcal{P}_\theta[\bm{s}|\bm{x}]=e^{-\mathcal{U}_\theta[\bm{s}, \bm{x}]}/\mathcal{Z}_\theta[\bm{x}]$ for $0\leq\theta\leq 1$. To ensure efficient convergence of the resulting Markov chain to its stationary distribution, the proposal kernel must balance two conflicting requirements.
To efficiently explore the state space per unit amount of CPU time, the proposed trajectory $\bm{s}^\prime$ must be sufficiently different from the original trajectory $\bm{s}$, while at the same time it should not be so radically different that the acceptance probability is drastically reduced.
Thus, the design of the proposal kernel is crucial for an efficient MCMC sampler, and we will discuss various strategies to create trial trajectories. 
Since different types of trial moves can easily be combined in a Metropolis-Hastings algorithm, the most efficient samplers often incorporate multiple complementary proposal strategies to improve the exploration speed of the trajectory space.

The simplest (and naïve) proposal kernel is to generate an entirely new trajectory $\bm{s}^\prime$ independent of $\bm{s}$, by sampling directly from $\mathcal{P}[\bm{s}]$ using the SSA. 
Hence, the transition kernel is given by $T(\bm{s}\rightarrow\bm{s}^\prime)=\mathcal{P}[\bm{s}^\prime]$ and a proposal $\bm{s}\rightarrow\bm{s}^\prime$ is accepted with probability
\begin{equation}
\begin{aligned}
    A(\bm{s}^\prime,\bm{s}) &= \min\left( 1, e^{\mathcal{U}_\theta[\bm{s}, \bm{x}] - \mathcal{U}_\theta[\bm{s}^\prime, \bm{x}]}\frac{\mathcal{P}[\bm{s}]}{\mathcal{P}[\bm{s}^\prime]} \right) \\
    &= \min\left( 1, \frac{\mathcal{P}[\bm{x}|\bm{s}^\prime]^\theta}{\mathcal{P}[\bm{x}|\bm{s}]^\theta} \right)
\end{aligned}
\label{eq:acceptance-rate}
\end{equation}
where the second line follows by inserting the definition of $\mathcal{U}_\theta[\bm{s},\bm{x}]$ given in \cref{eq:ti-hamiltonian}. Although this simple scheme to completely regenerate an entire trajectory and accepting/rejecting according to $A(\bm{s}^\prime,\bm{s})$ creates correctly distributed trajectories, it should not be used in simulations to compute $\mathcal{P}[\bm{x}]$. Indeed, we get a better estimate of $\mathcal{P}[\bm{x}]$ by just using the same number of independent sample trajectories from $\mathcal{P}[\bm{s}]$ and using the brute-force scheme in \cref{eq:marginal-naive} without taking the detour of thermodynamic integration to estimate the normalization constant.

Instead, an idea from transition path sampling is to only regenerate a part of the old trajectory as part of the proposal kernel \cite{1998a.Dellago}. By not regenerating the entire trajectory, the new trajectory $\bm{s}^\prime$ is going to be correlated with the original trajectory $\bm{s}$, and correlation in general improves the acceptance rate. 
The simplest way to generate trial trajectories using a partial update is a move termed \emph{forward shooting} in which a time point $\tau$ along the existing trajectory $\bm{s}$ is randomly selected, and a new trajectory segment is regrown from this point to the end, resulting in the proposal $\bm{s}^\prime$. Since the new segment is generated according to the input statistics given by $\mathcal{P}[\bm{s}_{[T-\tau,T]}]$, the acceptance probability for the proposed trajectory is given by \cref{eq:acceptance-rate}. 
If the input dynamics given by $\mathcal{P}[\bm{s}]$ are time-reversible, we can also perform a \emph{backward shooting} move. Here, the beginning of $\bm{s}$ is replaced by a new segment that is generated backwards in time.
Assuming that the initial condition is the input's steady state distribution, the corresponding acceptance probability of the backward shooting move is again given by \cref{eq:acceptance-rate}.
Using these two moves we create an MCMC sampler where both ends of the trajectory are flexible, and thus if the trajectory is not too long, the chain will quickly relax to its stationary distribution. This is indeed the MCMC sampler used to obtain the TI-PWS results for the coupled birth-death process in \cref{sec:birth-death}.

For long trajectories it can prove to be a problem that the middle section is too inflexible when the proposal moves only regenerate either the beginning or the end of a trajectory. Therefore, one could additionally try to incorporate mid-section regrowth to make sure that also the middle parts of the trajectory become flexible.
To regrow a middle segment with duration $\tau$ of a trajectory $\bm{s}$, we have to generate a new segment of duration $\tau$ according to the stochastic dynamics given by $\mathcal{P}[\bm{s}]$ but with the additional condition that we have to connect \emph{both} endpoints of the new segment to the existing trajectory. Although the starting point of the segment can be freely chosen, the challenge is to ensure that the end point of the new segment satisfies the end-point constraint. Stochastic processes that generate trajectories under the condition of hitting a specific point after a given duration $\tau$ are called stochastic bridging processes. 

The simplest way to generate trajectories from a bridging process is by generating a trajectory segment of length $\tau$ from the normal stochastic process and rejecting the segment if it does not hit the correct end point \cite{2009.Hobolth}. Clearly, this strategy is only feasible for very short segments and when the state space is discrete, as otherwise almost every generated segment will be rejected due to not hitting the correct end point.
To avoid this problem, more efficient algorithms have been developed to simulate stochastic bridges for some types of stochastic processes.
For diffusion processes, bridges can be simulated efficiently by introducing a guiding term into the corresponding Langevin equation \cite{2017.Meulen}. 
For jump processes, bridges can be simulated using particle filters \cite{2015.Golightly}, by a weighted stochastic simulation algorithm (wSSA) \cite{2019.Gillespie}, or using random time-discretization (uniformization) \cite{2009.Hobolth}.

Further techniques to create a trajectory space MCMC samplers have been developed in the literature. \citet{2000.Crooks} describes a scheme to create MCMC moves for trajectories evolving in non-equilibrium dynamics, by making MCMC moves to change the trajectories' noise histories. In the Particle Markov Chain Monte Carlo (PMCMC) algorithm, proposal trajectories are generated using a particle filter and accepted with an appropriate Metropolis criterion \cite{2010.Andrieu}. Another class of efficient samplers for Markov jump processes can be built using uniformization \cite{2013.Rao}.

\section{Dealing with Feedback\label{sec:feedback}}

Although so far, we have assumed the stochastic dynamics of the input to be independent of the generated output trajectories, in principle all physical information processing systems exhibit feedback.
The physical interaction needed to measure the input signal necessarily affects the incoming signal, and indeed, it follows that no information can be extracted from the input signal without any perturbation of the input dynamics. Often, it is assumed that the amplitude of such perturbations is comparatively small and thus that the feedback can safely be ignored. Above, the PWS scheme was derived with this assumption. In this section, we drop the assumption and will explicitly consider systems where the produced output perturbs the input, i.e. systems where the output feeds back onto the input. 
In the following we will first discuss the additional problems that arise when computing the mutual information for a system with feedback, and subsequently we present a modified version of PWS that can be used to compute the trajectory mutual information for these systems.

\subsection{Computing the Mutual Information with Feedback between Input and Output\label{sec:mi-feedback}}

All PWS schemes presented above require the computation of the trajectory likelihood $\mathcal{P}[\bm{x}|\bm{s}]$, a quantity that is not readily available for systems with feedback.
Indeed, as already mentioned in \cref{sec:likelihood}, for a given input trajectory $\bm{s}$, the output dynamics are no longer described by a Markov process in a system with feedback, and therefore we cannot find an expression for $\mathcal{P}[\bm{x}|\bm{s}]$ based on the master equation.
This implies that for systems with feedback, PWS schemes cannot be used without modification.
While it is generally not possible to derive an expression for the conditional probability $\mathcal{P}[\bm{x}|\bm{s}]$ in systems with feedback,
we often still can compute the joint probability density $\mathcal{P}[\bm{s},\bm{x}]$ instead.
Based on this quantity, we will present a modified PWS scheme to compute the mutual information for systems with feedback.

Specifically, since PWS is a model-based approach to compute the mutual information, when there is feedback from the output back to the input, we require a complete model of the combined system. Specifically, such a model must provide an expression for the joint probability $\mathcal{P}[\bm{s},\bm{x}]$, describing the input dynamics and the interaction between input and output, including the feedback. 

An estimate of the mutual information that only relies on the computation of joint probability densities $\mathcal{P}[\bm{s},\bm{x}]$ can be obtained by
expressing the mutual information as
\begin{equation}
    \mathrm{I}(\mathcal{S}, \mathcal{X}) = \int\mathcal{D}[\bm{s}]\int\mathcal{D}[\bm{x}]\ \mathcal{P}[\bm{s},\bm{x}] \ln \frac{\mathcal{P}[\bm{s}, \bm{x}]}{\mathcal{P}[\bm{s}]\,\mathcal{P}[\bm{x}]}\,.
\end{equation}
Thus, the PWS scheme with feedback consists of the computation of
\begin{equation}
    \mathrm{I}(\mathcal{S}, \mathcal{X}) = \left\langle
    \ln\frac{\mathcal{P}[\bm{s}, \bm{x}]}{\mathcal{P}[\bm{s}]\,\mathcal{P}[\bm{x}]}
    \right\rangle_{\mathcal{P}[\bm{s},\bm{x}]}
    \label{eq:mi-with-feedback}
\end{equation}
which we want to estimate via a Monte Carlo average using samples from $\mathcal{P}[\bm{s}, \bm{x}]$. We see that while we don't need to evaluate the likelihood $\mathcal{P}[\bm{x}|\bm{s}]$, we now need to explicitly compute the joint density $\mathcal{P}[\bm{s}, \bm{x}]$, and two marginal densities, $\mathcal{P}[\bm{s}]$ and $\mathcal{P}[\bm{x}]$, for each Monte Carlo sample $(\bm{s}, \bm{x})\sim\mathcal{P}[\bm{s},\bm{x}]$. While the joint density can be evaluated directly by assumption, each of the marginalized densities can only be computed using a nested Monte Carlo estimate.

Specifically, for PWS with feedback, we need to compute \emph{two} marginalization integrals per Monte Carlo sample:
\begin{equation}
    \mathcal{P}[\bm{s}] = \int\mathcal{D}[\bm{x}]\ \mathcal{P}[\bm{s}, \bm{x}]\,,
    \label{eq:marg1}
\end{equation}
and
\begin{equation}
    \mathcal{P}[\bm{x}] = \int\mathcal{D}[\bm{s}]\ \mathcal{P}[\bm{s}, \bm{x}] \,.
    \label{eq:marg2}
\end{equation}
However, these marginalization integrals cannot be directly computed with the techniques described so far.
Note that while in \cref{sec:improvements} we discussed in detail how to compute such marginalization integrals, all methods presented there themselves require the evaluation of the likelihood $\mathcal{P}[\bm{x}|\bm{s}]$ and cannot be used directly.
Therefore, in the following subsection, we discuss how to compute marginalization integrals for systems with feedback.

Additionally, as discussed in \cref{sec:integrating-out}, we may also need to integrate out internal components of the master equation even when the output feeds back onto these internal components.
The technique discussed below can also be used in this case as a way to compute the marginalization integral in \cref{eq:marginalization_integral}.

\subsection{Marginalization Integrals for Systems with Feedback\label{sec:marginalization-feedback}}

Computing marginalization integrals in systems with feedback is harder than it is in the case without feedback.
Specifically, we will show that it is not obvious how apply the brute force Monte Carlo estimate \cref{eq:marginal-naive} or the other, more advanced techniques from \cref{sec:marginalization} to systems with feedback.
Nevertheless, if the system with feedback can be decomposed into a non-interacting part and an interacting part that includes the feedback, it is often still possible to compute marginalization integrals.
Below, we sketch the steps that are necessary in order to compute marginalization integrals for systems with feedback using such a decomposition.

For concreteness, we discuss how to compute
\begin{equation}
    \mathcal{P}[\bm{x}]=\int\mathcal{D}[\bm{s}]\ \mathcal{P}[\bm{s},\bm{x}]
    \label{eq:feedback-marginalization-integral}
\end{equation}
as the prototype for a marginalization integral we want to compute. Unlike in \cref{sec:marginalization}, we now assume that $\bm{x}$ feeds back onto $\bm{s}$. That means that we have access to the joint distribution's density $\mathcal{P}[\bm{s},\bm{x}]$, but not to the marginal density $\mathcal{P}[\bm{s}]$ or the conditional density $\mathcal{P}[\bm{x}|\bm{s}]$.

Formulated in the language of statistical physics, all of the techniques of \cref{sec:marginalization} are estimators of the free-energy difference $\Delta\mathcal{F}[\bm{x}]=\mathcal{F}[\bm{x}]-\mathcal{F}_0[\bm{x}]$ between two ensembles described by potentials $\mathcal{U}[\bm{s},\bm{x}]$ and $\mathcal{U}_0[\bm{s},\bm{x}]$. Previously, for systems without feedback, we chose these potentials to be $\mathcal{U}_0[\bm{s},\bm{x}]=-\ln\mathcal{P}[\bm{s}]$ and $\mathcal{U}[\bm{s},\bm{x}]=-\ln\mathcal{P}[\bm{s}, \bm{x}]$ with the idea that $\mathcal{U}$ is the potential corresponding to the actual system and $\mathcal{U}_0$ is the potential of a reference system with known free energy. Then, by computing the free-energy difference between the reference system and the actual system, we could compute the marginal probability $\mathcal{P}[\bm{x}]$. 

However, in systems with feedback we face a problem. Note that the actual system is still described by the potential $\mathcal{U}[\bm{s},\bm{x}]=-\ln\mathcal{P}[\bm{s}, \bm{x}]$, even with feedback. Yet, for the reference system described by $\mathcal{U}_0[\bm{s},\bm{x}]$ we cannot make the same choice as before, because the previous choice involved the marginal probability $\mathcal{P}[\bm{s}]$ which is not available with feedback. 

Instead, we have to find an alternative expression for $\mathcal{U}_0[\bm{s},\bm{x}]$.
To construct a suitable reference potential, we can use a
decomposition of the full potential into three parts
\begin{equation}
    \mathcal{U}[\bm{s}, \bm{x}] = \mathcal{U}_S[\bm{s}] + \mathcal{U}_X[\bm{x}] + \Delta\mathcal{U}[\bm{s}, \bm{x}]
    \label{eq:hamiltonian-decomposition}
\end{equation}
where $\Delta\mathcal{U}[\bm{s}, \bm{x}]$ describes the features of the system that induce interaction, or correlation, between $\bm{s}$ and $\bm{x}$.
The first two terms of the potential above, $\mathcal{U}_S[\bm{s}] + \mathcal{U}_X[\bm{x}]$, therefore describe a \emph{non-interacting} version of the system, where the input and output are fully independent of each other. We want to use the potential of that non-interacting version as our expression for $\mathcal{U}_0$, i.e. $\mathcal{U}_0[\bm{s}, \bm{x}] = \mathcal{U}_S[\bm{s}] + \mathcal{U}_X[\bm{x}]$.
To be able to do so, we require that the partition function (normalization constant)
\begin{equation}
    \mathcal{Z}_0[\bm{x}] = \int\mathcal{D}[\bm{s}]\ e^{-\mathcal{U}_0[\bm{s}, \bm{x}]}
    \label{eq:z0}
\end{equation}
is known. In other words, we need to choose the decomposition in \cref{eq:hamiltonian-decomposition} such that the partition function \cref{eq:z0} is known either analytically or numerically. If such a decomposition is found, we can compute the marginal probability $\mathcal{P}[\bm{x}]$ from the difference in free energy $\Delta\mathcal{F}[\bm{x}]$ between $\mathcal{U}$ and $\mathcal{U}_0$:
\begin{equation}
    -\ln\mathcal{P}[\bm{x}] = \mathcal{F}[\bm{x}] = \mathcal{F}_0[\bm{x}] + \Delta\mathcal{F}[\bm{x}]
\end{equation}
where $\mathcal{F}_0 = -\ln\mathcal{Z}_0[\bm{x}]$ is known.
Because we have a known expression for $\mathcal{U}_0[\bm{s},\bm{x}]$, the free-energy difference $\Delta\mathcal{F}[\bm{x}]$ can now be computed using any of the techniques described in \cref{sec:marginalization}.

As an example for finding a decomposition like \cref{eq:hamiltonian-decomposition}, let us consider the case where the joint system of input and output is described by a single master equation, i.e. we have a master equation with two components, $S$ which represents the input, and $X$ which represents the output. In such a system, information is transmitted if there exist transitions that change the copy number of $X$ with a rate that depends on the copy number of $S$. In terms of chemical reactions, $S\rightarrow S+X$ is an example for such a transition. In turn, this system exhibits feedback if at least one of the transitions that change the copy number of $S$ has a rate that depends on $X$, as for example with the reaction $S + X\rightarrow X$. Note that with such reactions, the dynamics of $S$ depend on the current copy number of $X$, and therefore we cannot evolve $S$ trajectories independently of $X$ trajectories, a consequence of feedback. 
Both of the reactions $S\rightarrow S+X$ and $S + X\rightarrow X$ introduce correlations between the $S$ and $X$ trajectories. 

In a non-interacting system, such interactions between the input and output must be absent. Thus, a non-interacting version of the reaction system contains no single reaction that involves both $S$ and $X$.
We will now describe how we can use that non-interacting version of the reaction system, to obtain the reference potential $\mathcal{U}_0[\bm{s}, \bm{x}]$.
Since the input and output trajectories are completely independent in the non-interacting system, we can express the joint distribution's probability density as the product of the individual component's trajectory densities, $\mathcal{P}_0[\bm{s}, \bm{x}]=\mathcal{P}_0[\bm{s}]\ \mathcal{P}_0[\bm{x}]$. 
Note that $\mathcal{P}_0[\bm{s}]$ and $\mathcal{P}_0[\bm{x}]$ should not be confused with the marginal probabilities $\mathcal{P}[\bm{s}]$ and $\mathcal{P}[\bm{x}]$ of the \emph{interacting} version of the reaction system, which must be computed using a marginalization integral.
Since in the non-interacting version both, $S$ and $X$ obey independent dynamics which are characterized by individual master equations, both $\mathcal{P}_0[\bm{s}]$ and $\mathcal{P}_0[\bm{x}]$ can be individually computed using \cref{eq:traj-prob-master-eq}. 
Thus, in this case, the non-interacting potential is $\mathcal{U}_0[\bm{s}, \bm{x}] = -\ln\mathcal{P}_0[\bm{s}]-\ln\mathcal{P}_0[\bm{x}]$ and, since the probability densities $\mathcal{P}_0[\bm{s}]$ and $\mathcal{P}_0[\bm{x}]$ are normalized, the corresponding partition function is $\mathcal{Z}_0=1$. 
Hence, for this reaction system, we can straightforwardly define a non-interacting version that can be used to obtain the reference potential $\mathcal{U}_0[\bm{s}, \bm{x}]$. Using the techniques described in \cref{sec:marginalization}, we can then compute the free-energy difference between $\mathcal{U}_0[\bm{s}, \bm{x}]$ and $\mathcal{U}[\bm{s}, \bm{x}]=-\ln\mathcal{P}[\bm{s}, \bm{x}]$, where the latter potential describes the dynamics of the fully interacting system. Specifically, we can compute the marginal probabilities $\mathcal{P}[\bm{s}]$, $\mathcal{P}[\bm{x}]$ pertaining to the interacting system which are required for the mutual information estimate in \cref{eq:mi-with-feedback}.

In summary, for systems with feedback, we can compute marginalization integrals by specifying a reference potential $\mathcal{U}_0[\bm{s}, \bm{x}]$ by finding a non-interacting version of the system. However, barring a decomposition into interacting and non-interacting potentials, there is generally no unambiguous choice of the reference potential $\mathcal{U}_0[\bm{s},\bm{x}]$ to compute the marginal probability $\mathcal{P}[\bm{x}]$. Still, if a suitable expression for $\mathcal{U}_0[\bm{s},\bm{x}]$ can be found, we can make use of the techniques developed in \cref{sec:marginalization} to compute marginal probability $\mathcal{P}[\bm{x}]$. Thus, the specific choice of $\mathcal{U}_0[\bm{s},\bm{x}]$ is system-specific.

\section{Stochastic Chemotaxis Model\label{sec:chemotaxis-model}}

\begin{table*}[t]
    \centering
    \begin{tabular}{c|Sl|p{9cm}}
    \textbf{parameter} & \multicolumn{2}{c|}{\textbf{value}} & \textbf{description} \\
    \hline
        $a_v$ & 157.1 & \si{\micro\meter\squared\per\second\squared} & variance of up-gradient velocity \cite{2021.Mattinglyg3a} \\
        $\lambda$ & 0.862 & \si{\per\second} & velocity correlation decay constant \cite{2021.Mattinglyg3a} \\
        $c_0$ & 100 & \si{\micro\Molar} & mean ligand concentration \\
    \hline
        $N$ & 6 & & number of receptor units per cluster \cite{2010.Shimizu} \\
        $N_c$ & 400 & & number of receptor clusters \cite{2004.Li} \\
        $M$ & 4 & & number of methylation sites per receptor \cite{2010.Shimizu} \\
        $N_Y$ & 10 000 & & total copy number of CheY proteins (phosphorylated and unphosphorylated) \cite{2004.Li} \\
    \hline
        $K_a$ & 2900 & \si{\micro\Molar} & ligand dissociation constant of active receptors \cite{2020.Kamino} \\
        $K_i$ & 18 & \si{\micro\Molar} & ligand dissociation constant of inactive receptors \cite{2020.Kamino} \\
    \hline
        $k_R$ & 0.1 & \si{\per\second} & methylation rate \cite{2010.Shimizu,2021.Mattinglyg3a} \\
        $k_B$ & 0.2 & \si{\per\second} & demethylation rate \cite{2010.Shimizu,2021.Mattinglyg3a} \\
        $k_A$ & 0.015 & \si{\per\second} & phosphorylation rate \cite{2002.Sourjik,2002.Sourjikt5} \\
        $k_Z$ & 10.0 & \si{\per\second} & dephosphorylation rate \cite{2002.Sourjik,2002.Sourjikt5} \\
        $\phi_Y$ & 0.17 & & steady-state fraction of phosphorylated CheY \cite{2002.Sourjik} \\
    \hline
        $m_0 / N$ & 0.5 & & receptor methylation level at zero ligand concentration \cite{2010.Shimizu} \\
        $\dfm$ & -2.0 & $k_\mathrm{B}T$ & free energy change of active conformation from attachment of one methyl group \cite{2010.Shimizu} \\
    \end{tabular}
    \caption{\b{The parameters required for the chemotaxis model, based on literature values. 
    These are the parameters used in the so-called literature-based model.
    In the fitted model (see main text) the same parameter values are chosen, except for $N=15$ and $N_c=9$, which were obtained by fitting to the data of \citet{2021.Mattinglyg3a};
    we note that changing $N$ and $N_c$ also requires updating $k_A$ to keep the fraction $\phi_Y$ of phosphorylated CheY constant.}}
    \label{tab:chemotaxis-parameters}
\end{table*}

We developed a stochastic chemotaxis model that describes individual reactions using a master equation framework.
In our model, receptors are organized in clusters. 
\b{To each cluster we assign a probability of being active that depends on the ligand concentration.
Additionally, we explicitly model the methylation and demethylation events of each receptor which affect the activity of a cluster.
The cluster activity, in turn, determines its ability to phosphorylate the protein CheY. Phosphorylated CheY binds to the molecular motors driving the flagella, which alter the cell's tumbling rate. However, we do not model this downstream effect in our model.}

\subsection{MWC Model}

Receptors are organized in clusters on the cell surface. In our model, each cluster consists of $N$ receptors. The ligand binding dynamics to a cluster is cooperative, and in spirit of the Monod-Wyman-Changeux (MWC) model \cite{1965.Monod,2004.Sourjik} we model this cooperativity by coupling the ligand-binding dynamics to conformational switching dynamics of the receptors. 
Moreover, the energetic cost of two receptors in the same cluster being in different conformational states is prohibitively large.
This means that all receptors within a cluster of size $N$ switch conformations in concert, so that we can meaningfully speak of an active or an inactive cluster. 
\b{A typical value for the cluster size is reported to be $N=6$ by \citet{2010.Shimizu}.}
Detailed balance requires that the ligand binding affinity depends on whether a cluster is in the active or inactive state.
Consequently, we have a dissociation constant $K_a$ for a ligand bound to an active receptor and another dissociation constant $K_i$ for a ligand bound to an inactive receptor.
For chemotaxis, $K_a\gg K_i$, i.e. the ligand binding affinity is higher for the inactive state.

Additionally, each receptor monomer has $M$ methylation sites that can affect its conformation and therefore the kinase activity. \b{The aspartate receptor Tar has $M=4$ methylation sites \cite{2010.Shimizu}.} Methyl groups can be attached to a receptor by the protein CheR and are removed by the protein CheB.
We model the receptors' methylation dynamics following the model of \citet{1997.Barkai}, where CheB can only demethylate active receptors. 
Additionally, to ensure exact adaptation, in our model CheR can only attach methyl groups to inactive receptors, as in Ref.~\cite{1999.Morton-Firth}.

\b{In an environment with ligand concentration $c$, the probability of a receptor cluster with $m$ methylated sites to be active, $p_a(c, m)$, is determined by the free-energy difference between the active and inactive receptor states
\begin{equation}
    p_a(c, m) = \frac{1}{1 + e^{-f(c, m)}} \label{eq:prob_active}
\end{equation}
where
\begin{equation}
    f(c, m) = N \ln\left( \frac{1 + c/K_i}{1 + c/K_a} \right) + \dfm (m - m_0) \,. \label{eq:free_energy_active}
\end{equation}
Here, the number of methylated sites of a cluster (not receptor) is denoted by $m$, ranging from $0$ to $NM$. The parameters are again taken from \citet{2010.Shimizu}. Their experimental results indicate that $\delta\!f_m = -2 k_\mathrm{B}T,\ m_0=-N/2$. \citet{2020.Kamino} report ligand dissociation constants of $K_a = \SI{2900}{\micro\Molar}$ for active receptors and $K_i = \SI{18}{\micro\Molar}$ for inactive Tar receptors (for MeASP). Note that in the equations we assume units such that $k_\mathrm{B} T=1$.}

The dynamics of methylation in our model are described by the following mean-field equation
\begin{equation}
    \frac{\mathrm{d}m}{\mathrm{d}t} = (1-p_a(c, m)) k_R - p_a(c, m) k_B \,.
    \label{eq:methylation_dynamics}
\end{equation}
The system reaches a steady state for the adapted activity $p_a(c, m)=a_0$ where
\begin{equation}
    a_0 = \frac{k_R}{k_R + k_B} \,.
    \label{eq:adapted_activity}
\end{equation}
\b{The steady-state methylation $m^\star$ can be obtained from \cref{eq:prob_active,eq:free_energy_active} by solving $p_a(c, m^\star)=a_0$:
\begin{equation}
    m^\star = m_0 + \frac{N \ln\left( \frac{1 + c/K_i}{1 + c/K_a} \right)+ \ln\left( \frac{1-a_0}{a_0} \right)}{-\dfm}\,. 
\end{equation}}

\b{To characterize the methylation timescale, we linearize the dynamics of $m(t)$ around the steady state (at constant ligand concentration $c(t)=c_0$).
To first order, we can write
\begin{equation}
    \frac{\mathrm{d} m}{\mathrm{d} t} = -\frac{m(t) - m^\star}{\tau_m}\,.
\end{equation}
where $\tau_m$ is the characteristic timescale of the methylation dynamics.
We find $\tau_m$ by expanding $p_a$ (\cref{eq:prob_active}) around $m=m^\star$:
\begin{equation}
\begin{aligned}
    p_a(c, m) &= p_a(c, m^\star) + \frac{\partial p_a}{\partial m} \Bigg|_{m^\star} (m - m^\star) + \mathcal{O}(m^2) \\
    &= a_0 [1 - \dfm (1-a_0) (m-m^\star)] + \mathcal{O}(m^2)
    \,,
\end{aligned}
\end{equation}
and then plugging this first-order expansion into \cref{eq:methylation_dynamics} to get
\begin{equation}
    \frac{\mathrm{d} m}{\mathrm{d} t} =   
     \frac{\dfm (m - m^\star)}{k_R^{-1} + k_B^{-1}} \,.
\end{equation}
So, we find that for small perturbations, the timescale for methylation to approach steady state is given by
\begin{equation}
    \tau_m = \frac{k^{-1}_R + k^{-1}_B}{-\dfm} \,.
\end{equation}
Thus, the parameters $k_R$ and $k_R$ control two important characteristics of the methylation system: the adapted activity $a_0$ and the methylation time scale $\tau_m$.
\citet{2010.Shimizu} report an adapted activity of $a_0 = 1/3$ and based on experimental data \cite{2021.Mattinglyg3a,2010.Shimizu} we assume a methylation time scale of $\tau_m = \SI{10}{\second}$. Our parameter choice, which is consistent with both of these observations, is $k_R=\SI{0.075}{\per\second}$ and $k_B=\SI{0.15}{\per\second}$.}

CheY is phosphorylated by CheA, the receptor-associated kinase. The kinase activity is directly linked to the activity of a receptor cluster. Therefore, we assume that CheY is phosphorylated by active receptor clusters. Dephosphorylation of CheY-p is catalyzed by the phosphatase CheZ, which we assume to be present at a constant concentration. 
\b{The CheZ-catalyzed dephosphorylation rate was reported to be $\SI{2.2}{\per\second}$ for attractant response and $\SI{22}{\per\second}$ for repellent response \cite{2002.Sourjikt5}. Based on this data, we use the approximate dephosphorylation rate $k_Z = \SI{10}{\per\second}$ in our model. 
In the fully adapted state the fraction of active receptors is $a_0$ and therefore the mean fraction of phosphorylated CheY, $\phi_Y = [\text{CheYp}]/([\text{CheY}] + [\text{CheYp}])$, is given by
\begin{equation}
    \phi_Y = \frac{a_0 N_c k_A}{k_Z + a_0 N_c k_A} \,.
\end{equation}
In the fully adapted state the phosphorylated fraction was found to be $\phi_Y \approx 0.16$ \cite{2002.Sourjik}. Hence, we infer a phosphorylation rate of $k_A = k_Z \phi_Y / (a_0 N_c (1 - \phi_Y)) = \SI{0.015}{\per\second}$ for the literature-based model. Accordingly, for the ``fitted'' model,  based on fitting $K(t)$ and $N(t)$ to those measured by \citet{2021.Mattinglyg3a}, we use a larger phosphorylation rate due to the smaller number of clusters $N_c$.}


\subsection{Reaction Kinetics}

\b{Since the timescale of conformational switching of active and inactive receptors and ligand binding is much faster \cite{2013.Ortega} than the timescale of phosphorylation or methylation, we don't explicitly model ligand (un)binding and conformational switching. 
Each cluster is characterized by its methylation state $m$.
This ranges from 0 to the total number of methylation sites, which equals the number of sites per receptor $M$ times the number of receptors per cluster $N$.
In our Gillespie simulation, each possible state of a cluster is its own species, i.e., we have species $\mathrm{C}_m$ for $m=0,\ldots,NM$. Overall, our chemotaxis model consists of four types of reactions that describe 
(a) the methylation of a receptor $\mathrm{C}_m\to\mathrm{C}_{m+1}$, 
(b) the demethylation of a receptor $\mathrm{C}_m\to\mathrm{C}_{m-1}$, 
(c) the phosphorylation of CheY $\mathrm{C}_m + \mathrm{Y} \to \mathrm{C}_m + \mathrm{Y_p}$, and 
(d) the single dephosphorylation reaction $\mathrm{Y_p} \to \mathrm{Y}$.
Thus, due to the combinatorial explosion of receptor states, the system has a total number of $3NM+2$ elementary reactions (which amounts to 75 reactions in the literature-based model and 182 reactions in the fitted model).}

The ligand-concentration dependent methylation rate for $\mathrm{C}_m\rightarrow \mathrm{C}_{m+1}$ is given by
\begin{equation}
    k_{m+}(c,m) = (1-p_a(c,m)) k_R \,.
\end{equation}
The term $1-p_a(c,m)$ is needed because only inactive receptors can be methylated.
The demethylation rate for $\mathrm{C}_{m}\rightarrow \mathrm{C}_{m-1}$ is given by
\begin{equation}
    k_{m-}(c, m) = p_a(c,m) k_B 
\end{equation}
where only active receptors can be demethylated.
These zero-order dynamics of (de)methylation of receptors lead to the adaptive behavior of the chemotaxis system as described above.

Similarly, only active receptors can phosphorylate the CheY protein using the receptor-associated kinase CheA, therefore we model phosphorylation as a reaction $\mathrm{C}_{m} + \mathrm{Y} \rightarrow \mathrm{C}_{m} + \mathrm{Y_p}$ with propensity
\begin{equation}
    k_{\mathrm{Y}\rightarrow\mathrm{Y_p}}(c,m) = p_a(c, m) k_A
\end{equation}
where $k_A$ is a constant that represents the phosphorylation rate of an active cluster. The dephosphorylation $\mathrm{Y_p}\rightarrow \mathrm{Y}$ is carried out by the phosphatase CheZ at a constant rate $k_Z=\SI{10}{\per\second}$.

\section{\label{sec:chemotaxis_input}Stochastic Dynamics of the Input Signal for Chemotaxis}

\b{We assume an \emph{Escherichia coli} bacterium is swimming in a static nutrient concentration gradient $c(x)$. Following \citet{2021.Mattinglyg3a}, an exponential gradient $c(x)=c_0 e^{gx}$ with gradient steepness $g$ is assumed.}

\b{
In a shallow gradient, the speed $v_x(t)$ of \emph{E. coli} along the gradient axis can be considered as a stochastic process that fluctuates around the net chemotactic drift velocity. Following \citet{2021.Mattinglyg3a}, we assume that in a shallow gradient the bacterial swimming dynamics are, to a good approximation, the same as in the absence of a gradient. Their experimental evidence shows that the velocity fluctuations in absence of a gradient are described by an exponentially decaying auto-correlation function:
\begin{equation}
    V(t) = \langle v_x(0) v_x(t) \rangle = a_v e^{-\lambda |t|} \,.
\end{equation}
Therefore, in a shallow gradient, the gradient-climbing speed can be modeled as a zero-mean Ornstein-Uhlenbeck process
\begin{equation}
    \frac{\mathrm{d}v_x}{\mathrm{d}t} = -\lambda v_x + \sigma \xi(t)
\end{equation}
where $\sigma = \sqrt{2 a_v \lambda}$, and $\xi(t)$ is white noise with $\langle \xi(t) \xi(t^\prime) \rangle = \delta (t - t^\prime)$. The $x$-position of the bacterium is given by the integral of the velocity, i.e., $x(t)=\int^t_0\mathrm{d}\tau\, v_x(\tau)$. Thus, when projected onto the gradient axis, the bacterium performs a 1D random walk described by the Langevin equation
\begin{equation}
    \frac{\mathrm{d}^2 x}{\mathrm{d}t^2} = -\lambda \frac{\mathrm{d}x}{\mathrm{d}t} + \sigma \xi(t) \,.
    \label{eq:random_walk}
\end{equation}
}

\b{
Since the bacterium moves in a static concentration gradient described by $c(x)$, the concentration dynamics that the cell observes are generated directly from its own movement dynamics. At time $t$ the cell is at position $x(t)$ and thus measures the concentration $c(t)=c(x(t))$. We find the stochastic dynamics of $c$ by differentiating using the chain rule
\begin{equation}
    \frac{\mathrm{d}c}{\mathrm{d}t} = \frac{\partial c}{\partial x} \frac{\partial x}{\partial t} = g c(t)\,v_x(t)\,.
    \label{eq:concentration_dynamics}
\end{equation}
The concentration dynamics are thus fully determined by the stochastic dynamics of the cell's swimming velocity $v_x(t)$ in the absence of a gradient and by the shape of the concentration gradient $c(x)$.}

\b{
In the PWS simulations we use $c(t)$ directly as the input to our system. Yet, for the Gaussian approximation we need to use a different input signal because the chemotaxis system does not respond linearly to $c(t)$. Instead, \citet{2021.Mattinglyg3a} show that the chemotaxis system responds approximately linear to an input $s(t)$ defined by
\begin{equation}
    s(t) = \frac{\mathrm{d}}{\mathrm{d}t} \ln c(t) = gv_x(t) \,.
\end{equation}
The correlation function of $s(t)$ is given by
\begin{equation}
    C_{ss}(t) = \langle s(\tau) s(t + \tau) \rangle = g^2 V(t)\,.
\end{equation}
The power spectral density of this signal is given by the Fourier transform of its correlation function:
\begin{equation}
P_{ss}(\omega) = g^2 V(\omega) = g^2 \frac{2a_v \lambda}{\omega^2 + \lambda^2}\,.
\end{equation}}

\b{
We use this same input below in \cref{sec:lna} to compute the Gaussian approximation of the mutual information rate. 
As discussed in more detail in the main text, we note that the mutual information between the output and the input trajectory $c(t)$, as measured in the PWS simulations, is identical to that between the output and the input trajectory $s(t)$, as computed in the Gaussian model because of the deterministic and monotonic mapping between $c(t)$ and $s(t)$.}

\section{Mutual Information Rate for the Gaussian Chemotaxis System \label{sec:lna}}


\b{In Ref.~\cite{2010.Tostevin} it is shown that the mutual information for a discrete-time Gaussian system can be computed using
\begin{equation}
    \mathrm{I}(\mathcal{S}, \mathcal{X}) = \frac{1}{2} \ln \left[ \frac{|\Sigma_{ss}| |\Sigma_{xx}|}{|Z|} \right] \label{eq:gaussian_mi}
\end{equation}
with
\begin{equation}
    Z = \left( \begin{array}{cc}
        \Sigma_{ss} & \Sigma_{xs} \\
        \Sigma_{sx} & \Sigma_{xx}
    \end{array} \right)\,.
\end{equation}
Here $\Sigma_{ss}$ and $\Sigma_{xx}$ are the (auto-)covariance matrices of the input and the output, respectively, whereas $\Sigma_{sx}$ and $\Sigma_{xs}$ contain the cross-covariances. The matrix elements are thus given by $\Sigma^{ij}_{\alpha\beta} = \langle \alpha(t_i) \beta(t_j) \rangle = C_{\alpha\beta}(t_i - t_j)$ where $C_{\alpha\beta}(t)$ denote the (cross-)correlation functions of the system's input and output variables. }

\b{
In continuous time, the information transmission rate $R(\mathcal{S}, \mathcal{X})$ of a Gaussian system in steady state can be computed exactly from the spectral density functions of the system:
\begin{equation}
    R(\mathcal{S}, \mathcal{X}) = -\frac{1}{4\pi} \int^\infty_{-\infty} \mathrm{d}\omega\ \ln\left[1 - \frac{|P_{sx}(\omega)|^2}{P_{ss}(\omega)P_{xx}(\omega)}\right] \,.
    \label{eq:info_rate_gaussian}
\end{equation}
Here, the power spectral density $P_{\alpha\beta}(\omega)$ is the Fourier transform of $C_{\alpha\beta}(t)$, defined as
\begin{equation}
    P_{\alpha\beta}(\omega) = \int^\infty_{-\infty} \mathrm{d}t\ e^{-i\omega t} C_{\alpha\beta}(t) \,.
\end{equation}
}

\b{
The information rate in the Gaussian framework can thus be computed by obtaining the required \hbox{(cross-)}correlation functions.
In their experiments with \emph{E. coli} bacteria, \citet{2021.Mattinglyg3a} don't obtain these correlation functions directly, however. Instead, they obtain three kernels, $V(t)$, $K(t)$ and $N(t)$, from which the correlation functions can be inferred. We proceed by discussing the three kernels individually.}

\b{$V(t)$ denotes the autocorrelation function of the swimming velocity of bacteria, i.e., $V(t)=\langle v_x(\tau) v_x(\tau + t)\rangle$. As explained in \cref{sec:chemotaxis_input}, the swimming dynamics of the bacteria determine the statistics of the input signal $s(t) = \frac{\mathrm{d}}{\mathrm{d}t}\ln c(t)$, where $c(t)$ is the ligand concentration as experienced by the bacterium and $g$ is the gradient steepness. The input signal correlation function, denoted by $C_{ss}(t)$, can then be expressed as $C_{ss}(t) = g^2 V(t)$.}

\b{
The response kernel, denoted by $K(t)$, represents the time evolution of the average activity of the receptors in response to an instantaneous step change in the input concentration.
More precisely, $K(t)$ is defined as
\begin{equation}
    K(t) = \theta(t)  \langle a(t) - a_0 \rangle \ln\frac{c_s}{c_0}
\end{equation}
where we assume the input concentration jumps instantaneously from $c_0$ to $c_s$ at time $t=0$.
$\theta(t)$ is the Heaviside step function.
Note that because the signal $s(t)$ is defined as the time-derivative of the concentration $c(t)$, a step-change in $c(t)$ corresponds to a delta impulse in $s(t)$.
Thus, $K(t)$ describes the deterministic dynamics of the system after being subjected to a unit stimulus $s(t) = \delta(t)$, i.e. $K(t)$ is the Green's function of the system.
The stochastic response $a(t)$ to an arbitrary time-dependent signal $s(t)$ can be written as a convolution of $K(t)$ with $s(t)$
\begin{equation}
    a(t) = a_0 + \int^t_{-\infty} \mathrm{d}t^\prime\ K(t-t^\prime) s(t^\prime) + \eta_a(t)
\end{equation}
where $\eta_a(t)$ is the receptor activity noise.
We define the response $x(t)=a(t) - a_0$.
Assuming the input statistics are stationary and described by the correlation function $C_{ss}(t)$, it is easy to show that the cross-correlation between $s(t)$ and $x(t)$ is given by
\begin{equation}
    C_{sx}(t) = \langle s(\tau) x(\tau + t) \rangle = \int^t_{-\infty} \mathrm{d}t^\prime\ K(t-t^\prime) C_{ss}(t^\prime) \,.
    \label{eq:conv_correlation}
\end{equation}
In other words, the cross-correlation between $s(t)$ and $x(t)$ is given by the convolution of the response kernel with the input correlation function.
}

\b{
The noise kernel $N(t)$ describes the autocorrelation of the activity fluctuations in the absence of an input stimulus. In particular, $N(t) = \langle x(\tau) x(\tau + t)\rangle = \langle \eta_a(\tau) \eta_a(\tau + t) \rangle$ where we assume that $s(t)=0$.
}

\b{
We now rewrite \cref{eq:info_rate_gaussian} for the mutual information rate in terms of the three kernels described above.
So we need to express the power spectra $P_{\alpha\beta}(\omega)$ in terms of the Fourier-transformed kernels $V(\omega)$, $K(\omega)$, and $N(\omega)$.
In \cref{sec:chemotaxis_input} we already showed that $P_{ss}(\omega) = g^2 V(\omega)$.
The cross power spectrum is given by $P_{sx}(\omega) = K(\omega) P_{ss}(\omega)$ which follows from \cref{eq:conv_correlation}. 
Finally, from Ref.~\cite{2009.Tostevin} we use the identity $P_{xx}(\omega) = P_{ss}(\omega) |K(\omega)|^2 + N(\omega)$ to express the output power spectrum. We insert these expressions into \cref{eq:info_rate_gaussian} which yields
\begin{equation}
    R(\mathcal{S}, \mathcal{X}) = \frac{1}{4\pi} \int^\infty_{-\infty} \mathrm{d}\omega\ \ln\left(1+\frac{g^2 V(\omega) |K(\omega)|^2}{N(\omega)}\right).
\end{equation}
Then, for shallow gradients, we can make a Taylor approximation to obtain
\begin{equation}
    R(\mathcal{S}, \mathcal{X}) = \frac{g^2}{4\pi} \int^\infty_{-\infty} \mathrm{d}\omega\ \frac{V(\omega)|K(\omega)|^2}{N(\omega)} + \mathcal{O}(g^4) \,.
    \label{eq:gauss_mi_approx}
\end{equation}
This result shows that the information rate in shallow gradients is proportional to $g^2$ and the proportionality constant is determined by the measured kernels.
\citet{2021.Mattinglyg3a} obtain the relevant kernels $V(\omega)$, $K(\omega)$, and $N(\omega)$ from experiments by fitting phenomenological models to their single-cell data.
How we obtain these kernels for our chemotaxis model is described in \cref{sec:comparison}.
}

\begin{figure*}[t]
    \centering
    \includegraphics{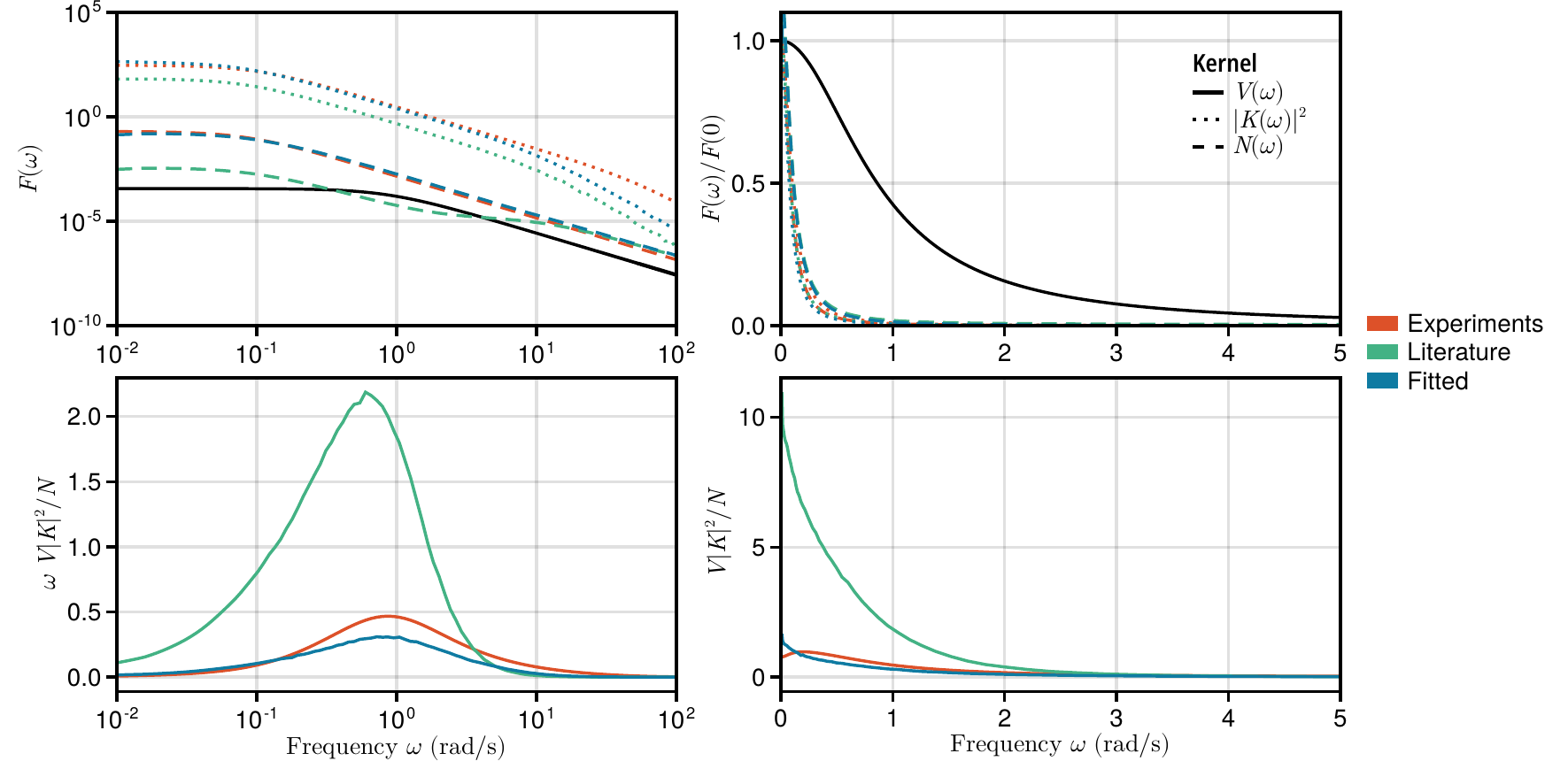}
    \caption{Fourier representation of the kernels for computing the information transmission rate in the Gaussian approximation, the velocity power spectrum $V(\omega)$ (units $(\si{\milli\meter\per\second})^2$), the squared frequency response $|K(\omega)|^2$, and the noise power spectrum $N(\omega)$. The top-left panel shows the individual Fourier kernels as a function of frequency $\omega$ for the different models. On the top-right the normalized kernels are shown with linear axis scales. In the bottom panels the integrand for computing the mutual information rate in the Gaussian approximation is shown. In the bottom right, the area under the curves represents the proportionality between the squared gradient steepness $g^2$ and the information rate (units \si{\bit\per\second \per \milli\metre\squared}). In the bottom left plot, the integrand is multiplied by $\omega$, so that with log scaling of the axes the area under the curve is equal to the integral. }
    \label{fig:fourier_kernels}
\end{figure*}

\b{
\Cref{fig:fourier_kernels} shows the Fourier representations of the relevant kernels, $V(\omega)$, $K(\omega)$, and $N(\omega)$. We computed these kernels for the three different systems: the literature-based model, the fitted model, and the experimental system of \citet{2021.Mattinglyg3a}. Because the kernels are different, so are the Gaussian information rates that we obtain. The results are shown and discussed in the main text.
}

\section{Quantitative Comparison with Experiments\label{sec:comparison}}

\b{We wanted to test whether our theoretical model (``literature-based model'') reproduces the behavior of the experimental system studied by \citet{2021.Mattinglyg3a}.
To do so, we measured the same kernels for our model as were measured experimentally. Indeed, the simulation protocol we used for measuring the kernels was directly modeled after the experimental protocol \cite{2021.Mattinglyg3a}.}

\subsection{Measuring the Response Kernel}

\b{To compute the response kernel, we record the response of our chemotaxis model to a step input. 
To do this we first adapt the system to the ligand concentration $c_0=\SI{100}{\micro\Molar}$ for $t_0 = \SI{50}{\second}$ and then instantaneously increase the concentration to from $c_0$ to $c_s=c_0 + 0.1 c_0$.
We then record the response of the system in the \SI{200}{\second} following this step-increase at a time resolution of \SI{0.01}{\second}.
Note that we don't directly obtain the average receptor activity $a(t)$ from the simulations.
\b{Since in the PWS simulations we compute the mutual information between the ligand trajectory $c(t)$ and the output trajectory of the phosphorylated CheY, $y_p(t)$, we estimate $a(t)$ from the phosphorylation level of CheY.}
Specifically, we record the fraction $f(t)$ between phosphorylated and unphosphorylated CheY, i.e. $f(t)=[\mathrm{Y_p}]/[\mathrm{Y}]$.
This fraction serves as a proxy for the activity $a(t)$. Indeed, because the copy number of CheY is relatively large, we can estimate the activity as
\begin{equation}
    a(t) = \frac{k_Z}{k_A N_c} f(t) \,.
    \label{eq:frac_to_act}
\end{equation}
We then obtain an estimate for the response Kernel $K(t) = \ln (c_s/c_0) \langle a(t-t_0)-a(t_0) \rangle$ by averaging the recorded activity traces over $10^5$ recorded trajectories.}

\subsection{Output Noise Statistics}

\b{
We can similarly obtain the noise statistics of the output from simulations of our chemotaxis model. In this case, we stochastically evolve the chemotaxis model at constant ligand concentration $c_0=\SI{100}{\micro\Molar}$ for a very long time of \SI{1e4}{\second}. The result is a time trace of the activity $a(t)$, which we again obtain from the fraction $f(t)$ using \cref{eq:frac_to_act}. We discretize this time trace at a resolution of \SI{0.01}{\second}. This results in a time series $\boldsymbol{a}=(a_1,\ldots,a_N)^T$ where $a_i=a(t_i)$. To estimate the correlations in the time series we subtract the mean activity from each data point and thus obtain the data vector $\boldsymbol{x}$ where $x_i=a_i - \sum^N_{j=1} a_j/N$. From $\boldsymbol{x}$ we estimate the auto-correlation function $C_{xx}(t) = \langle x(\tau) x(\tau + t) \rangle$ of the activity. To obtain precise results we average the correlation function for $10^5$ trajectories.
}

\subsection{Obtaining the Fourier Kernels using the FFT}

\b{To compute the Gaussian information rate, we need the frequency-space representations of the kernels $V(t)$, $K(t)$, and $N(t)$. We already derived the analytical form of $V(\omega)$ in \cref{sec:chemotaxis_input}. We obtain $K(\omega)$ and $N(\omega)$ numerically via a discrete Fourier transform of the corresponding measured kernel.}

\begin{figure*}[t]
    \centering
    \includegraphics[width=0.75\textwidth]{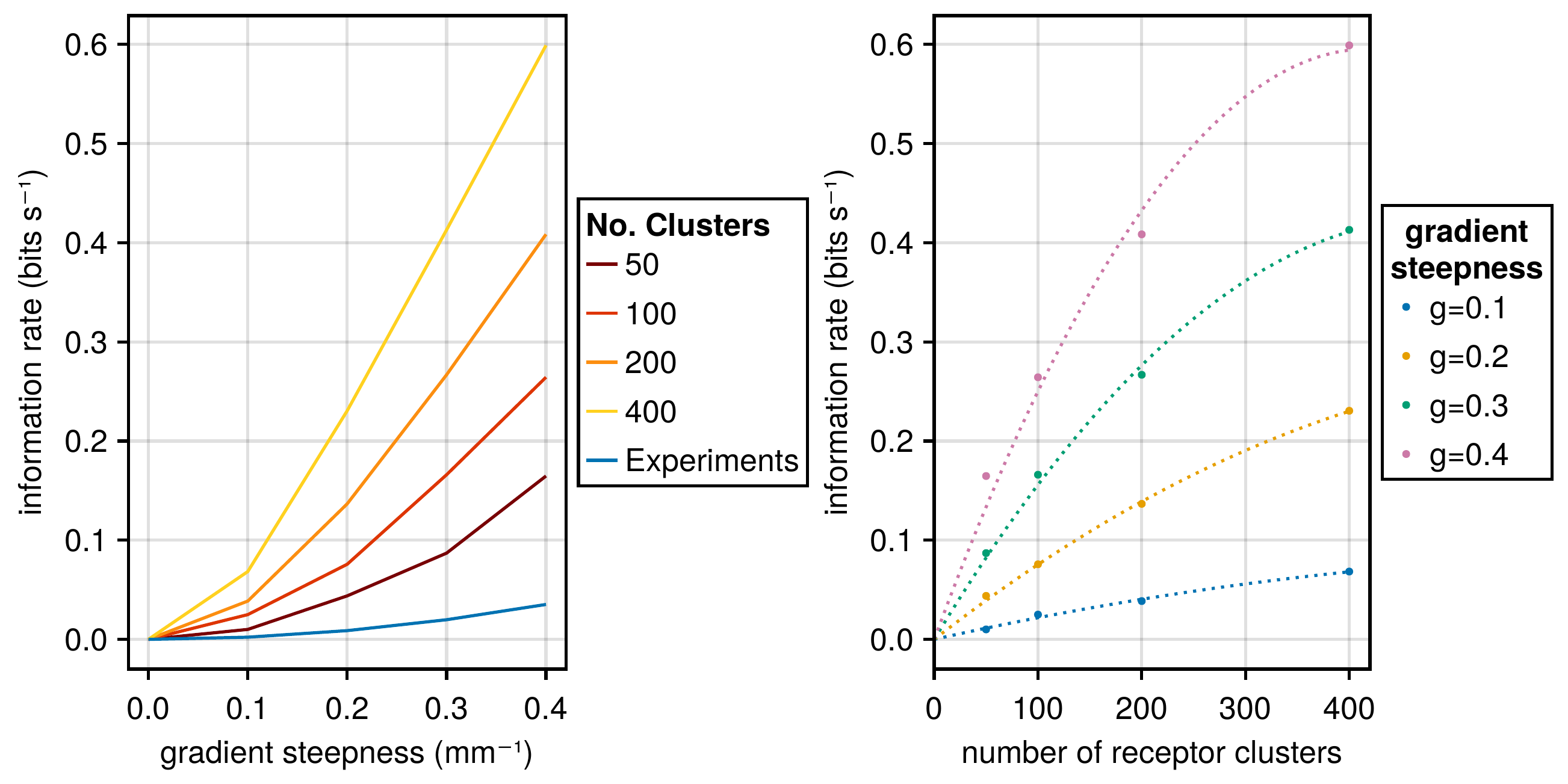}
    \caption{The information rate as a function of the number of receptor clusters $N_c$. 
    The cluster size is fixed at $N=15$.
    The left panel shows the increase of information rate as a function of gradient steepness for different values of $N_c$, including a line for the experimental data from \citet{2021.Mattinglyg3a}. 
    The right panel shows the same data but highlights the increase of the information rate and when increasing the number of receptor clusters.
    A quadratic fit (shown as dotted lines) is used to extrapolate the information rate.
    All results were obtained using RR-PWS.
 }
    \label{fig:num_clusters}
\end{figure*}

\b{
As explained above, we compute time-discretized kernels $K_i = K(t_i)$ and $N_i = N(t_i)$ from time traces obtained via stochastic simulations of our model. We take samples at times $t_0,\ldots,t_{N-1}$ at a sampling frequency of $f_s = 100\ \mathrm{s}^{-1}$. Then, we use the discrete Fourier transform (DFT) to obtain approximations for $K(\omega)$ and $N(\omega)$ as follows. The DFT coefficients $\tilde{K}_k$ of a the time discrete response kernel are given by
\begin{equation}
    \tilde{K}_k = \sum_{n=0}^{N-1} K_n e^{-i2\pi nk/N}
\end{equation}
where $k=0,1,\ldots,N-1$.
These DFT coefficients can be computed efficiently using the Fast Fourier Transform (FFT) algorithm. The DFT provides point estimates for the Fourier-domain kernel $K(\omega)$ at discrete frequencies
\begin{equation}
    \omega_k = \frac{2\pi f_s k}{N},\quad k=0,1,...,N-1 \,,
\end{equation}
i.e., $K(\omega_k) \approx \tilde{K}_k$. 
This approximation introduces some level of error, known as spectral leakage, due to the finite duration and sampling of the signal. This error can be reduced by multiplying the time-domain kernel with a window function.
Thus, we multiply the kernel with a Hanning window, which is a smooth function that tapers at the edges of the kernel, reducing the effect of discontinuities at the beginning and end of the time series. The Hanning window is defined as:
\begin{equation}
    h_n = \frac{1}{2}\left[1 - \cos\left(\frac{2\pi n}{N-1}\right)\right],\quad n=0,1,...,N-1 \,.
\end{equation}
 The windowed kernel $k_n$ is obtained by multiplying the time-domain kernel $K_n$ with the Hanning window $h_n$:
\begin{equation}
k_n = K_n h_n,\quad n=0,1,...,N-1
\end{equation}
Using the FFT algorithm we then compute the DFT coefficients $\tilde{k}_k$ of the windowed kernel.}

\b{
The  procedure described above to obtain the DFT coefficients $\tilde{k}_k$ from $K(t)$ is applied to $N(t)$ as well to obtain the coefficients $\tilde{n}_k$. We can then evaluate the information rate using \cref{eq:gauss_mi_approx} by discretizing the integral $\int\mathrm{d}\omega\,F(\omega) \to \sum_k \Delta\omega\,F(\omega_k)$ with $\Delta\omega = 2\pi f_s / N$. More precisely, we compute the Gaussian information rate as
\begin{equation}
    R(\mathcal{S}, \mathcal{X}) = \frac{g^2}{4\pi} \sum^{N-1}_{k=0} \Delta\omega\ \frac{V(\omega_k)|\tilde{k}_k|^2}{\tilde{n}_k} \,.
\end{equation}
}

\section{\label{sec:pws_fitted}PWS Estimate for the Fitted Chemotaxis Model}

\b{
In the main text, we described that a chemotaxis model with $N_c = 9$ receptor clusters, each containing $N=15$ receptors, matches the experimental kernels of \citet{2021.Mattinglyg3a}. We then computed the information rate for this model using both the exact PWS method and a Gaussian approximation. How the rate in the Gaussian model is computed is described in \cref{sec:lna}. Here, we describe briefly how we compute the exact rate using PWS.
}

\b{
While in principle the rate could be computed directly via PWS for the model with $N_c=9$ and $N=15$, the receptor activity noise was so large that obtaining this estimate directly in a single PWS simulation proved to be inefficient.
Instead, we computed the rate via an extrapolation procedure. 
In particular, we computed the rate for a series of models with $N=15$, yet with $N_c$ going down from 400 to 50.
The rate for the model of interest, with $N=15$ and $N_c=9$, was then obtained by fitting this data to a simple polynomial and then extrapolating to $N_c=9$.
}

\begin{table}[b]
    \centering
    \begin{tabular}{SSS}
        {$g$ (\si{\per\milli\meter})} & {$a$ (\si{\bit\per\second})} & {$b$ (\si{\bit\per\second})}  \\
        \hline
        0.1 & 0.234e-3 & 0.160e-6 \\
        0.2 & 0.814e-3 & 0.598e-6 \\
        0.3 & 1.74e-3 & 1.77e-6 \\
        0.4 & 2.84e-3 & 3.39e-6
    \end{tabular}
    \caption{Fit coefficients for the information rate as a function of the number of clusters $N_c$. These coefficients are for a quadratic function $R(N_c) = a N_c - b N^2_c$.}
    \label{tab:fit_coefficients}
\end{table}

\b{
In \cref{fig:num_clusters} we show the information rate for a range of values of $N_c$, and for different gradient steepnesses $g$. 
We see that the information rate increases non-linearly with the number $N_c$ of independent clusters.
Based on the assumption that the information rate is zero in the limit $N_c\to 0$, we fit a quadratic function $R(N_c) = a N_c - b N^2_c$ with positive coefficients $a, b$ to the data.
We provide the fit coefficients for different gradient steepnesses $g$ in \cref{tab:fit_coefficients}. From these fits we can obtain the extrapolated information rates for $N_c=9$ that are shown in the main text.
}

\bibliography{library.bib}

\begin{thebibliography}{112}%
\makeatletter
\providecommand \@ifxundefined [1]{%
 \@ifx{#1\undefined}
}%
\providecommand \@ifnum [1]{%
 \ifnum #1\expandafter \@firstoftwo
 \else \expandafter \@secondoftwo
 \fi
}%
\providecommand \@ifx [1]{%
 \ifx #1\expandafter \@firstoftwo
 \else \expandafter \@secondoftwo
 \fi
}%
\providecommand \natexlab [1]{#1}%
\providecommand \enquote  [1]{``#1''}%
\providecommand \bibnamefont  [1]{#1}%
\providecommand \bibfnamefont [1]{#1}%
\providecommand \citenamefont [1]{#1}%
\providecommand \href@noop [0]{\@secondoftwo}%
\providecommand \href [0]{\begingroup \@sanitize@url \@href}%
\providecommand \@href[1]{\@@startlink{#1}\@@href}%
\providecommand \@@href[1]{\endgroup#1\@@endlink}%
\providecommand \@sanitize@url [0]{\catcode `\\12\catcode `\$12\catcode
  `\&12\catcode `\#12\catcode `\^12\catcode `\_12\catcode `\%12\relax}%
\providecommand \@@startlink[1]{}%
\providecommand \@@endlink[0]{}%
\providecommand \url  [0]{\begingroup\@sanitize@url \@url }%
\providecommand \@url [1]{\endgroup\@href {#1}{\urlprefix }}%
\providecommand \urlprefix  [0]{URL }%
\providecommand \Eprint [0]{\href }%
\providecommand \doibase [0]{https://doi.org/}%
\providecommand \selectlanguage [0]{\@gobble}%
\providecommand \bibinfo  [0]{\@secondoftwo}%
\providecommand \bibfield  [0]{\@secondoftwo}%
\providecommand \translation [1]{[#1]}%
\providecommand \BibitemOpen [0]{}%
\providecommand \bibitemStop [0]{}%
\providecommand \bibitemNoStop [0]{.\EOS\space}%
\providecommand \EOS [0]{\spacefactor3000\relax}%
\providecommand \BibitemShut  [1]{\csname bibitem#1\endcsname}%
\let\auto@bib@innerbib\@empty
\bibitem [{\citenamefont {Tka\v{c}ik}\ and\ \citenamefont
  {Walczak}(2011)}]{2011.Tkacik}%
  \BibitemOpen
  \bibfield  {author} {\bibinfo {author} {\bibfnamefont {G.}~\bibnamefont
  {Tka\v{c}ik}}\ and\ \bibinfo {author} {\bibfnamefont {A.~M.}\ \bibnamefont
  {Walczak}},\ }\bibfield  {title} {\bibinfo {title} {{Information transmission
  in genetic regulatory networks: a review}},\ }\href
  {https://doi.org/10.1088/0953-8984/23/15/153102} {\bibfield  {journal}
  {\bibinfo  {journal} {Journal of Physics: Condensed Matter}\ }\textbf
  {\bibinfo {volume} {23}},\ \bibinfo {pages} {153102} (\bibinfo {year}
  {2011})},\ \Eprint {https://arxiv.org/abs/1101.4240} {1101.4240} \BibitemShut
  {NoStop}%
\bibitem [{\citenamefont {Tka\v{c}ik}\ and\ \citenamefont
  {Bialek}(2016)}]{2016.Tkacik}%
  \BibitemOpen
  \bibfield  {author} {\bibinfo {author} {\bibfnamefont {G.}~\bibnamefont
  {Tka\v{c}ik}}\ and\ \bibinfo {author} {\bibfnamefont {W.}~\bibnamefont
  {Bialek}},\ }\bibfield  {title} {\bibinfo {title} {{Information Processing in
  Living Systems}},\ }\href
  {https://doi.org/10.1146/annurev-conmatphys-031214-014803} {\bibfield
  {journal} {\bibinfo  {journal} {Annual Review of Condensed Matter Physics}\
  }\textbf {\bibinfo {volume} {7}},\ \bibinfo {pages} {89} (\bibinfo {year}
  {2016})}\BibitemShut {NoStop}%
\bibitem [{\citenamefont {MacKay}(2003)}]{2003.MacKay}%
  \BibitemOpen
  \bibfield  {author} {\bibinfo {author} {\bibfnamefont {D.}~\bibnamefont
  {MacKay}},\ }\href@noop {} {\emph {\bibinfo {title} {{Information theory,
  inference, and learning algorithms}}}}\ (\bibinfo  {publisher} {Cambridge
  University Press},\ \bibinfo {address} {Cambridge, UK},\ \bibinfo {year}
  {2003})\BibitemShut {NoStop}%
\bibitem [{\citenamefont {Shannon}(1948)}]{1948.Shannon}%
  \BibitemOpen
  \bibfield  {author} {\bibinfo {author} {\bibfnamefont {C.~E.}\ \bibnamefont
  {Shannon}},\ }\bibfield  {title} {\bibinfo {title} {{A Mathematical Theory of
  Communication}},\ }\href {https://doi.org/10.1002/j.1538-7305.1948.tb01338.x}
  {\bibfield  {journal} {\bibinfo  {journal} {Bell System Technical Journal}\
  }\textbf {\bibinfo {volume} {27}},\ \bibinfo {pages} {379} (\bibinfo {year}
  {1948})}\BibitemShut {NoStop}%
\bibitem [{\citenamefont {Cover}\ and\ \citenamefont
  {Thomas}(2006)}]{2006.Cover}%
  \BibitemOpen
  \bibfield  {author} {\bibinfo {author} {\bibfnamefont {T.~M.}\ \bibnamefont
  {Cover}}\ and\ \bibinfo {author} {\bibfnamefont {J.~A.}\ \bibnamefont
  {Thomas}},\ }\href@noop {} {\emph {\bibinfo {title} {{Elements of Information
  Theory}}}},\ \bibinfo {edition} {2nd}\ ed.\ (\bibinfo  {publisher} {John
  Wiley \& Sons},\ \bibinfo {year} {2006})\BibitemShut {NoStop}%
\bibitem [{\citenamefont {Tostevin}\ and\ \citenamefont {ten
  Wolde}(2009)}]{2009.Tostevin}%
  \BibitemOpen
  \bibfield  {author} {\bibinfo {author} {\bibfnamefont {F.}~\bibnamefont
  {Tostevin}}\ and\ \bibinfo {author} {\bibfnamefont {P.~R.}\ \bibnamefont {ten
  Wolde}},\ }\bibfield  {title} {\bibinfo {title} {{Mutual Information between
  Input and Output Trajectories of Biochemical Networks}},\ }\href
  {https://doi.org/10.1103/physrevlett.102.218101} {\bibfield  {journal}
  {\bibinfo  {journal} {Physical Review Letters}\ }\textbf {\bibinfo {volume}
  {102}},\ \bibinfo {pages} {218101} (\bibinfo {year} {2009})},\ \Eprint
  {https://arxiv.org/abs/0901.0280} {0901.0280} \BibitemShut {NoStop}%
\bibitem [{\citenamefont {Fiedor}(2014)}]{2014.Fiedor}%
  \BibitemOpen
  \bibfield  {author} {\bibinfo {author} {\bibfnamefont {P.}~\bibnamefont
  {Fiedor}},\ }\bibfield  {title} {\bibinfo {title} {{Networks in financial
  markets based on the mutual information rate}},\ }\href
  {https://doi.org/10.1103/physreve.89.052801} {\bibfield  {journal} {\bibinfo
  {journal} {Physical Review E}\ }\textbf {\bibinfo {volume} {89}},\ \bibinfo
  {pages} {052801} (\bibinfo {year} {2014})},\ \Eprint
  {https://arxiv.org/abs/1401.2548} {1401.2548} \BibitemShut {NoStop}%
\bibitem [{\citenamefont {Massey}(1990)}]{1990.Massey}%
  \BibitemOpen
  \bibfield  {author} {\bibinfo {author} {\bibfnamefont {J.~L.}\ \bibnamefont
  {Massey}},\ }\bibfield  {title} {\bibinfo {title} {{Causality, Feedback and
  Directed Information}},\ }\href@noop {} {\bibfield  {journal} {\bibinfo
  {journal} {Proceedings of the International Symposium on Information Theory
  and its Applications}\ ,\ \bibinfo {pages} {303–305}} (\bibinfo {year}
  {1990})}\BibitemShut {NoStop}%
\bibitem [{\citenamefont {Schreiber}(2000)}]{2000.Schreiber}%
  \BibitemOpen
  \bibfield  {author} {\bibinfo {author} {\bibfnamefont {T.}~\bibnamefont
  {Schreiber}},\ }\bibfield  {title} {\bibinfo {title} {{Measuring information
  transfer}},\ }\href {https://doi.org/10.1103/physrevlett.85.461} {\bibfield
  {journal} {\bibinfo  {journal} {Physical Review Letters}\ }\textbf {\bibinfo
  {volume} {85}},\ \bibinfo {pages} {461} (\bibinfo {year} {2000})},\ \Eprint
  {https://arxiv.org/abs/nlin/0001042} {nlin/0001042} \BibitemShut {NoStop}%
\bibitem [{\citenamefont {Block}\ \emph {et~al.}(1983)\citenamefont {Block},
  \citenamefont {Segall},\ and\ \citenamefont {Berg}}]{1983.Block}%
  \BibitemOpen
  \bibfield  {author} {\bibinfo {author} {\bibfnamefont {S.~M.}\ \bibnamefont
  {Block}}, \bibinfo {author} {\bibfnamefont {J.~E.}\ \bibnamefont {Segall}},\
  and\ \bibinfo {author} {\bibfnamefont {H.~C.}\ \bibnamefont {Berg}},\
  }\bibfield  {title} {\bibinfo {title} {{Adaptation kinetics in bacterial
  chemotaxis}},\ }\href {https://doi.org/10.1128/jb.154.1.312-323.1983}
  {\bibfield  {journal} {\bibinfo  {journal} {Journal of Bacteriology}\
  }\textbf {\bibinfo {volume} {154}},\ \bibinfo {pages} {312} (\bibinfo {year}
  {1983})}\BibitemShut {NoStop}%
\bibitem [{\citenamefont {Segall}\ \emph {et~al.}(1986)\citenamefont {Segall},
  \citenamefont {Block},\ and\ \citenamefont {Berg}}]{1986.Segall}%
  \BibitemOpen
  \bibfield  {author} {\bibinfo {author} {\bibfnamefont {J.~E.}\ \bibnamefont
  {Segall}}, \bibinfo {author} {\bibfnamefont {S.~M.}\ \bibnamefont {Block}},\
  and\ \bibinfo {author} {\bibfnamefont {H.~C.}\ \bibnamefont {Berg}},\
  }\bibfield  {title} {\bibinfo {title} {{Temporal comparisons in bacterial
  chemotaxis}},\ }\href {https://doi.org/10.1073/pnas.83.23.8987} {\bibfield
  {journal} {\bibinfo  {journal} {Proceedings of the National Academy of
  Sciences}\ }\textbf {\bibinfo {volume} {83}},\ \bibinfo {pages} {8987}
  (\bibinfo {year} {1986})}\BibitemShut {NoStop}%
\bibitem [{\citenamefont {Marshall}(1995)}]{1995.Marshall}%
  \BibitemOpen
  \bibfield  {author} {\bibinfo {author} {\bibfnamefont {C.}~\bibnamefont
  {Marshall}},\ }\bibfield  {title} {\bibinfo {title} {{Specificity of receptor
  tyrosine kinase signaling: Transient versus sustained extracellular
  signal-regulated kinase activation}},\ }\href
  {https://doi.org/10.1016/0092-8674(95)90401-8} {\bibfield  {journal}
  {\bibinfo  {journal} {Cell}\ }\textbf {\bibinfo {volume} {80}},\ \bibinfo
  {pages} {179} (\bibinfo {year} {1995})}\BibitemShut {NoStop}%
\bibitem [{\citenamefont {Purvis}\ and\ \citenamefont
  {Lahav}(2013)}]{2013.Purvis}%
  \BibitemOpen
  \bibfield  {author} {\bibinfo {author} {\bibfnamefont {J.}~\bibnamefont
  {Purvis}}\ and\ \bibinfo {author} {\bibfnamefont {G.}~\bibnamefont {Lahav}},\
  }\bibfield  {title} {\bibinfo {title} {{Encoding and Decoding Cellular
  Information through Signaling Dynamics}},\ }\href
  {https://doi.org/10.1016/j.cell.2013.02.005} {\bibfield  {journal} {\bibinfo
  {journal} {Cell}\ }\textbf {\bibinfo {volume} {152}},\ \bibinfo {pages} {945}
  (\bibinfo {year} {2013})}\BibitemShut {NoStop}%
\bibitem [{\citenamefont {Selimkhanov}\ \emph {et~al.}(2014)\citenamefont
  {Selimkhanov}, \citenamefont {Taylor}, \citenamefont {Yao}, \citenamefont
  {Pilko}, \citenamefont {Albeck}, \citenamefont {Hoffmann}, \citenamefont
  {Tsimring},\ and\ \citenamefont {Wollman}}]{2014.Selimkhanov}%
  \BibitemOpen
  \bibfield  {author} {\bibinfo {author} {\bibfnamefont {J.}~\bibnamefont
  {Selimkhanov}}, \bibinfo {author} {\bibfnamefont {B.}~\bibnamefont {Taylor}},
  \bibinfo {author} {\bibfnamefont {J.}~\bibnamefont {Yao}}, \bibinfo {author}
  {\bibfnamefont {A.}~\bibnamefont {Pilko}}, \bibinfo {author} {\bibfnamefont
  {J.}~\bibnamefont {Albeck}}, \bibinfo {author} {\bibfnamefont
  {A.}~\bibnamefont {Hoffmann}}, \bibinfo {author} {\bibfnamefont
  {L.}~\bibnamefont {Tsimring}},\ and\ \bibinfo {author} {\bibfnamefont
  {R.}~\bibnamefont {Wollman}},\ }\bibfield  {title} {\bibinfo {title}
  {{Accurate information transmission through dynamic biochemical signaling
  networks}},\ }\href {https://doi.org/10.1126/science.1254933} {\bibfield
  {journal} {\bibinfo  {journal} {Science}\ }\textbf {\bibinfo {volume}
  {346}},\ \bibinfo {pages} {1370} (\bibinfo {year} {2014})}\BibitemShut
  {NoStop}%
\bibitem [{\citenamefont {Granados}\ \emph {et~al.}(2018)\citenamefont
  {Granados}, \citenamefont {Pietsch}, \citenamefont {Cepeda-Humerez},
  \citenamefont {Farquhar}, \citenamefont {Tka\v{c}ik},\ and\ \citenamefont
  {Swain}}]{2018.Granados}%
  \BibitemOpen
  \bibfield  {author} {\bibinfo {author} {\bibfnamefont {A.~A.}\ \bibnamefont
  {Granados}}, \bibinfo {author} {\bibfnamefont {J.~M.~J.}\ \bibnamefont
  {Pietsch}}, \bibinfo {author} {\bibfnamefont {S.~A.}\ \bibnamefont
  {Cepeda-Humerez}}, \bibinfo {author} {\bibfnamefont {I.~L.}\ \bibnamefont
  {Farquhar}}, \bibinfo {author} {\bibfnamefont {G.}~\bibnamefont
  {Tka\v{c}ik}},\ and\ \bibinfo {author} {\bibfnamefont {P.~S.}\ \bibnamefont
  {Swain}},\ }\bibfield  {title} {\bibinfo {title} {{Distributed and dynamic
  intracellular organization of extracellular information}},\ }\href
  {https://doi.org/10.1073/pnas.1716659115} {\bibfield  {journal} {\bibinfo
  {journal} {Proceedings of the National Academy of Sciences}\ }\textbf
  {\bibinfo {volume} {115}},\ \bibinfo {pages} {201716659} (\bibinfo {year}
  {2018})}\BibitemShut {NoStop}%
\bibitem [{\citenamefont {Strong}\ \emph {et~al.}(1998)\citenamefont {Strong},
  \citenamefont {Koberle}, \citenamefont {de~Ruyter~van Steveninck},\ and\
  \citenamefont {Bialek}}]{1998.Strong}%
  \BibitemOpen
  \bibfield  {author} {\bibinfo {author} {\bibfnamefont {S.~P.}\ \bibnamefont
  {Strong}}, \bibinfo {author} {\bibfnamefont {R.}~\bibnamefont {Koberle}},
  \bibinfo {author} {\bibfnamefont {R.~R.}\ \bibnamefont {de~Ruyter~van
  Steveninck}},\ and\ \bibinfo {author} {\bibfnamefont {W.}~\bibnamefont
  {Bialek}},\ }\bibfield  {title} {\bibinfo {title} {{Entropy and Information
  in Neural Spike Trains}},\ }\href
  {https://doi.org/10.1103/physrevlett.80.197} {\bibfield  {journal} {\bibinfo
  {journal} {Physical Review Letters}\ }\textbf {\bibinfo {volume} {80}},\
  \bibinfo {pages} {197} (\bibinfo {year} {1998})}\BibitemShut {NoStop}%
\bibitem [{\citenamefont {Paninski}(2003)}]{2003.Paninski}%
  \BibitemOpen
  \bibfield  {author} {\bibinfo {author} {\bibfnamefont {L.}~\bibnamefont
  {Paninski}},\ }\bibfield  {title} {\bibinfo {title} {{Estimation of Entropy
  and Mutual Information}},\ }\href
  {https://doi.org/10.1162/089976603321780272} {\bibfield  {journal} {\bibinfo
  {journal} {Neural Computation}\ }\textbf {\bibinfo {volume} {15}},\ \bibinfo
  {pages} {1191} (\bibinfo {year} {2003})}\BibitemShut {NoStop}%
\bibitem [{\citenamefont {Cheong}\ \emph {et~al.}(2011)\citenamefont {Cheong},
  \citenamefont {Rhee}, \citenamefont {Wang}, \citenamefont {Nemenman},\ and\
  \citenamefont {Levchenko}}]{2011.Cheong}%
  \BibitemOpen
  \bibfield  {author} {\bibinfo {author} {\bibfnamefont {R.}~\bibnamefont
  {Cheong}}, \bibinfo {author} {\bibfnamefont {A.}~\bibnamefont {Rhee}},
  \bibinfo {author} {\bibfnamefont {C.~J.}\ \bibnamefont {Wang}}, \bibinfo
  {author} {\bibfnamefont {I.}~\bibnamefont {Nemenman}},\ and\ \bibinfo
  {author} {\bibfnamefont {A.}~\bibnamefont {Levchenko}},\ }\bibfield  {title}
  {\bibinfo {title} {{Information Transduction Capacity of Noisy Biochemical
  Signaling Networks}},\ }\href {https://doi.org/10.1126/science.1204553}
  {\bibfield  {journal} {\bibinfo  {journal} {Science}\ }\textbf {\bibinfo
  {volume} {334}},\ \bibinfo {pages} {354} (\bibinfo {year}
  {2011})}\BibitemShut {NoStop}%
\bibitem [{\citenamefont {Tka\v{c}ik}\ \emph {et~al.}(2008)\citenamefont
  {Tka\v{c}ik}, \citenamefont {Callan},\ and\ \citenamefont
  {Bialek}}]{2008.Tkacik}%
  \BibitemOpen
  \bibfield  {author} {\bibinfo {author} {\bibfnamefont {G.}~\bibnamefont
  {Tka\v{c}ik}}, \bibinfo {author} {\bibfnamefont {C.~G.}\ \bibnamefont
  {Callan}},\ and\ \bibinfo {author} {\bibfnamefont {W.}~\bibnamefont
  {Bialek}},\ }\bibfield  {title} {\bibinfo {title} {{Information flow and
  optimization in transcriptional regulation}},\ }\href
  {https://doi.org/10.1073/pnas.0806077105} {\bibfield  {journal} {\bibinfo
  {journal} {Proceedings of the National Academy of Sciences}\ }\textbf
  {\bibinfo {volume} {105}},\ \bibinfo {pages} {12265} (\bibinfo {year}
  {2008})},\ \Eprint {https://arxiv.org/abs/0705.0313} {0705.0313} \BibitemShut
  {NoStop}%
\bibitem [{\citenamefont {Tka\v{c}ik}\ \emph {et~al.}(2014)\citenamefont
  {Tka\v{c}ik}, \citenamefont {Dubuis}, \citenamefont {Petkova},\ and\
  \citenamefont {Gregor}}]{2014.Tkacik}%
  \BibitemOpen
  \bibfield  {author} {\bibinfo {author} {\bibfnamefont {G.}~\bibnamefont
  {Tka\v{c}ik}}, \bibinfo {author} {\bibfnamefont {J.~O.}\ \bibnamefont
  {Dubuis}}, \bibinfo {author} {\bibfnamefont {M.~D.}\ \bibnamefont
  {Petkova}},\ and\ \bibinfo {author} {\bibfnamefont {T.}~\bibnamefont
  {Gregor}},\ }\bibfield  {title} {\bibinfo {title} {{Positional Information,
  Positional Error, and Read-Out Precision in Morphogenesis: A Mathematical
  Framework}},\ }\href {https://doi.org/10.1534/genetics.114.171850} {\bibfield
   {journal} {\bibinfo  {journal} {Genetics}\ }\textbf {\bibinfo {volume}
  {199}},\ \bibinfo {pages} {genetics.114.171850} (\bibinfo {year}
  {2014})}\BibitemShut {NoStop}%
\bibitem [{\citenamefont {Meijers}\ \emph {et~al.}(2021)\citenamefont
  {Meijers}, \citenamefont {Ito},\ and\ \citenamefont {ten
  Wolde}}]{2021.Meijers}%
  \BibitemOpen
  \bibfield  {author} {\bibinfo {author} {\bibfnamefont {M.}~\bibnamefont
  {Meijers}}, \bibinfo {author} {\bibfnamefont {S.}~\bibnamefont {Ito}},\ and\
  \bibinfo {author} {\bibfnamefont {P.~R.}\ \bibnamefont {ten Wolde}},\
  }\bibfield  {title} {\bibinfo {title} {{Behavior of information flow near
  criticality}},\ }\href {https://doi.org/10.1103/physreve.103.l010102}
  {\bibfield  {journal} {\bibinfo  {journal} {Physical Review E}\ }\textbf
  {\bibinfo {volume} {103}},\ \bibinfo {pages} {L010102} (\bibinfo {year}
  {2021})},\ \Eprint {https://arxiv.org/abs/1906.00787} {1906.00787}
  \BibitemShut {NoStop}%
\bibitem [{\citenamefont {Rieke}\ \emph {et~al.}(1999)\citenamefont {Rieke},
  \citenamefont {Warland}, \citenamefont {de~Ruyter~van Steveninck},\ and\
  \citenamefont {Bialek}}]{1999.Rieke}%
  \BibitemOpen
  \bibfield  {author} {\bibinfo {author} {\bibfnamefont {F.}~\bibnamefont
  {Rieke}}, \bibinfo {author} {\bibfnamefont {D.}~\bibnamefont {Warland}},
  \bibinfo {author} {\bibfnamefont {R.}~\bibnamefont {de~Ruyter~van
  Steveninck}},\ and\ \bibinfo {author} {\bibfnamefont {W.}~\bibnamefont
  {Bialek}},\ }\href@noop {} {\emph {\bibinfo {title} {{Spikes: exploring the
  neural code}}}}\ (\bibinfo  {publisher} {MIT Press},\ \bibinfo {address}
  {Cambridge, Massachusetts},\ \bibinfo {year} {1999})\BibitemShut {NoStop}%
\bibitem [{\citenamefont {Kaiser}\ and\ \citenamefont
  {Schreiber}(2002)}]{2002.Kaiser}%
  \BibitemOpen
  \bibfield  {author} {\bibinfo {author} {\bibfnamefont {A.}~\bibnamefont
  {Kaiser}}\ and\ \bibinfo {author} {\bibfnamefont {T.}~\bibnamefont
  {Schreiber}},\ }\bibfield  {title} {\bibinfo {title} {{Information transfer
  in continuous processes}},\ }\href
  {https://doi.org/10.1016/s0167-2789(02)00432-3} {\bibfield  {journal}
  {\bibinfo  {journal} {Physica D: Nonlinear Phenomena}\ }\textbf {\bibinfo
  {volume} {166}},\ \bibinfo {pages} {43} (\bibinfo {year} {2002})}\BibitemShut
  {NoStop}%
\bibitem [{\citenamefont {Kraskov}\ \emph {et~al.}(2004)\citenamefont
  {Kraskov}, \citenamefont {St\"ogbauer},\ and\ \citenamefont
  {Grassberger}}]{2004.Kraskov}%
  \BibitemOpen
  \bibfield  {author} {\bibinfo {author} {\bibfnamefont {A.}~\bibnamefont
  {Kraskov}}, \bibinfo {author} {\bibfnamefont {H.}~\bibnamefont
  {St\"ogbauer}},\ and\ \bibinfo {author} {\bibfnamefont {P.}~\bibnamefont
  {Grassberger}},\ }\bibfield  {title} {\bibinfo {title} {{Estimating mutual
  information}},\ }\href {https://doi.org/10.1103/physreve.69.066138}
  {\bibfield  {journal} {\bibinfo  {journal} {Physical Review E}\ }\textbf
  {\bibinfo {volume} {69}},\ \bibinfo {pages} {066138} (\bibinfo {year}
  {2004})},\ \Eprint {https://arxiv.org/abs/cond-mat/0305641}
  {cond-mat/0305641} \BibitemShut {NoStop}%
\bibitem [{\citenamefont {Cepeda-Humerez}\ \emph {et~al.}(2019)\citenamefont
  {Cepeda-Humerez}, \citenamefont {Ruess},\ and\ \citenamefont
  {Tka\v{c}ik}}]{2019.Cepeda-Humerez}%
  \BibitemOpen
  \bibfield  {author} {\bibinfo {author} {\bibfnamefont {S.~A.}\ \bibnamefont
  {Cepeda-Humerez}}, \bibinfo {author} {\bibfnamefont {J.}~\bibnamefont
  {Ruess}},\ and\ \bibinfo {author} {\bibfnamefont {G.}~\bibnamefont
  {Tka\v{c}ik}},\ }\bibfield  {title} {\bibinfo {title} {{Estimating
  information in time-varying signals.}},\ }\href
  {https://doi.org/10.1371/journal.pcbi.1007290} {\bibfield  {journal}
  {\bibinfo  {journal} {PLoS computational biology}\ }\textbf {\bibinfo
  {volume} {15}},\ \bibinfo {pages} {e1007290} (\bibinfo {year}
  {2019})}\BibitemShut {NoStop}%
\bibitem [{\citenamefont {Gao}\ \emph {et~al.}(2008)\citenamefont {Gao},
  \citenamefont {Kontoyiannis},\ and\ \citenamefont {Bienenstock}}]{2008.Gao}%
  \BibitemOpen
  \bibfield  {author} {\bibinfo {author} {\bibfnamefont {Y.}~\bibnamefont
  {Gao}}, \bibinfo {author} {\bibfnamefont {I.}~\bibnamefont {Kontoyiannis}},\
  and\ \bibinfo {author} {\bibfnamefont {E.}~\bibnamefont {Bienenstock}},\
  }\bibfield  {title} {\bibinfo {title} {{Estimating the Entropy of Binary Time
  Series: Methodology, Some Theory and a Simulation Study}},\ }\href
  {https://doi.org/10.3390/entropy-e10020071} {\bibfield  {journal} {\bibinfo
  {journal} {Entropy}\ }\textbf {\bibinfo {volume} {10}},\ \bibinfo {pages}
  {71} (\bibinfo {year} {2008})},\ \Eprint {https://arxiv.org/abs/0802.4363}
  {0802.4363} \BibitemShut {NoStop}%
\bibitem [{\citenamefont {Borst}\ and\ \citenamefont
  {Theunissen}(1999)}]{1999.Borst}%
  \BibitemOpen
  \bibfield  {author} {\bibinfo {author} {\bibfnamefont {A.}~\bibnamefont
  {Borst}}\ and\ \bibinfo {author} {\bibfnamefont {F.~E.}\ \bibnamefont
  {Theunissen}},\ }\bibfield  {title} {\bibinfo {title} {{Information theory
  and neural coding}},\ }\href {https://doi.org/10.1038/14731} {\bibfield
  {journal} {\bibinfo  {journal} {Nature Neuroscience}\ }\textbf {\bibinfo
  {volume} {2}},\ \bibinfo {pages} {947} (\bibinfo {year} {1999})}\BibitemShut
  {NoStop}%
\bibitem [{\citenamefont {Hled\'ik}\ \emph {et~al.}(2019)\citenamefont
  {Hled\'ik}, \citenamefont {Sokolowski},\ and\ \citenamefont
  {Tka\v{c}ik}}]{2019.Hledik}%
  \BibitemOpen
  \bibfield  {author} {\bibinfo {author} {\bibfnamefont {M.}~\bibnamefont
  {Hled\'ik}}, \bibinfo {author} {\bibfnamefont {T.~R.}\ \bibnamefont
  {Sokolowski}},\ and\ \bibinfo {author} {\bibfnamefont {G.}~\bibnamefont
  {Tka\v{c}ik}},\ }\bibfield  {title} {\bibinfo {title} {{A Tight Upper Bound
  on Mutual Information}},\ }in\ \href
  {https://doi.org/10.1109/itw44776.2019.8989292} {\emph {\bibinfo {booktitle}
  {2019 IEEE Information Theory Workshop (ITW)}}}\ (\bibinfo {year} {2019})\
  pp.\ \bibinfo {pages} {1--5}\BibitemShut {NoStop}%
\bibitem [{\citenamefont {Thomas}\ and\ \citenamefont
  {Eckford}(2016)}]{2016.Thomas}%
  \BibitemOpen
  \bibfield  {author} {\bibinfo {author} {\bibfnamefont {P.~J.}\ \bibnamefont
  {Thomas}}\ and\ \bibinfo {author} {\bibfnamefont {A.~W.}\ \bibnamefont
  {Eckford}},\ }\bibfield  {title} {\bibinfo {title} {{Capacity of a Simple
  Intercellular Signal Transduction Channel}},\ }\href
  {https://doi.org/10.1109/tit.2016.2599178} {\bibfield  {journal} {\bibinfo
  {journal} {IEEE Transactions on Information Theory}\ }\textbf {\bibinfo
  {volume} {62}},\ \bibinfo {pages} {7358} (\bibinfo {year} {2016})},\ \Eprint
  {https://arxiv.org/abs/1305.2245} {1305.2245} \BibitemShut {NoStop}%
\bibitem [{\citenamefont {Duso}\ and\ \citenamefont
  {Zechner}(2019)}]{2019.Duso}%
  \BibitemOpen
  \bibfield  {author} {\bibinfo {author} {\bibfnamefont {L.}~\bibnamefont
  {Duso}}\ and\ \bibinfo {author} {\bibfnamefont {C.}~\bibnamefont {Zechner}},\
  }\bibfield  {title} {\bibinfo {title} {Path mutual information for a class of
  biochemical reaction networks},\ }in\ \href
  {https://doi.org/10.1109/CDC40024.2019.9029316} {\emph {\bibinfo {booktitle}
  {2019 IEEE 58th Conference on Decision and Control (CDC)}}}\ (\bibinfo
  {publisher} {IEEE},\ \bibinfo {address} {Piscataway, N.J.},\ \bibinfo {year}
  {2019})\ pp.\ \bibinfo {pages} {6610--6615},\ \Eprint
  {https://arxiv.org/abs/1904.01988} {1904.01988} \BibitemShut {NoStop}%
\bibitem [{\citenamefont {Moor}\ and\ \citenamefont
  {Zechner}(2023)}]{2023.Moor}%
  \BibitemOpen
  \bibfield  {author} {\bibinfo {author} {\bibfnamefont {A.-L.}\ \bibnamefont
  {Moor}}\ and\ \bibinfo {author} {\bibfnamefont {C.}~\bibnamefont {Zechner}},\
  }\bibfield  {title} {\bibinfo {title} {{Dynamic information transfer in
  stochastic biochemical networks}},\ }\href
  {https://doi.org/10.1103/physrevresearch.5.013032} {\bibfield  {journal}
  {\bibinfo  {journal} {Physical Review Research}\ }\textbf {\bibinfo {volume}
  {5}},\ \bibinfo {pages} {013032} (\bibinfo {year} {2023})}\BibitemShut
  {NoStop}%
\bibitem [{\citenamefont {Delbr\"uck}(1940)}]{1940.Delbrueck}%
  \BibitemOpen
  \bibfield  {author} {\bibinfo {author} {\bibfnamefont {M.}~\bibnamefont
  {Delbr\"uck}},\ }\bibfield  {title} {\bibinfo {title} {{Statistical
  Fluctuations in Autocatalytic Reactions}},\ }\href
  {https://doi.org/10.1063/1.1750549} {\bibfield  {journal} {\bibinfo
  {journal} {The Journal of Chemical Physics}\ }\textbf {\bibinfo {volume}
  {8}},\ \bibinfo {pages} {120} (\bibinfo {year} {1940})}\BibitemShut {NoStop}%
\bibitem [{\citenamefont {McQuarrie}(1963)}]{1963.McQuarrie}%
  \BibitemOpen
  \bibfield  {author} {\bibinfo {author} {\bibfnamefont {D.~A.}\ \bibnamefont
  {McQuarrie}},\ }\bibfield  {title} {\bibinfo {title} {{Kinetics of Small
  Systems. I}},\ }\href {https://doi.org/10.1063/1.1733676} {\bibfield
  {journal} {\bibinfo  {journal} {The Journal of Chemical Physics}\ }\textbf
  {\bibinfo {volume} {38}},\ \bibinfo {pages} {433} (\bibinfo {year}
  {1963})}\BibitemShut {NoStop}%
\bibitem [{\citenamefont {McQuarrie}\ \emph {et~al.}(1964)\citenamefont
  {McQuarrie}, \citenamefont {Jachimowski},\ and\ \citenamefont
  {Russell}}]{1964.McQuarrie}%
  \BibitemOpen
  \bibfield  {author} {\bibinfo {author} {\bibfnamefont {D.~A.}\ \bibnamefont
  {McQuarrie}}, \bibinfo {author} {\bibfnamefont {C.~J.}\ \bibnamefont
  {Jachimowski}},\ and\ \bibinfo {author} {\bibfnamefont {M.~E.}\ \bibnamefont
  {Russell}},\ }\bibfield  {title} {\bibinfo {title} {{Kinetics of Small
  Systems. II}},\ }\href {https://doi.org/10.1063/1.1724926} {\bibfield
  {journal} {\bibinfo  {journal} {The Journal of Chemical Physics}\ }\textbf
  {\bibinfo {volume} {40}},\ \bibinfo {pages} {2914} (\bibinfo {year}
  {1964})}\BibitemShut {NoStop}%
\bibitem [{\citenamefont {Elowitz}\ and\ \citenamefont
  {Leibler}(2000)}]{2000.Elowitz}%
  \BibitemOpen
  \bibfield  {author} {\bibinfo {author} {\bibfnamefont {M.~B.}\ \bibnamefont
  {Elowitz}}\ and\ \bibinfo {author} {\bibfnamefont {S.}~\bibnamefont
  {Leibler}},\ }\bibfield  {title} {\bibinfo {title} {{A synthetic oscillatory
  network of transcriptional regulators}},\ }\href
  {https://doi.org/10.1038/35002125} {\bibfield  {journal} {\bibinfo  {journal}
  {Nature}\ }\textbf {\bibinfo {volume} {403}},\ \bibinfo {pages} {335}
  (\bibinfo {year} {2000})}\BibitemShut {NoStop}%
\bibitem [{\citenamefont {Feller}(1939)}]{1939.Feller}%
  \BibitemOpen
  \bibfield  {author} {\bibinfo {author} {\bibfnamefont {W.}~\bibnamefont
  {Feller}},\ }\bibfield  {title} {\bibinfo {title} {{Die Grundlagen der
  Volterraschen Theorie des Kampfes ums Dasein in
  wahrscheinlichkeitstheoretischer Behandlung}},\ }\href
  {https://doi.org/10.1007/bf01602932} {\bibfield  {journal} {\bibinfo
  {journal} {Acta Biotheoretica}\ }\textbf {\bibinfo {volume} {5}},\ \bibinfo
  {pages} {11} (\bibinfo {year} {1939})}\BibitemShut {NoStop}%
\bibitem [{\citenamefont {Park}\ \emph {et~al.}(2010)\citenamefont {Park},
  \citenamefont {Mu\~noz},\ and\ \citenamefont {Deem}}]{2010.Park}%
  \BibitemOpen
  \bibfield  {author} {\bibinfo {author} {\bibfnamefont {J.-M.}\ \bibnamefont
  {Park}}, \bibinfo {author} {\bibfnamefont {E.}~\bibnamefont {Mu\~noz}},\ and\
  \bibinfo {author} {\bibfnamefont {M.~W.}\ \bibnamefont {Deem}},\ }\bibfield
  {title} {\bibinfo {title} {{Quasispecies theory for finite populations}},\
  }\href {https://doi.org/10.1103/physreve.81.011902} {\bibfield  {journal}
  {\bibinfo  {journal} {Physical Review E}\ }\textbf {\bibinfo {volume} {81}},\
  \bibinfo {pages} {011902} (\bibinfo {year} {2010})},\ \Eprint
  {https://arxiv.org/abs/1002.3837} {1002.3837} \BibitemShut {NoStop}%
\bibitem [{\citenamefont {Cremer}\ \emph {et~al.}(2012)\citenamefont {Cremer},
  \citenamefont {Melbinger},\ and\ \citenamefont {Frey}}]{2012.Cremer}%
  \BibitemOpen
  \bibfield  {author} {\bibinfo {author} {\bibfnamefont {J.}~\bibnamefont
  {Cremer}}, \bibinfo {author} {\bibfnamefont {A.}~\bibnamefont {Melbinger}},\
  and\ \bibinfo {author} {\bibfnamefont {E.}~\bibnamefont {Frey}},\ }\bibfield
  {title} {\bibinfo {title} {{Growth dynamics and the evolution of cooperation
  in microbial populations}},\ }\href {https://doi.org/10.1038/srep00281}
  {\bibfield  {journal} {\bibinfo  {journal} {Scientific Reports}\ }\textbf
  {\bibinfo {volume} {2}},\ \bibinfo {pages} {281} (\bibinfo {year} {2012})},\
  \Eprint {https://arxiv.org/abs/1203.5863} {1203.5863} \BibitemShut {NoStop}%
\bibitem [{\citenamefont {Weidlich}\ and\ \citenamefont
  {Braun}(1992)}]{1992.Weidlich}%
  \BibitemOpen
  \bibfield  {author} {\bibinfo {author} {\bibfnamefont {W.}~\bibnamefont
  {Weidlich}}\ and\ \bibinfo {author} {\bibfnamefont {M.}~\bibnamefont
  {Braun}},\ }\bibfield  {title} {\bibinfo {title} {{The master equation
  approach to nonlinear economics}},\ }\href
  {https://doi.org/10.1007/bf01202420} {\bibfield  {journal} {\bibinfo
  {journal} {Journal of Evolutionary Economics}\ }\textbf {\bibinfo {volume}
  {2}},\ \bibinfo {pages} {233} (\bibinfo {year} {1992})}\BibitemShut {NoStop}%
\bibitem [{\citenamefont {Lux}(1995)}]{1995.Lux}%
  \BibitemOpen
  \bibfield  {author} {\bibinfo {author} {\bibfnamefont {T.}~\bibnamefont
  {Lux}},\ }\bibfield  {title} {\bibinfo {title} {{Herd Behaviour, Bubbles and
  Crashes}},\ }\href {https://doi.org/10.2307/2235156} {\bibfield  {journal}
  {\bibinfo  {journal} {The Economic Journal}\ }\textbf {\bibinfo {volume}
  {105}},\ \bibinfo {pages} {881} (\bibinfo {year} {1995})}\BibitemShut
  {NoStop}%
\bibitem [{\citenamefont {Helbing}(2001)}]{2001.Helbing}%
  \BibitemOpen
  \bibfield  {author} {\bibinfo {author} {\bibfnamefont {D.}~\bibnamefont
  {Helbing}},\ }\bibfield  {title} {\bibinfo {title} {{Traffic and related
  self-driven many-particle systems}},\ }\href
  {https://doi.org/10.1103/revmodphys.73.1067} {\bibfield  {journal} {\bibinfo
  {journal} {Reviews of Modern Physics}\ }\textbf {\bibinfo {volume} {73}},\
  \bibinfo {pages} {1067} (\bibinfo {year} {2001})},\ \Eprint
  {https://arxiv.org/abs/cond-mat/0012229} {cond-mat/0012229} \BibitemShut
  {NoStop}%
\bibitem [{\citenamefont {Castellano}\ \emph {et~al.}(2009)\citenamefont
  {Castellano}, \citenamefont {Fortunato},\ and\ \citenamefont
  {Loreto}}]{2009.Castellano}%
  \BibitemOpen
  \bibfield  {author} {\bibinfo {author} {\bibfnamefont {C.}~\bibnamefont
  {Castellano}}, \bibinfo {author} {\bibfnamefont {S.}~\bibnamefont
  {Fortunato}},\ and\ \bibinfo {author} {\bibfnamefont {V.}~\bibnamefont
  {Loreto}},\ }\bibfield  {title} {\bibinfo {title} {{Statistical physics of
  social dynamics}},\ }\href {https://doi.org/10.1103/revmodphys.81.591}
  {\bibfield  {journal} {\bibinfo  {journal} {Reviews of Modern Physics}\
  }\textbf {\bibinfo {volume} {81}},\ \bibinfo {pages} {591} (\bibinfo {year}
  {2009})},\ \Eprint {https://arxiv.org/abs/0710.3256} {0710.3256} \BibitemShut
  {NoStop}%
\bibitem [{\citenamefont {Siepmann}(1990)}]{1990.Siepmann}%
  \BibitemOpen
  \bibfield  {author} {\bibinfo {author} {\bibfnamefont {J.~I.}\ \bibnamefont
  {Siepmann}},\ }\bibfield  {title} {\bibinfo {title} {{A method for the direct
  calculation of chemical potentials for dense chain systems}},\ }\href
  {https://doi.org/10.1080/00268979000101591} {\bibfield  {journal} {\bibinfo
  {journal} {Molecular Physics}\ }\textbf {\bibinfo {volume} {70}},\ \bibinfo
  {pages} {1145} (\bibinfo {year} {1990})}\BibitemShut {NoStop}%
\bibitem [{\citenamefont {Grassberger}(1997)}]{1997.Grassberger}%
  \BibitemOpen
  \bibfield  {author} {\bibinfo {author} {\bibfnamefont {P.}~\bibnamefont
  {Grassberger}},\ }\bibfield  {title} {\bibinfo {title} {{Pruned-enriched
  Rosenbluth method: Simulations of $\theta$ polymers of chain length up to 1
  000 000}},\ }\href {https://doi.org/10.1103/physreve.56.3682} {\bibfield
  {journal} {\bibinfo  {journal} {Physical Review E}\ }\textbf {\bibinfo
  {volume} {56}},\ \bibinfo {pages} {3682} (\bibinfo {year}
  {1997})}\BibitemShut {NoStop}%
\bibitem [{\citenamefont {Frenkel}\ and\ \citenamefont
  {Smit}(2002)}]{2002.Frenkel}%
  \BibitemOpen
  \bibfield  {author} {\bibinfo {author} {\bibfnamefont {D.}~\bibnamefont
  {Frenkel}}\ and\ \bibinfo {author} {\bibfnamefont {B.}~\bibnamefont {Smit}},\
  }\href {https://doi.org/10.1016/b978-0-12-267351-1.x5000-7} {\emph {\bibinfo
  {title} {{Understanding Molecular Simulation}}}},\ \bibinfo {edition} {2nd}\
  ed.\ (\bibinfo  {publisher} {Academic Press},\ \bibinfo {address} {San
  Diego},\ \bibinfo {year} {2002})\BibitemShut {NoStop}%
\bibitem [{\citenamefont {Frenkel}\ and\ \citenamefont
  {Ladd}(1984)}]{1984.Frenkel}%
  \BibitemOpen
  \bibfield  {author} {\bibinfo {author} {\bibfnamefont {D.}~\bibnamefont
  {Frenkel}}\ and\ \bibinfo {author} {\bibfnamefont {A.~J.~C.}\ \bibnamefont
  {Ladd}},\ }\bibfield  {title} {\bibinfo {title} {{New Monte Carlo method to
  compute the free energy of arbitrary solids. Application to the fcc and hcp
  phases of hard spheres}},\ }\href {https://doi.org/10.1063/1.448024}
  {\bibfield  {journal} {\bibinfo  {journal} {The Journal of Chemical Physics}\
  }\textbf {\bibinfo {volume} {81}},\ \bibinfo {pages} {3188} (\bibinfo {year}
  {1984})}\BibitemShut {NoStop}%
\bibitem [{\citenamefont {Gelman}\ and\ \citenamefont
  {Meng}(1998)}]{1998.Gelman}%
  \BibitemOpen
  \bibfield  {author} {\bibinfo {author} {\bibfnamefont {A.}~\bibnamefont
  {Gelman}}\ and\ \bibinfo {author} {\bibfnamefont {X.-L.}\ \bibnamefont
  {Meng}},\ }\bibfield  {title} {\bibinfo {title} {{Simulating normalizing
  constants: from importance sampling to bridge sampling to path sampling}},\
  }\href {https://doi.org/10.1214/ss/1028905934} {\bibfield  {journal}
  {\bibinfo  {journal} {Statistical Science}\ }\textbf {\bibinfo {volume}
  {13}},\ \bibinfo {pages} {163} (\bibinfo {year} {1998})}\BibitemShut
  {NoStop}%
\bibitem [{\citenamefont {Neal}(2001)}]{2001.Neal}%
  \BibitemOpen
  \bibfield  {author} {\bibinfo {author} {\bibfnamefont {R.~M.}\ \bibnamefont
  {Neal}},\ }\bibfield  {title} {\bibinfo {title} {{Annealed importance
  sampling}},\ }\href {https://doi.org/10.1023/a:1008923215028} {\bibfield
  {journal} {\bibinfo  {journal} {Statistics and Computing}\ }\textbf {\bibinfo
  {volume} {11}},\ \bibinfo {pages} {125} (\bibinfo {year} {2001})}\BibitemShut
  {NoStop}%
\bibitem [{\citenamefont {Bolhuis}\ \emph {et~al.}(2002)\citenamefont
  {Bolhuis}, \citenamefont {Chandler}, \citenamefont {Dellago},\ and\
  \citenamefont {Geissler}}]{2002.Bolhuis}%
  \BibitemOpen
  \bibfield  {author} {\bibinfo {author} {\bibfnamefont {P.~G.}\ \bibnamefont
  {Bolhuis}}, \bibinfo {author} {\bibfnamefont {D.}~\bibnamefont {Chandler}},
  \bibinfo {author} {\bibfnamefont {C.}~\bibnamefont {Dellago}},\ and\ \bibinfo
  {author} {\bibfnamefont {P.~L.}\ \bibnamefont {Geissler}},\ }\bibfield
  {title} {\bibinfo {title} {{TRANSITION PATH SAMPLING: Throwing Ropes Over
  Rough Mountain Passes, in the Dark}},\ }\href
  {https://doi.org/10.1146/annurev.physchem.53.082301.113146} {\bibfield
  {journal} {\bibinfo  {journal} {Annual Review of Physical Chemistry}\
  }\textbf {\bibinfo {volume} {53}},\ \bibinfo {pages} {291} (\bibinfo {year}
  {2002})}\BibitemShut {NoStop}%
\bibitem [{\citenamefont {Mattingly}\ \emph {et~al.}(2021)\citenamefont
  {Mattingly}, \citenamefont {Kamino}, \citenamefont {Machta},\ and\
  \citenamefont {Emonet}}]{2021.Mattinglyg3a}%
  \BibitemOpen
  \bibfield  {author} {\bibinfo {author} {\bibfnamefont {H.~H.}\ \bibnamefont
  {Mattingly}}, \bibinfo {author} {\bibfnamefont {K.}~\bibnamefont {Kamino}},
  \bibinfo {author} {\bibfnamefont {B.~B.}\ \bibnamefont {Machta}},\ and\
  \bibinfo {author} {\bibfnamefont {T.}~\bibnamefont {Emonet}},\ }\bibfield
  {title} {\bibinfo {title} {{Escherichia coli chemotaxis is information
  limited}},\ }\href {https://doi.org/10.1038/s41567-021-01380-3} {\bibfield
  {journal} {\bibinfo  {journal} {Nature Physics}\ }\textbf {\bibinfo {volume}
  {17}},\ \bibinfo {pages} {1426} (\bibinfo {year} {2021})}\BibitemShut
  {NoStop}%
\bibitem [{\citenamefont {van Kampen}(2007)}]{2007.vanKampen}%
  \BibitemOpen
  \bibfield  {author} {\bibinfo {author} {\bibfnamefont {N.~G.}\ \bibnamefont
  {van Kampen}},\ }\href {https://doi.org/10.1016/b978-0-444-52965-7.x5000-4}
  {\emph {\bibinfo {title} {{Stochastic Processes in Physics and
  Chemistry}}}},\ \bibinfo {edition} {3rd}\ ed.\ (\bibinfo  {publisher}
  {Elsevier},\ \bibinfo {address} {Amsterdam},\ \bibinfo {year}
  {2007})\BibitemShut {NoStop}%
\bibitem [{\citenamefont {Weber}\ and\ \citenamefont
  {Frey}(2017)}]{2017.Weber}%
  \BibitemOpen
  \bibfield  {author} {\bibinfo {author} {\bibfnamefont {M.~F.}\ \bibnamefont
  {Weber}}\ and\ \bibinfo {author} {\bibfnamefont {E.}~\bibnamefont {Frey}},\
  }\bibfield  {title} {\bibinfo {title} {{Master equations and the theory of
  stochastic path integrals}},\ }\href
  {https://doi.org/10.1088/1361-6633/aa5ae2} {\bibfield  {journal} {\bibinfo
  {journal} {Reports on Progress in Physics}\ }\textbf {\bibinfo {volume}
  {80}},\ \bibinfo {pages} {046601} (\bibinfo {year} {2017})},\ \Eprint
  {https://arxiv.org/abs/1609.02849} {1609.02849} \BibitemShut {NoStop}%
\bibitem [{\citenamefont {Walczak}\ \emph {et~al.}(2009)\citenamefont
  {Walczak}, \citenamefont {Mugler},\ and\ \citenamefont
  {Wiggins}}]{2009.Walczak}%
  \BibitemOpen
  \bibfield  {author} {\bibinfo {author} {\bibfnamefont {A.~M.}\ \bibnamefont
  {Walczak}}, \bibinfo {author} {\bibfnamefont {A.}~\bibnamefont {Mugler}},\
  and\ \bibinfo {author} {\bibfnamefont {C.~H.}\ \bibnamefont {Wiggins}},\
  }\bibfield  {title} {\bibinfo {title} {{A stochastic spectral analysis of
  transcriptional regulatory cascades}},\ }\href
  {https://doi.org/10.1073/pnas.0811999106} {\bibfield  {journal} {\bibinfo
  {journal} {Proceedings of the National Academy of Sciences}\ }\textbf
  {\bibinfo {volume} {106}},\ \bibinfo {pages} {6529} (\bibinfo {year}
  {2009})},\ \Eprint {https://arxiv.org/abs/0811.4149} {0811.4149} \BibitemShut
  {NoStop}%
\bibitem [{\citenamefont {Kim}\ and\ \citenamefont {Wang}(2007)}]{2007.Kim}%
  \BibitemOpen
  \bibfield  {author} {\bibinfo {author} {\bibfnamefont {K.-Y.}\ \bibnamefont
  {Kim}}\ and\ \bibinfo {author} {\bibfnamefont {J.}~\bibnamefont {Wang}},\
  }\bibfield  {title} {\bibinfo {title} {{Potential Energy Landscape and
  Robustness of a Gene Regulatory Network: Toggle Switch}},\ }\href
  {https://doi.org/10.1371/journal.pcbi.0030060} {\bibfield  {journal}
  {\bibinfo  {journal} {PLoS Computational Biology}\ }\textbf {\bibinfo
  {volume} {3}},\ \bibinfo {pages} {e60} (\bibinfo {year} {2007})}\BibitemShut
  {NoStop}%
\bibitem [{\citenamefont {Gillespie}(1976)}]{1976.Gillespie}%
  \BibitemOpen
  \bibfield  {author} {\bibinfo {author} {\bibfnamefont {D.~T.}\ \bibnamefont
  {Gillespie}},\ }\bibfield  {title} {\bibinfo {title} {{A general method for
  numerically simulating the stochastic time evolution of coupled chemical
  reactions}},\ }\href {https://doi.org/10.1016/0021-9991(76)90041-3}
  {\bibfield  {journal} {\bibinfo  {journal} {Journal of Computational
  Physics}\ }\textbf {\bibinfo {volume} {22}},\ \bibinfo {pages} {403}
  (\bibinfo {year} {1976})}\BibitemShut {NoStop}%
\bibitem [{\citenamefont {Prados}\ \emph {et~al.}(1997)\citenamefont {Prados},
  \citenamefont {Brey},\ and\ \citenamefont {S\'{a}nchez-Rey}}]{1997.Prados}%
  \BibitemOpen
  \bibfield  {author} {\bibinfo {author} {\bibfnamefont {A.}~\bibnamefont
  {Prados}}, \bibinfo {author} {\bibfnamefont {J.~J.}\ \bibnamefont {Brey}},\
  and\ \bibinfo {author} {\bibfnamefont {B.}~\bibnamefont {S\'{a}nchez-Rey}},\
  }\bibfield  {title} {\bibinfo {title} {{A dynamical monte carlo algorithm for
  master equations with time-dependent transition rates}},\ }\href
  {https://doi.org/10.1007/bf02765541} {\bibfield  {journal} {\bibinfo
  {journal} {Journal of Statistical Physics}\ }\textbf {\bibinfo {volume}
  {89}},\ \bibinfo {pages} {709} (\bibinfo {year} {1997})}\BibitemShut
  {NoStop}%
\bibitem [{\citenamefont {Thanh}\ and\ \citenamefont
  {Priami}(2015)}]{2015.Thanh}%
  \BibitemOpen
  \bibfield  {author} {\bibinfo {author} {\bibfnamefont {V.~H.}\ \bibnamefont
  {Thanh}}\ and\ \bibinfo {author} {\bibfnamefont {C.}~\bibnamefont {Priami}},\
  }\bibfield  {title} {\bibinfo {title} {{Simulation of biochemical reactions
  with time-dependent rates by the rejection-based algorithm}},\ }\href
  {https://doi.org/10.1063/1.4927916} {\bibfield  {journal} {\bibinfo
  {journal} {The Journal of Chemical Physics}\ }\textbf {\bibinfo {volume}
  {143}},\ \bibinfo {pages} {054104} (\bibinfo {year} {2015})}\BibitemShut
  {NoStop}%
\bibitem [{\citenamefont {Kloeden}\ and\ \citenamefont
  {Platen}(1992)}]{1992.Kloeden}%
  \BibitemOpen
  \bibfield  {author} {\bibinfo {author} {\bibfnamefont {P.~E.}\ \bibnamefont
  {Kloeden}}\ and\ \bibinfo {author} {\bibfnamefont {E.}~\bibnamefont
  {Platen}},\ }\href {https://doi.org/10.1007/978-3-662-12616-5} {\emph
  {\bibinfo {title} {{Numerical Solution of Stochastic Differential
  Equations}}}}\ (\bibinfo  {publisher} {Springer},\ \bibinfo {address}
  {Berlin, Heidelberg},\ \bibinfo {year} {1992})\BibitemShut {NoStop}%
\bibitem [{Note1()}]{Note1}%
  \BibitemOpen
  \bibinfo {note} {Indeed, the mutual information $\protect \mathrm
  {I}(\protect \mathcal {S}, \protect \mathcal {X})$ precisely quantifies how
  strong the statistical dependence is between the trajectory-valued random
  variables $\protect \mathcal {S}$ and $\protect \mathcal {X}$. From its
  definition $\protect \mathrm {I}(\protect \mathcal {S}, \protect \mathcal
  {X})=\protect \mathrm {H}(\protect \mathcal {S}) - \protect \mathrm
  {H}(\protect \mathcal {S}|\protect \mathcal {X})$ we can understand more
  clearly how this affects the efficiency of the Monte Carlo estimate. Roughly
  speaking, $\protect \mathrm {H}(\protect \mathcal {S})$ is related to the
  number of distinct trajectories $\protect \bm {s}$ that can arise from the
  dynamics given by $\protect \mathcal {P}[\protect \bm {s}]$, while $\protect
  \mathrm {H}(\protect \mathcal {S}|\protect \mathcal {X})$ is related to the
  number of distinct trajectories $\protect \bm {s}$ that could have lead to a
  specific output $\protect \bm {x}$, on average. Therefore, if the mutual
  information is very large, the difference between these two numbers is very
  large, and consequently the number of overall distinct trajectories is much
  larger than the number of distinct trajectories compatible with output
  $\protect \bm {x}$. Now, if we generate trajectories according to the
  dynamics given by $\protect \mathcal {P}[\protect \bm {s}]$, with
  overwhelming probability we generate a trajectory $\protect \bm {s}$ which is
  not compatible with the output trajectory $\protect \bm {x}$, and therefore
  $\protect \mathcal {P}[\protect \bm {x}|\protect \bm {s}]\approx 0$. Hence,
  the effective number of samples $M_\protect \text {eff}$ is much smaller than
  the actual number of generated trajectories $M$, i.e. $M_\protect \text {eff}
  \ll M$. We therefore only expect the estimate in \protect \cref
  {eq:marginal-naive} to be reliable when computing the mutual information for
  systems where it is not too high. Thus, strikingly, the difficulty of
  computing the mutual information is proportional to the magnitude of the
  mutual information itself.}\BibitemShut {Stop}%
\bibitem [{\citenamefont {M\"uller}\ and\ \citenamefont
  {Paul}(1994)}]{1994.Mueller}%
  \BibitemOpen
  \bibfield  {author} {\bibinfo {author} {\bibfnamefont {M.}~\bibnamefont
  {M\"uller}}\ and\ \bibinfo {author} {\bibfnamefont {W.}~\bibnamefont
  {Paul}},\ }\bibfield  {title} {\bibinfo {title} {{Measuring the chemical
  potential of polymer solutions and melts in computer simulations}},\ }\href
  {https://doi.org/10.1063/1.466937} {\bibfield  {journal} {\bibinfo  {journal}
  {The Journal of Chemical Physics}\ }\textbf {\bibinfo {volume} {100}},\
  \bibinfo {pages} {719} (\bibinfo {year} {1994})}\BibitemShut {NoStop}%
\bibitem [{\citenamefont {Rosenbluth}\ and\ \citenamefont
  {Rosenbluth}(1955)}]{1955.Rosenbluth}%
  \BibitemOpen
  \bibfield  {author} {\bibinfo {author} {\bibfnamefont {M.~N.}\ \bibnamefont
  {Rosenbluth}}\ and\ \bibinfo {author} {\bibfnamefont {A.~W.}\ \bibnamefont
  {Rosenbluth}},\ }\bibfield  {title} {\bibinfo {title} {{Monte Carlo
  Calculation of the Average Extension of Molecular Chains}},\ }\href
  {https://doi.org/10.1063/1.1741967} {\bibfield  {journal} {\bibinfo
  {journal} {The Journal of Chemical Physics}\ }\textbf {\bibinfo {volume}
  {23}},\ \bibinfo {pages} {356} (\bibinfo {year} {1955})}\BibitemShut
  {NoStop}%
\bibitem [{\citenamefont {Gordon}\ \emph {et~al.}(1993)\citenamefont {Gordon},
  \citenamefont {Salmond},\ and\ \citenamefont {Smith}}]{1993.Gordon}%
  \BibitemOpen
  \bibfield  {author} {\bibinfo {author} {\bibfnamefont {N.~J.}\ \bibnamefont
  {Gordon}}, \bibinfo {author} {\bibfnamefont {D.~J.}\ \bibnamefont
  {Salmond}},\ and\ \bibinfo {author} {\bibfnamefont {A.~F.~M.}\ \bibnamefont
  {Smith}},\ }\bibfield  {title} {\bibinfo {title} {{Novel approach to
  nonlinear/non-Gaussian Bayesian state estimation}},\ }\href
  {https://doi.org/10.1049/ip-f-2.1993.0015} {\bibfield  {journal} {\bibinfo
  {journal} {IEE Proceedings F Radar and Signal Processing}\ }\textbf {\bibinfo
  {volume} {140}},\ \bibinfo {pages} {107} (\bibinfo {year}
  {1993})}\BibitemShut {NoStop}%
\bibitem [{\citenamefont {Prellberg}\ and\ \citenamefont
  {Krawczyk}(2004)}]{2004.Prellberg}%
  \BibitemOpen
  \bibfield  {author} {\bibinfo {author} {\bibfnamefont {T.}~\bibnamefont
  {Prellberg}}\ and\ \bibinfo {author} {\bibfnamefont {J.}~\bibnamefont
  {Krawczyk}},\ }\bibfield  {title} {\bibinfo {title} {{Flat Histogram Version
  of the Pruned and Enriched Rosenbluth Method}},\ }\href
  {https://doi.org/10.1103/physrevlett.92.120602} {\bibfield  {journal}
  {\bibinfo  {journal} {Physical Review Letters}\ }\textbf {\bibinfo {volume}
  {92}},\ \bibinfo {pages} {120602} (\bibinfo {year} {2004})}\BibitemShut
  {NoStop}%
\bibitem [{\citenamefont {Allen}\ \emph {et~al.}(2006)\citenamefont {Allen},
  \citenamefont {Frenkel},\ and\ \citenamefont {ten Wolde}}]{2006.Allen}%
  \BibitemOpen
  \bibfield  {author} {\bibinfo {author} {\bibfnamefont {R.~J.}\ \bibnamefont
  {Allen}}, \bibinfo {author} {\bibfnamefont {D.}~\bibnamefont {Frenkel}},\
  and\ \bibinfo {author} {\bibfnamefont {P.~R.}\ \bibnamefont {ten Wolde}},\
  }\bibfield  {title} {\bibinfo {title} {{Simulating rare events in equilibrium
  or nonequilibrium stochastic systems}},\ }\href
  {https://doi.org/10.1063/1.2140273} {\bibfield  {journal} {\bibinfo
  {journal} {The Journal of Chemical Physics}\ }\textbf {\bibinfo {volume}
  {124}},\ \bibinfo {pages} {024102} (\bibinfo {year} {2006})},\ \Eprint
  {https://arxiv.org/abs/cond-mat/0509499} {cond-mat/0509499} \BibitemShut
  {NoStop}%
\bibitem [{\citenamefont {Becker}\ \emph {et~al.}(2012)\citenamefont {Becker},
  \citenamefont {Allen},\ and\ \citenamefont {ten Wolde}}]{2012.Becker}%
  \BibitemOpen
  \bibfield  {author} {\bibinfo {author} {\bibfnamefont {N.~B.}\ \bibnamefont
  {Becker}}, \bibinfo {author} {\bibfnamefont {R.~J.}\ \bibnamefont {Allen}},\
  and\ \bibinfo {author} {\bibfnamefont {P.~R.}\ \bibnamefont {ten Wolde}},\
  }\bibfield  {title} {\bibinfo {title} {{Non-stationary forward flux
  sampling}},\ }\href {https://doi.org/10.1063/1.4704810} {\bibfield  {journal}
  {\bibinfo  {journal} {The Journal of Chemical Physics}\ }\textbf {\bibinfo
  {volume} {136}},\ \bibinfo {pages} {174118} (\bibinfo {year} {2012})},\
  \Eprint {https://arxiv.org/abs/1201.3823} {1201.3823} \BibitemShut {NoStop}%
\bibitem [{\citenamefont {Martino}\ \emph {et~al.}(2017)\citenamefont
  {Martino}, \citenamefont {Elvira},\ and\ \citenamefont
  {Louzada}}]{2017.Martino}%
  \BibitemOpen
  \bibfield  {author} {\bibinfo {author} {\bibfnamefont {L.}~\bibnamefont
  {Martino}}, \bibinfo {author} {\bibfnamefont {V.}~\bibnamefont {Elvira}},\
  and\ \bibinfo {author} {\bibfnamefont {F.}~\bibnamefont {Louzada}},\
  }\bibfield  {title} {\bibinfo {title} {{Effective sample size for importance
  sampling based on discrepancy measures}},\ }\href
  {https://doi.org/10.1016/j.sigpro.2016.08.025} {\bibfield  {journal}
  {\bibinfo  {journal} {Signal Processing}\ }\textbf {\bibinfo {volume}
  {131}},\ \bibinfo {pages} {386} (\bibinfo {year} {2017})},\ \Eprint
  {https://arxiv.org/abs/1602.03572} {1602.03572} \BibitemShut {NoStop}%
\bibitem [{\citenamefont {Doucet}\ and\ \citenamefont
  {Johansen}(2011)}]{2011.Doucet}%
  \BibitemOpen
  \bibfield  {author} {\bibinfo {author} {\bibfnamefont {A.}~\bibnamefont
  {Doucet}}\ and\ \bibinfo {author} {\bibfnamefont {A.~M.}\ \bibnamefont
  {Johansen}},\ }\bibfield  {title} {\bibinfo {title} {{A tutorial on particle
  filtering and smoothing: fifteen years later}},\ }in\ \href@noop {} {\emph
  {\bibinfo {booktitle} {Oxford Handbook of Nonlinear Filtering}}},\ \bibinfo
  {editor} {edited by\ \bibinfo {editor} {\bibfnamefont {D.}~\bibnamefont
  {Crisan}}\ and\ \bibinfo {editor} {\bibfnamefont {B.}~\bibnamefont
  {Rozovskii}}}\ (\bibinfo  {publisher} {Oxford University Press},\ \bibinfo
  {year} {2011})\BibitemShut {NoStop}%
\bibitem [{\citenamefont {Metropolis}\ \emph {et~al.}(1953)\citenamefont
  {Metropolis}, \citenamefont {Rosenbluth}, \citenamefont {Rosenbluth},
  \citenamefont {Teller},\ and\ \citenamefont {Teller}}]{1953.Metropolis}%
  \BibitemOpen
  \bibfield  {author} {\bibinfo {author} {\bibfnamefont {N.}~\bibnamefont
  {Metropolis}}, \bibinfo {author} {\bibfnamefont {A.~W.}\ \bibnamefont
  {Rosenbluth}}, \bibinfo {author} {\bibfnamefont {M.~N.}\ \bibnamefont
  {Rosenbluth}}, \bibinfo {author} {\bibfnamefont {A.~H.}\ \bibnamefont
  {Teller}},\ and\ \bibinfo {author} {\bibfnamefont {E.}~\bibnamefont
  {Teller}},\ }\bibfield  {title} {\bibinfo {title} {{Equation of State
  Calculations by Fast Computing Machines}},\ }\href
  {https://doi.org/10.1063/1.1699114} {\bibfield  {journal} {\bibinfo
  {journal} {The Journal of Chemical Physics}\ }\textbf {\bibinfo {volume}
  {21}},\ \bibinfo {pages} {1087} (\bibinfo {year} {1953})}\BibitemShut
  {NoStop}%
\bibitem [{\citenamefont {Bezanson}\ \emph {et~al.}(2017)\citenamefont
  {Bezanson}, \citenamefont {Edelman}, \citenamefont {Karpinski},\ and\
  \citenamefont {Shah}}]{2017.Bezanson}%
  \BibitemOpen
  \bibfield  {author} {\bibinfo {author} {\bibfnamefont {J.}~\bibnamefont
  {Bezanson}}, \bibinfo {author} {\bibfnamefont {A.}~\bibnamefont {Edelman}},
  \bibinfo {author} {\bibfnamefont {S.}~\bibnamefont {Karpinski}},\ and\
  \bibinfo {author} {\bibfnamefont {V.~B.}\ \bibnamefont {Shah}},\ }\bibfield
  {title} {\bibinfo {title} {{Julia: A Fresh Approach to Numerical
  Computing}},\ }\href {https://doi.org/10.1137/141000671} {\bibfield
  {journal} {\bibinfo  {journal} {SIAM Review}\ }\textbf {\bibinfo {volume}
  {59}},\ \bibinfo {pages} {65} (\bibinfo {year} {2017})}\BibitemShut {NoStop}%
\bibitem [{\citenamefont {Reinhardt}(2021)}]{manuel_reinhardt_2021_6334035}%
  \BibitemOpen
  \bibfield  {author} {\bibinfo {author} {\bibfnamefont {M.}~\bibnamefont
  {Reinhardt}},\ }\href {https://doi.org/10.5281/zenodo.6334035} {\bibinfo
  {title} {{PathWeightSampling.jl (v0.1.0)}}} (\bibinfo {year} {2021}),\
  \bibinfo {note} {{Zenodo}}\BibitemShut {NoStop}%
\bibitem [{\citenamefont {Reinhardt}()}]{pws_github}%
  \BibitemOpen
  \bibfield  {author} {\bibinfo {author} {\bibfnamefont {M.}~\bibnamefont
  {Reinhardt}},\ }\href {https://github.com/manuel-rhdt/PathWeightSampling.jl}
  {\bibinfo {title} {{manuel-rhdt/PathWeightSampling.jl}}},\ \bibinfo {note}
  {{GitHub}}\BibitemShut {NoStop}%
\bibitem [{\citenamefont {Rackauckas}\ and\ \citenamefont
  {Nie}(2017)}]{2017.Rackauckas}%
  \BibitemOpen
  \bibfield  {author} {\bibinfo {author} {\bibfnamefont {C.}~\bibnamefont
  {Rackauckas}}\ and\ \bibinfo {author} {\bibfnamefont {Q.}~\bibnamefont
  {Nie}},\ }\bibfield  {title} {\bibinfo {title} {{DifferentialEquations.jl –
  A Performant and Feature-Rich Ecosystem for Solving Differential Equations in
  Julia}},\ }\href {https://doi.org/10.5334/jors.151} {\bibfield  {journal}
  {\bibinfo  {journal} {Journal of Open Research Software}\ }\textbf {\bibinfo
  {volume} {5}},\ \bibinfo {pages} {15} (\bibinfo {year} {2017})}\BibitemShut
  {NoStop}%
\bibitem [{\citenamefont {Ma}\ \emph {et~al.}(2021)\citenamefont {Ma},
  \citenamefont {Gowda}, \citenamefont {Anantharaman}, \citenamefont
  {Laughman}, \citenamefont {Shah},\ and\ \citenamefont
  {Rackauckas}}]{2021.Ma}%
  \BibitemOpen
  \bibfield  {author} {\bibinfo {author} {\bibfnamefont {Y.}~\bibnamefont
  {Ma}}, \bibinfo {author} {\bibfnamefont {S.}~\bibnamefont {Gowda}}, \bibinfo
  {author} {\bibfnamefont {R.}~\bibnamefont {Anantharaman}}, \bibinfo {author}
  {\bibfnamefont {C.}~\bibnamefont {Laughman}}, \bibinfo {author}
  {\bibfnamefont {V.}~\bibnamefont {Shah}},\ and\ \bibinfo {author}
  {\bibfnamefont {C.}~\bibnamefont {Rackauckas}},\ }\bibfield  {title}
  {\bibinfo {title} {{ModelingToolkit: A Composable Graph Transformation System
  For Equation-Based Modeling}},\ }\href@noop {} {\bibfield  {journal}
  {\bibinfo  {journal} {arXiv}\ } (\bibinfo {year} {2021})},\ \Eprint
  {https://arxiv.org/abs/2103.05244} {2103.05244} \BibitemShut {NoStop}%
\bibitem [{Note2()}]{Note2}%
  \BibitemOpen
  \bibinfo {note} {\protect \url
  {https://github.com/zechnerlab/PathMI/tree/302f03e51ad195adc6be39fa9618886c76590cc4}}\BibitemShut
  {NoStop}%
\bibitem [{\citenamefont {Berg}\ and\ \citenamefont
  {Purcell}(1977)}]{1977.Berg}%
  \BibitemOpen
  \bibfield  {author} {\bibinfo {author} {\bibfnamefont {H.}~\bibnamefont
  {Berg}}\ and\ \bibinfo {author} {\bibfnamefont {E.}~\bibnamefont {Purcell}},\
  }\bibfield  {title} {\bibinfo {title} {{Physics of chemoreception}},\ }\href
  {https://doi.org/10.1016/s0006-3495(77)85544-6} {\bibfield  {journal}
  {\bibinfo  {journal} {Biophysical Journal}\ }\textbf {\bibinfo {volume}
  {20}},\ \bibinfo {pages} {193} (\bibinfo {year} {1977})}\BibitemShut
  {NoStop}%
\bibitem [{\citenamefont {Barkai}\ and\ \citenamefont
  {Leibler}(1997)}]{1997.Barkai}%
  \BibitemOpen
  \bibfield  {author} {\bibinfo {author} {\bibfnamefont {N.}~\bibnamefont
  {Barkai}}\ and\ \bibinfo {author} {\bibfnamefont {S.}~\bibnamefont
  {Leibler}},\ }\bibfield  {title} {\bibinfo {title} {{Robustness in simple
  biochemical networks}},\ }\href {https://doi.org/10.1038/43199} {\bibfield
  {journal} {\bibinfo  {journal} {Nature}\ }\textbf {\bibinfo {volume} {387}},\
  \bibinfo {pages} {913} (\bibinfo {year} {1997})}\BibitemShut {NoStop}%
\bibitem [{\citenamefont {Endres}\ and\ \citenamefont
  {Wingreen}(2006)}]{2006.Endres}%
  \BibitemOpen
  \bibfield  {author} {\bibinfo {author} {\bibfnamefont {R.~G.}\ \bibnamefont
  {Endres}}\ and\ \bibinfo {author} {\bibfnamefont {N.~S.}\ \bibnamefont
  {Wingreen}},\ }\bibfield  {title} {\bibinfo {title} {{Precise adaptation in
  bacterial chemotaxis through “assistance neighborhoods”}},\ }\href
  {https://doi.org/10.1073/pnas.0603101103} {\bibfield  {journal} {\bibinfo
  {journal} {Proceedings of the National Academy of Sciences}\ }\textbf
  {\bibinfo {volume} {103}},\ \bibinfo {pages} {13040} (\bibinfo {year}
  {2006})}\BibitemShut {NoStop}%
\bibitem [{\citenamefont {Lan}\ \emph {et~al.}(2012)\citenamefont {Lan},
  \citenamefont {Sartori}, \citenamefont {Neumann}, \citenamefont {Sourjik},\
  and\ \citenamefont {Tu}}]{2012.Lan}%
  \BibitemOpen
  \bibfield  {author} {\bibinfo {author} {\bibfnamefont {G.}~\bibnamefont
  {Lan}}, \bibinfo {author} {\bibfnamefont {P.}~\bibnamefont {Sartori}},
  \bibinfo {author} {\bibfnamefont {S.}~\bibnamefont {Neumann}}, \bibinfo
  {author} {\bibfnamefont {V.}~\bibnamefont {Sourjik}},\ and\ \bibinfo {author}
  {\bibfnamefont {Y.}~\bibnamefont {Tu}},\ }\bibfield  {title} {\bibinfo
  {title} {{The energy–speed–accuracy trade-off in sensory adaptation}},\
  }\href {https://doi.org/10.1038/nphys2276} {\bibfield  {journal} {\bibinfo
  {journal} {Nature Physics}\ }\textbf {\bibinfo {volume} {8}},\ \bibinfo
  {pages} {422} (\bibinfo {year} {2012})}\BibitemShut {NoStop}%
\bibitem [{\citenamefont {Sartori}\ and\ \citenamefont
  {Tu}(2015)}]{2015.Sartori}%
  \BibitemOpen
  \bibfield  {author} {\bibinfo {author} {\bibfnamefont {P.}~\bibnamefont
  {Sartori}}\ and\ \bibinfo {author} {\bibfnamefont {Y.}~\bibnamefont {Tu}},\
  }\bibfield  {title} {\bibinfo {title} {{Free Energy Cost of Reducing Noise
  while Maintaining a High Sensitivity}},\ }\href
  {https://doi.org/10.1103/physrevlett.115.118102} {\bibfield  {journal}
  {\bibinfo  {journal} {Physical Review Letters}\ }\textbf {\bibinfo {volume}
  {115}},\ \bibinfo {pages} {118102} (\bibinfo {year} {2015})},\ \Eprint
  {https://arxiv.org/abs/1505.07413} {1505.07413} \BibitemShut {NoStop}%
\bibitem [{\citenamefont {Monod}\ \emph {et~al.}(1965)\citenamefont {Monod},
  \citenamefont {Wyman},\ and\ \citenamefont {Changeux}}]{1965.Monod}%
  \BibitemOpen
  \bibfield  {author} {\bibinfo {author} {\bibfnamefont {J.}~\bibnamefont
  {Monod}}, \bibinfo {author} {\bibfnamefont {J.}~\bibnamefont {Wyman}},\ and\
  \bibinfo {author} {\bibfnamefont {J.-P.}\ \bibnamefont {Changeux}},\
  }\bibfield  {title} {\bibinfo {title} {{On the nature of allosteric
  transitions: A plausible model}},\ }\href
  {https://doi.org/10.1016/s0022-2836(65)80285-6} {\bibfield  {journal}
  {\bibinfo  {journal} {Journal of Molecular Biology}\ }\textbf {\bibinfo
  {volume} {12}},\ \bibinfo {pages} {88} (\bibinfo {year} {1965})}\BibitemShut
  {NoStop}%
\bibitem [{\citenamefont {Shimizu}\ \emph {et~al.}(2010)\citenamefont
  {Shimizu}, \citenamefont {Tu},\ and\ \citenamefont {Berg}}]{2010.Shimizu}%
  \BibitemOpen
  \bibfield  {author} {\bibinfo {author} {\bibfnamefont {T.~S.}\ \bibnamefont
  {Shimizu}}, \bibinfo {author} {\bibfnamefont {Y.}~\bibnamefont {Tu}},\ and\
  \bibinfo {author} {\bibfnamefont {H.~C.}\ \bibnamefont {Berg}},\ }\bibfield
  {title} {\bibinfo {title} {{A modular gradient‐sensing network for
  chemotaxis in Escherichia coli revealed by responses to time‐varying
  stimuli}},\ }\href {https://doi.org/10.1038/msb.2010.37} {\bibfield
  {journal} {\bibinfo  {journal} {Molecular Systems Biology}\ }\textbf
  {\bibinfo {volume} {6}},\ \bibinfo {pages} {382} (\bibinfo {year}
  {2010})}\BibitemShut {NoStop}%
\bibitem [{\citenamefont {Kamino}\ \emph {et~al.}(2020)\citenamefont {Kamino},
  \citenamefont {Keegstra}, \citenamefont {Long}, \citenamefont {Emonet},\ and\
  \citenamefont {Shimizu}}]{2020.Kamino}%
  \BibitemOpen
  \bibfield  {author} {\bibinfo {author} {\bibfnamefont {K.}~\bibnamefont
  {Kamino}}, \bibinfo {author} {\bibfnamefont {J.~M.}\ \bibnamefont
  {Keegstra}}, \bibinfo {author} {\bibfnamefont {J.}~\bibnamefont {Long}},
  \bibinfo {author} {\bibfnamefont {T.}~\bibnamefont {Emonet}},\ and\ \bibinfo
  {author} {\bibfnamefont {T.~S.}\ \bibnamefont {Shimizu}},\ }\bibfield
  {title} {\bibinfo {title} {{Adaptive tuning of cell sensory diversity without
  changes in gene expression}},\ }\href
  {https://doi.org/10.1126/sciadv.abc1087} {\bibfield  {journal} {\bibinfo
  {journal} {Science Advances}\ }\textbf {\bibinfo {volume} {6}},\ \bibinfo
  {pages} {eabc1087} (\bibinfo {year} {2020})}\BibitemShut {NoStop}%
\bibitem [{\citenamefont {Li}\ and\ \citenamefont
  {Hazelbauer}(2004)}]{2004.Li}%
  \BibitemOpen
  \bibfield  {author} {\bibinfo {author} {\bibfnamefont {M.}~\bibnamefont
  {Li}}\ and\ \bibinfo {author} {\bibfnamefont {G.~L.}\ \bibnamefont
  {Hazelbauer}},\ }\bibfield  {title} {\bibinfo {title} {{Cellular
  Stoichiometry of the Components of the Chemotaxis Signaling Complex}},\
  }\href {https://doi.org/10.1128/jb.186.12.3687-3694.2004} {\bibfield
  {journal} {\bibinfo  {journal} {Journal of Bacteriology}\ }\textbf {\bibinfo
  {volume} {186}},\ \bibinfo {pages} {3687} (\bibinfo {year}
  {2004})}\BibitemShut {NoStop}%
\bibitem [{\citenamefont {Lazova}\ \emph {et~al.}(2011)\citenamefont {Lazova},
  \citenamefont {Ahmed}, \citenamefont {Bellomo}, \citenamefont {Stocker},\
  and\ \citenamefont {Shimizu}}]{2011.Lazova}%
  \BibitemOpen
  \bibfield  {author} {\bibinfo {author} {\bibfnamefont {M.~D.}\ \bibnamefont
  {Lazova}}, \bibinfo {author} {\bibfnamefont {T.}~\bibnamefont {Ahmed}},
  \bibinfo {author} {\bibfnamefont {D.}~\bibnamefont {Bellomo}}, \bibinfo
  {author} {\bibfnamefont {R.}~\bibnamefont {Stocker}},\ and\ \bibinfo {author}
  {\bibfnamefont {T.~S.}\ \bibnamefont {Shimizu}},\ }\bibfield  {title}
  {\bibinfo {title} {{Response rescaling in bacterial chemotaxis}},\ }\href
  {https://doi.org/10.1073/pnas.1108608108} {\bibfield  {journal} {\bibinfo
  {journal} {Proceedings of the National Academy of Sciences}\ }\textbf
  {\bibinfo {volume} {108}},\ \bibinfo {pages} {13870} (\bibinfo {year}
  {2011})}\BibitemShut {NoStop}%
\bibitem [{\citenamefont {Mello}\ and\ \citenamefont {Tu}(2007)}]{2007.Mello}%
  \BibitemOpen
  \bibfield  {author} {\bibinfo {author} {\bibfnamefont {B.~A.}\ \bibnamefont
  {Mello}}\ and\ \bibinfo {author} {\bibfnamefont {Y.}~\bibnamefont {Tu}},\
  }\bibfield  {title} {\bibinfo {title} {{Effects of Adaptation in Maintaining
  High Sensitivity over a Wide Range of Backgrounds for Escherichia coli
  Chemotaxis}},\ }\href {https://doi.org/10.1529/biophysj.106.097808}
  {\bibfield  {journal} {\bibinfo  {journal} {Biophysical Journal}\ }\textbf
  {\bibinfo {volume} {92}},\ \bibinfo {pages} {2329} (\bibinfo {year}
  {2007})}\BibitemShut {NoStop}%
\bibitem [{\citenamefont {Keegstra}\ \emph {et~al.}(2023)\citenamefont
  {Keegstra}, \citenamefont {Avgidis}, \citenamefont {Mulla}, \citenamefont
  {Parkinson},\ and\ \citenamefont {Shimizu}}]{2023.Keegstra}%
  \BibitemOpen
  \bibfield  {author} {\bibinfo {author} {\bibfnamefont {J.~M.}\ \bibnamefont
  {Keegstra}}, \bibinfo {author} {\bibfnamefont {F.}~\bibnamefont {Avgidis}},
  \bibinfo {author} {\bibfnamefont {Y.}~\bibnamefont {Mulla}}, \bibinfo
  {author} {\bibfnamefont {J.~S.}\ \bibnamefont {Parkinson}},\ and\ \bibinfo
  {author} {\bibfnamefont {T.~S.}\ \bibnamefont {Shimizu}},\ }\bibfield
  {title} {\bibinfo {title} {{Near-critical tuning of cooperativity revealed by
  spontaneous switching in a protein signalling array}},\ }\href
  {https://doi.org/10.1101/2022.12.04.518992} {\bibfield  {journal} {\bibinfo
  {journal} {bioRxiv}\ ,\ \bibinfo {pages} {2022.12.04.518992}} (\bibinfo
  {year} {2023})}\BibitemShut {NoStop}%
\bibitem [{\citenamefont {Tka\v{c}ik}\ \emph {et~al.}(2015)\citenamefont
  {Tka\v{c}ik}, \citenamefont {Mora}, \citenamefont {Marre}, \citenamefont
  {Amodei}, \citenamefont {Palmer}, \citenamefont {Berry},\ and\ \citenamefont
  {Bialek}}]{2015.Tkacik}%
  \BibitemOpen
  \bibfield  {author} {\bibinfo {author} {\bibfnamefont {G.}~\bibnamefont
  {Tka\v{c}ik}}, \bibinfo {author} {\bibfnamefont {T.}~\bibnamefont {Mora}},
  \bibinfo {author} {\bibfnamefont {O.}~\bibnamefont {Marre}}, \bibinfo
  {author} {\bibfnamefont {D.}~\bibnamefont {Amodei}}, \bibinfo {author}
  {\bibfnamefont {S.~E.}\ \bibnamefont {Palmer}}, \bibinfo {author}
  {\bibfnamefont {M.~J.}\ \bibnamefont {Berry}},\ and\ \bibinfo {author}
  {\bibfnamefont {W.}~\bibnamefont {Bialek}},\ }\bibfield  {title} {\bibinfo
  {title} {{Thermodynamics and signatures of criticality in a network of
  neurons}},\ }\href {https://doi.org/10.1073/pnas.1514188112} {\bibfield
  {journal} {\bibinfo  {journal} {Proceedings of the National Academy of
  Sciences}\ }\textbf {\bibinfo {volume} {112}},\ \bibinfo {pages} {11508}
  (\bibinfo {year} {2015})}\BibitemShut {NoStop}%
\bibitem [{\citenamefont {Skoge}\ \emph {et~al.}(2011)\citenamefont {Skoge},
  \citenamefont {Meir},\ and\ \citenamefont {Wingreen}}]{2011.Skoge}%
  \BibitemOpen
  \bibfield  {author} {\bibinfo {author} {\bibfnamefont {M.}~\bibnamefont
  {Skoge}}, \bibinfo {author} {\bibfnamefont {Y.}~\bibnamefont {Meir}},\ and\
  \bibinfo {author} {\bibfnamefont {N.~S.}\ \bibnamefont {Wingreen}},\
  }\bibfield  {title} {\bibinfo {title} {{Dynamics of Cooperativity in Chemical
  Sensing among Cell-Surface Receptors}},\ }\href
  {https://doi.org/10.1103/physrevlett.107.178101} {\bibfield  {journal}
  {\bibinfo  {journal} {Physical Review Letters}\ }\textbf {\bibinfo {volume}
  {107}},\ \bibinfo {pages} {178101} (\bibinfo {year} {2011})},\ \Eprint
  {https://arxiv.org/abs/1109.4160} {1109.4160} \BibitemShut {NoStop}%
\bibitem [{\citenamefont {Onsager}\ and\ \citenamefont
  {Machlup}(1953)}]{1953.Onsager}%
  \BibitemOpen
  \bibfield  {author} {\bibinfo {author} {\bibfnamefont {L.}~\bibnamefont
  {Onsager}}\ and\ \bibinfo {author} {\bibfnamefont {S.}~\bibnamefont
  {Machlup}},\ }\bibfield  {title} {\bibinfo {title} {{Fluctuations and
  Irreversible Processes}},\ }\href {https://doi.org/10.1103/physrev.91.1505}
  {\bibfield  {journal} {\bibinfo  {journal} {Physical Review}\ }\textbf
  {\bibinfo {volume} {91}},\ \bibinfo {pages} {1505} (\bibinfo {year}
  {1953})}\BibitemShut {NoStop}%
\bibitem [{\citenamefont {Adib}(2008)}]{2008.Adib}%
  \BibitemOpen
  \bibfield  {author} {\bibinfo {author} {\bibfnamefont {A.~B.}\ \bibnamefont
  {Adib}},\ }\bibfield  {title} {\bibinfo {title} {{Stochastic Actions for
  Diffusive Dynamics: Reweighting, Sampling, and Minimization}},\ }\href
  {https://doi.org/10.1021/jp0751458} {\bibfield  {journal} {\bibinfo
  {journal} {The Journal of Physical Chemistry B}\ }\textbf {\bibinfo {volume}
  {112}},\ \bibinfo {pages} {5910} (\bibinfo {year} {2008})},\ \Eprint
  {https://arxiv.org/abs/0712.1255} {0712.1255} \BibitemShut {NoStop}%
\bibitem [{\citenamefont {Baum}\ and\ \citenamefont
  {Petrie}(1966)}]{1966.Baum}%
  \BibitemOpen
  \bibfield  {author} {\bibinfo {author} {\bibfnamefont {L.~E.}\ \bibnamefont
  {Baum}}\ and\ \bibinfo {author} {\bibfnamefont {T.}~\bibnamefont {Petrie}},\
  }\bibfield  {title} {\bibinfo {title} {{Statistical Inference for
  Probabilistic Functions of Finite State Markov Chains}},\ }\href
  {https://doi.org/10.1214/aoms/1177699147} {\bibfield  {journal} {\bibinfo
  {journal} {The Annals of Mathematical Statistics}\ }\textbf {\bibinfo
  {volume} {37}},\ \bibinfo {pages} {1554} (\bibinfo {year}
  {1966})}\BibitemShut {NoStop}%
\bibitem [{\citenamefont {Paninski}\ \emph {et~al.}(2010)\citenamefont
  {Paninski}, \citenamefont {Ahmadian}, \citenamefont {Ferreira}, \citenamefont
  {Koyama}, \citenamefont {Rad}, \citenamefont {Vidne}, \citenamefont
  {Vogelstein},\ and\ \citenamefont {Wu}}]{2010.Paninski}%
  \BibitemOpen
  \bibfield  {author} {\bibinfo {author} {\bibfnamefont {L.}~\bibnamefont
  {Paninski}}, \bibinfo {author} {\bibfnamefont {Y.}~\bibnamefont {Ahmadian}},
  \bibinfo {author} {\bibfnamefont {D.~G.}\ \bibnamefont {Ferreira}}, \bibinfo
  {author} {\bibfnamefont {S.}~\bibnamefont {Koyama}}, \bibinfo {author}
  {\bibfnamefont {K.~R.}\ \bibnamefont {Rad}}, \bibinfo {author} {\bibfnamefont
  {M.}~\bibnamefont {Vidne}}, \bibinfo {author} {\bibfnamefont
  {J.}~\bibnamefont {Vogelstein}},\ and\ \bibinfo {author} {\bibfnamefont
  {W.}~\bibnamefont {Wu}},\ }\bibfield  {title} {\bibinfo {title} {{A new look
  at state-space models for neural data}},\ }\href
  {https://doi.org/10.1007/s10827-009-0179-x} {\bibfield  {journal} {\bibinfo
  {journal} {Journal of Computational Neuroscience}\ }\textbf {\bibinfo
  {volume} {29}},\ \bibinfo {pages} {107} (\bibinfo {year} {2010})}\BibitemShut
  {NoStop}%
\bibitem [{\citenamefont {Tang}\ \emph {et~al.}(2021)\citenamefont {Tang},
  \citenamefont {Adelaja}, \citenamefont {Ye}, \citenamefont {Deeds},
  \citenamefont {Wollman},\ and\ \citenamefont {Hoffmann}}]{2021.Tang}%
  \BibitemOpen
  \bibfield  {author} {\bibinfo {author} {\bibfnamefont {Y.}~\bibnamefont
  {Tang}}, \bibinfo {author} {\bibfnamefont {A.}~\bibnamefont {Adelaja}},
  \bibinfo {author} {\bibfnamefont {F.~X.-F.}\ \bibnamefont {Ye}}, \bibinfo
  {author} {\bibfnamefont {E.}~\bibnamefont {Deeds}}, \bibinfo {author}
  {\bibfnamefont {R.}~\bibnamefont {Wollman}},\ and\ \bibinfo {author}
  {\bibfnamefont {A.}~\bibnamefont {Hoffmann}},\ }\bibfield  {title} {\bibinfo
  {title} {{Quantifying information accumulation encoded in the dynamics of
  biochemical signaling}},\ }\href {https://doi.org/10.1038/s41467-021-21562-0}
  {\bibfield  {journal} {\bibinfo  {journal} {Nature Communications}\ }\textbf
  {\bibinfo {volume} {12}},\ \bibinfo {pages} {1272} (\bibinfo {year}
  {2021})}\BibitemShut {NoStop}%
\bibitem [{\citenamefont {Siepmann}\ and\ \citenamefont
  {Frenkel}(1992)}]{1992.Siepmann}%
  \BibitemOpen
  \bibfield  {author} {\bibinfo {author} {\bibfnamefont {J.~I.}\ \bibnamefont
  {Siepmann}}\ and\ \bibinfo {author} {\bibfnamefont {D.}~\bibnamefont
  {Frenkel}},\ }\bibfield  {title} {\bibinfo {title} {{Configurational bias
  Monte Carlo: a new sampling scheme for flexible chains}},\ }\href
  {https://doi.org/10.1080/00268979200100061} {\bibfield  {journal} {\bibinfo
  {journal} {Molecular Physics}\ }\textbf {\bibinfo {volume} {75}},\ \bibinfo
  {pages} {59} (\bibinfo {year} {1992})}\BibitemShut {NoStop}%
\bibitem [{\citenamefont {van Erp}\ \emph {et~al.}(2003)\citenamefont {van
  Erp}, \citenamefont {Moroni},\ and\ \citenamefont {Bolhuis}}]{2003.Erp}%
  \BibitemOpen
  \bibfield  {author} {\bibinfo {author} {\bibfnamefont {T.~S.}\ \bibnamefont
  {van Erp}}, \bibinfo {author} {\bibfnamefont {D.}~\bibnamefont {Moroni}},\
  and\ \bibinfo {author} {\bibfnamefont {P.~G.}\ \bibnamefont {Bolhuis}},\
  }\bibfield  {title} {\bibinfo {title} {{A novel path sampling method for the
  calculation of rate constants}},\ }\href {https://doi.org/10.1063/1.1562614}
  {\bibfield  {journal} {\bibinfo  {journal} {The Journal of Chemical Physics}\
  }\textbf {\bibinfo {volume} {118}},\ \bibinfo {pages} {7762} (\bibinfo {year}
  {2003})},\ \Eprint {https://arxiv.org/abs/cond-mat/0210614}
  {cond-mat/0210614} \BibitemShut {NoStop}%
\bibitem [{\citenamefont {Vlugt}\ and\ \citenamefont
  {Smit}(2001)}]{2001.Vlugt}%
  \BibitemOpen
  \bibfield  {author} {\bibinfo {author} {\bibfnamefont {T.~J.~H.}\
  \bibnamefont {Vlugt}}\ and\ \bibinfo {author} {\bibfnamefont
  {B.}~\bibnamefont {Smit}},\ }\bibfield  {title} {\bibinfo {title} {{On the
  efficient sampling of pathways in the transition path ensemble}},\ }\href
  {https://doi.org/10.1039/b009865p} {\bibfield  {journal} {\bibinfo  {journal}
  {PhysChemComm}\ }\textbf {\bibinfo {volume} {4}},\ \bibinfo {pages} {11}
  (\bibinfo {year} {2001})}\BibitemShut {NoStop}%
\bibitem [{\citenamefont {Del~Moral}(1997)}]{1997.Moral}%
  \BibitemOpen
  \bibfield  {author} {\bibinfo {author} {\bibfnamefont {P.}~\bibnamefont
  {Del~Moral}},\ }\bibfield  {title} {\bibinfo {title} {{Nonlinear filtering:
  Interacting particle resolution}},\ }\href
  {https://doi.org/10.1016/s0764-4442(97)84778-7} {\bibfield  {journal}
  {\bibinfo  {journal} {Comptes Rendus de l'Académie des Sciences - Series I -
  Mathematics}\ }\textbf {\bibinfo {volume} {325}},\ \bibinfo {pages} {653}
  (\bibinfo {year} {1997})}\BibitemShut {NoStop}%
\bibitem [{\citenamefont {Smith}\ and\ \citenamefont
  {Gelfand}(1992)}]{1992.Smith}%
  \BibitemOpen
  \bibfield  {author} {\bibinfo {author} {\bibfnamefont {A.~F.~M.}\
  \bibnamefont {Smith}}\ and\ \bibinfo {author} {\bibfnamefont {A.~E.}\
  \bibnamefont {Gelfand}},\ }\bibfield  {title} {\bibinfo {title} {{Bayesian
  Statistics without Tears: A Sampling-Resampling Perspective}},\ }\href
  {https://doi.org/10.2307/2684170} {\bibfield  {journal} {\bibinfo  {journal}
  {The American Statistician}\ }\textbf {\bibinfo {volume} {46}},\ \bibinfo
  {pages} {84} (\bibinfo {year} {1992})}\BibitemShut {NoStop}%
\bibitem [{\citenamefont {Dellago}\ \emph {et~al.}(1998)\citenamefont
  {Dellago}, \citenamefont {Bolhuis},\ and\ \citenamefont
  {Chandler}}]{1998a.Dellago}%
  \BibitemOpen
  \bibfield  {author} {\bibinfo {author} {\bibfnamefont {C.}~\bibnamefont
  {Dellago}}, \bibinfo {author} {\bibfnamefont {P.~G.}\ \bibnamefont
  {Bolhuis}},\ and\ \bibinfo {author} {\bibfnamefont {D.}~\bibnamefont
  {Chandler}},\ }\bibfield  {title} {\bibinfo {title} {{Efficient transition
  path sampling: Application to Lennard-Jones cluster rearrangements}},\ }\href
  {https://doi.org/10.1063/1.476378} {\bibfield  {journal} {\bibinfo  {journal}
  {The Journal of Chemical Physics}\ }\textbf {\bibinfo {volume} {108}},\
  \bibinfo {pages} {9236} (\bibinfo {year} {1998})}\BibitemShut {NoStop}%
\bibitem [{\citenamefont {Hobolth}\ and\ \citenamefont
  {Stone}(2009)}]{2009.Hobolth}%
  \BibitemOpen
  \bibfield  {author} {\bibinfo {author} {\bibfnamefont {A.}~\bibnamefont
  {Hobolth}}\ and\ \bibinfo {author} {\bibfnamefont {E.~A.}\ \bibnamefont
  {Stone}},\ }\bibfield  {title} {\bibinfo {title} {{Simulation from
  endpoint-conditioned, continuous-time Markov chains on a finite state space,
  with applications to molecular evolution}},\ }\href
  {https://doi.org/10.1214/09-aoas247} {\bibfield  {journal} {\bibinfo
  {journal} {The Annals of Applied Statistics}\ }\textbf {\bibinfo {volume}
  {3}},\ \bibinfo {pages} {1204} (\bibinfo {year} {2009})},\ \Eprint
  {https://arxiv.org/abs/0910.1683} {0910.1683} \BibitemShut {NoStop}%
\bibitem [{\citenamefont {van~der Meulen}\ and\ \citenamefont
  {Schauer}(2017)}]{2017.Meulen}%
  \BibitemOpen
  \bibfield  {author} {\bibinfo {author} {\bibfnamefont {F.}~\bibnamefont
  {van~der Meulen}}\ and\ \bibinfo {author} {\bibfnamefont {M.}~\bibnamefont
  {Schauer}},\ }\bibfield  {title} {\bibinfo {title} {{Bayesian estimation of
  discretely observed multi-dimensional diffusion processes using guided
  proposals}},\ }\href {https://doi.org/10.1214/17-ejs1290} {\bibfield
  {journal} {\bibinfo  {journal} {Electronic Journal of Statistics}\ }\textbf
  {\bibinfo {volume} {11}},\ \bibinfo {pages} {2358} (\bibinfo {year}
  {2017})},\ \Eprint {https://arxiv.org/abs/1406.4704} {1406.4704} \BibitemShut
  {NoStop}%
\bibitem [{\citenamefont {Golightly}\ and\ \citenamefont
  {Wilkinson}(2015)}]{2015.Golightly}%
  \BibitemOpen
  \bibfield  {author} {\bibinfo {author} {\bibfnamefont {A.}~\bibnamefont
  {Golightly}}\ and\ \bibinfo {author} {\bibfnamefont {D.~J.}\ \bibnamefont
  {Wilkinson}},\ }\bibfield  {title} {\bibinfo {title} {{Bayesian inference for
  Markov jump processes with informative observations}},\ }\href
  {https://doi.org/10.1515/sagmb-2014-0070} {\bibfield  {journal} {\bibinfo
  {journal} {Statistical Applications in Genetics and Molecular Biology}\
  }\textbf {\bibinfo {volume} {14}},\ \bibinfo {pages} {169} (\bibinfo {year}
  {2015})}\BibitemShut {NoStop}%
\bibitem [{\citenamefont {Gillespie}\ and\ \citenamefont
  {Golightly}(2019)}]{2019.Gillespie}%
  \BibitemOpen
  \bibfield  {author} {\bibinfo {author} {\bibfnamefont {C.~S.}\ \bibnamefont
  {Gillespie}}\ and\ \bibinfo {author} {\bibfnamefont {A.}~\bibnamefont
  {Golightly}},\ }\bibfield  {title} {\bibinfo {title} {{Guided proposals for
  efficient weighted stochastic simulation}},\ }\href
  {https://doi.org/10.1063/1.5090979} {\bibfield  {journal} {\bibinfo
  {journal} {The Journal of Chemical Physics}\ }\textbf {\bibinfo {volume}
  {150}},\ \bibinfo {pages} {224103} (\bibinfo {year} {2019})}\BibitemShut
  {NoStop}%
\bibitem [{\citenamefont {Crooks}(2000)}]{2000.Crooks}%
  \BibitemOpen
  \bibfield  {author} {\bibinfo {author} {\bibfnamefont {G.~E.}\ \bibnamefont
  {Crooks}},\ }\bibfield  {title} {\bibinfo {title} {{Path-ensemble averages in
  systems driven far from equilibrium}},\ }\href
  {https://doi.org/10.1103/physreve.61.2361} {\bibfield  {journal} {\bibinfo
  {journal} {Physical Review E}\ }\textbf {\bibinfo {volume} {61}},\ \bibinfo
  {pages} {2361} (\bibinfo {year} {2000})},\ \Eprint
  {https://arxiv.org/abs/cond-mat/9908420} {cond-mat/9908420} \BibitemShut
  {NoStop}%
\bibitem [{\citenamefont {Andrieu}\ \emph {et~al.}(2010)\citenamefont
  {Andrieu}, \citenamefont {Doucet},\ and\ \citenamefont
  {Holenstein}}]{2010.Andrieu}%
  \BibitemOpen
  \bibfield  {author} {\bibinfo {author} {\bibfnamefont {C.}~\bibnamefont
  {Andrieu}}, \bibinfo {author} {\bibfnamefont {A.}~\bibnamefont {Doucet}},\
  and\ \bibinfo {author} {\bibfnamefont {R.}~\bibnamefont {Holenstein}},\
  }\bibfield  {title} {\bibinfo {title} {{Particle Markov chain Monte Carlo
  methods}},\ }\href {https://doi.org/10.1111/j.1467-9868.2009.00736.x}
  {\bibfield  {journal} {\bibinfo  {journal} {Journal of the Royal Statistical
  Society: Series B (Statistical Methodology)}\ }\textbf {\bibinfo {volume}
  {72}},\ \bibinfo {pages} {269} (\bibinfo {year} {2010})}\BibitemShut
  {NoStop}%
\bibitem [{\citenamefont {Rao}\ and\ \citenamefont {Teh}(2013)}]{2013.Rao}%
  \BibitemOpen
  \bibfield  {author} {\bibinfo {author} {\bibfnamefont {V.}~\bibnamefont
  {Rao}}\ and\ \bibinfo {author} {\bibfnamefont {Y.~W.}\ \bibnamefont {Teh}},\
  }\bibfield  {title} {\bibinfo {title} {{Fast MCMC sampling for Markov jump
  processes and extensions}},\ }\href {http://jmlr.org/papers/v14/rao13a.html}
  {\bibfield  {journal} {\bibinfo  {journal} {Journal of Machine Learning
  Research}\ }\textbf {\bibinfo {volume} {14}},\ \bibinfo {pages} {3295}
  (\bibinfo {year} {2013})}\BibitemShut {NoStop}%
\bibitem [{\citenamefont {Sourjik}\ and\ \citenamefont
  {Berg}(2002{\natexlab{a}})}]{2002.Sourjik}%
  \BibitemOpen
  \bibfield  {author} {\bibinfo {author} {\bibfnamefont {V.}~\bibnamefont
  {Sourjik}}\ and\ \bibinfo {author} {\bibfnamefont {H.~C.}\ \bibnamefont
  {Berg}},\ }\bibfield  {title} {\bibinfo {title} {{Receptor sensitivity in
  bacterial chemotaxis}},\ }\href {https://doi.org/10.1073/pnas.011589998}
  {\bibfield  {journal} {\bibinfo  {journal} {Proceedings of the National
  Academy of Sciences}\ }\textbf {\bibinfo {volume} {99}},\ \bibinfo {pages}
  {123} (\bibinfo {year} {2002}{\natexlab{a}})}\BibitemShut {NoStop}%
\bibitem [{\citenamefont {Sourjik}\ and\ \citenamefont
  {Berg}(2002{\natexlab{b}})}]{2002.Sourjikt5}%
  \BibitemOpen
  \bibfield  {author} {\bibinfo {author} {\bibfnamefont {V.}~\bibnamefont
  {Sourjik}}\ and\ \bibinfo {author} {\bibfnamefont {H.~C.}\ \bibnamefont
  {Berg}},\ }\bibfield  {title} {\bibinfo {title} {{Binding of the Escherichia
  coli response regulator CheY to its target measured in vivo by fluorescence
  resonance energy transfer}},\ }\href {https://doi.org/10.1073/pnas.192463199}
  {\bibfield  {journal} {\bibinfo  {journal} {Proceedings of the National
  Academy of Sciences}\ }\textbf {\bibinfo {volume} {99}},\ \bibinfo {pages}
  {12669} (\bibinfo {year} {2002}{\natexlab{b}})}\BibitemShut {NoStop}%
\bibitem [{\citenamefont {Sourjik}\ and\ \citenamefont
  {Berg}(2004)}]{2004.Sourjik}%
  \BibitemOpen
  \bibfield  {author} {\bibinfo {author} {\bibfnamefont {V.}~\bibnamefont
  {Sourjik}}\ and\ \bibinfo {author} {\bibfnamefont {H.~C.}\ \bibnamefont
  {Berg}},\ }\bibfield  {title} {\bibinfo {title} {{Functional interactions
  between receptors in bacterial chemotaxis}},\ }\href
  {https://doi.org/10.1038/nature02406} {\bibfield  {journal} {\bibinfo
  {journal} {Nature}\ }\textbf {\bibinfo {volume} {428}},\ \bibinfo {pages}
  {437} (\bibinfo {year} {2004})}\BibitemShut {NoStop}%
\bibitem [{\citenamefont {Morton-Firth}\ \emph {et~al.}(1999)\citenamefont
  {Morton-Firth}, \citenamefont {Shimizu},\ and\ \citenamefont
  {Bray}}]{1999.Morton-Firth}%
  \BibitemOpen
  \bibfield  {author} {\bibinfo {author} {\bibfnamefont {C.~J.}\ \bibnamefont
  {Morton-Firth}}, \bibinfo {author} {\bibfnamefont {T.~S.}\ \bibnamefont
  {Shimizu}},\ and\ \bibinfo {author} {\bibfnamefont {D.}~\bibnamefont
  {Bray}},\ }\bibfield  {title} {\bibinfo {title} {{A free-energy-based
  stochastic simulation of the tar receptor complex}},\ }\href
  {https://doi.org/10.1006/jmbi.1999.2535} {\bibfield  {journal} {\bibinfo
  {journal} {Journal of Molecular Biology}\ }\textbf {\bibinfo {volume}
  {286}},\ \bibinfo {pages} {1059} (\bibinfo {year} {1999})}\BibitemShut
  {NoStop}%
\bibitem [{\citenamefont {Ortega}\ \emph {et~al.}(2013)\citenamefont {Ortega},
  \citenamefont {Yang}, \citenamefont {Ames}, \citenamefont {Baudry},
  \citenamefont {Parkinson},\ and\ \citenamefont {Zhulin}}]{2013.Ortega}%
  \BibitemOpen
  \bibfield  {author} {\bibinfo {author} {\bibfnamefont {D.~R.}\ \bibnamefont
  {Ortega}}, \bibinfo {author} {\bibfnamefont {C.}~\bibnamefont {Yang}},
  \bibinfo {author} {\bibfnamefont {P.}~\bibnamefont {Ames}}, \bibinfo {author}
  {\bibfnamefont {J.}~\bibnamefont {Baudry}}, \bibinfo {author} {\bibfnamefont
  {J.~S.}\ \bibnamefont {Parkinson}},\ and\ \bibinfo {author} {\bibfnamefont
  {I.~B.}\ \bibnamefont {Zhulin}},\ }\bibfield  {title} {\bibinfo {title} {{A
  phenylalanine rotameric switch for signal-state control in bacterial
  chemoreceptors}},\ }\href {https://doi.org/10.1038/ncomms3881} {\bibfield
  {journal} {\bibinfo  {journal} {Nature Communications}\ }\textbf {\bibinfo
  {volume} {4}},\ \bibinfo {pages} {2881} (\bibinfo {year} {2013})}\BibitemShut
  {NoStop}%
\bibitem [{\citenamefont {Tostevin}\ and\ \citenamefont {ten
  Wolde}(2010)}]{2010.Tostevin}%
  \BibitemOpen
  \bibfield  {author} {\bibinfo {author} {\bibfnamefont {F.}~\bibnamefont
  {Tostevin}}\ and\ \bibinfo {author} {\bibfnamefont {P.~R.}\ \bibnamefont {ten
  Wolde}},\ }\bibfield  {title} {\bibinfo {title} {{Mutual information in
  time-varying biochemical systems}},\ }\href
  {https://doi.org/10.1103/physreve.81.061917} {\bibfield  {journal} {\bibinfo
  {journal} {Physical Review E}\ }\textbf {\bibinfo {volume} {81}},\ \bibinfo
  {pages} {061917} (\bibinfo {year} {2010})},\ \Eprint
  {https://arxiv.org/abs/1002.4273} {1002.4273} \BibitemShut {NoStop}%
\end{thebibliography}%

\end{document}